{}
{}

\documentclass[11pt,a4paper]{article}

\newif\ifpublic\publictrue
\newif\iffancy\fancytrue

\usepackage[a4paper,text={160mm,247mm},centering]{geometry}
\usepackage{setspace}
\usepackage{rotating}
\usepackage{amsmath,amssymb, amsfonts, geometry}
\makeatletter
\providecommand*{\shuffle}{%
  \mathbin{\mathpalette\shuffle@{}}%
}
\newcommand*{\shuffle@}[2]{%
  \sbox0{$#1\vcenter{}$}%
  \kern .15\ht0 
  \rlap{\vrule height .25\ht0 depth 0pt width 2.5\ht0}%
  \raise.1\ht0\hbox to 2.5\ht0{%
    \vrule height 1.75\ht0 depth -.1\ht0 width .17\ht0 %
    \hfill
    \vrule height 1.75\ht0 depth -.1\ht0 width .17\ht0 %
    \hfill
    \vrule height 1.75\ht0 depth -.1\ht0 width .17\ht0 %
  }%
  \kern .15\ht0 
}
\makeatother

\usepackage{xparse}

\ExplSyntaxOn
\NewDocumentCommand{\Tgen}{m m}
 {
  \TL\left[\begin{smallmatrix}
  \Tgen_print:n {#1} \\
  \Tgen_print:n {#2}
  \end{smallmatrix}\right]
 }

\seq_new:N \l_Tgen_list_seq

\cs_new_protected:Npn \Tgen_print:n #1
{
  \seq_set_split:Nnn \l_Tgen_list_seq { , } { #1 }
  \seq_use:Nn \l_Tgen_list_seq { , & }
}
\ExplSyntaxOff

\ExplSyntaxOn
\NewDocumentCommand{\omwb}{m m}
{
 \omm\left(\begin{smallmatrix}
 \omwb_print:n {#1} \\
 \omwb_print:n {#2}
 \end{smallmatrix}\right)
}
\seq_new:N \l_omwb_list_seq
\cs_new_protected:Npn \omwb_print:n #1
{
  \seq_set_split:Nnn \l_omwb_list_seq { , } { #1 }
  \seq_use:Nn \l_omwb_list_seq { , & }
}
\ExplSyntaxOff

\ExplSyntaxOn
\NewDocumentCommand{\omwbc}{m m}
{
 \ommz\left(\begin{smallmatrix}
 \omwbc_print:n {#1} \\
 \omwbc_print:n {#2}
 \end{smallmatrix}\right)
}
\seq_new:N \l_omwbc_list_seq
\cs_new_protected:Npn \omwbc_print:n #1
{
  \seq_set_split:Nnn \l_omwbc_list_seq { , } { #1 }
  \seq_use:Nn \l_omwbc_list_seq { , & }
}
\ExplSyntaxOff

\ExplSyntaxOn
\NewDocumentCommand{\cgomwb}{m m}
{
 c\left(\begin{smallmatrix}
 \cgomwb_print:n {#1} \\
 \cgomwb_print:n {#2}
 \end{smallmatrix}\right)
}
\seq_new:N \l_cgomwb_list_seq
\cs_new_protected:Npn \cgomwb_print:n #1
{
  \seq_set_split:Nnn \l_cgomwb_list_seq { , } { #1 }
  \seq_use:Nn \l_cgomwb_list_seq { , & }
}
\ExplSyntaxOff

\ExplSyntaxOn
\NewDocumentCommand{\TgenR}{m m}
{
  \TL^R\left[\begin{smallmatrix}
 \TgenR_print:n {#1} \\
 \TgenR_print:n {#2}
 \end{smallmatrix}\right]
}
\seq_new:N \l_TgenR_list_seq
\cs_new_protected:Npn \TgenR_print:n #1
{
  \seq_set_split:Nnn \l_TgenR_list_seq { , } { #1 }
  \seq_use:Nn \l_TgenR_list_seq { , & }
}
\ExplSyntaxOff

\ExplSyntaxOn
\NewDocumentCommand{\TgenRE}{m m}
{
 \TRE\left[\begin{smallmatrix}
 \TgenRE_print:n {#1} \\
 \TgenRE_print:n {#2}
 \end{smallmatrix}\right]
}
\seq_new:N \l_TgenRE_list_seq
\cs_new_protected:Npn \TgenRE_print:n #1
{
  \seq_set_split:Nnn \l_TgenRE_list_seq { , } { #1 }
  \seq_use:Nn \l_TgenRE_list_seq { , & }
}
\ExplSyntaxOff

\ExplSyntaxOn
\NewDocumentCommand{\TgenTRE}{m m}
{
 \TTRE\left[\begin{smallmatrix}
 \TgenTRE_print:n {#1} \\
 \TgenTRE_print:n {#2}
 \end{smallmatrix}\right]
}
\seq_new:N \l_TgenTRE_list_seq
\cs_new_protected:Npn \TgenTRE_print:n #1
{
  \seq_set_split:Nnn \l_TgenTRE_list_seq { , } { #1 }
  \seq_use:Nn \l_TgenTRE_list_seq { , & }
}
\ExplSyntaxOff

\ExplSyntaxOn
\NewDocumentCommand{\TgenTREz}{m m m}
{
 \TTRE\left[\begin{smallmatrix}
 \TgenTREz_print:n {#1} \\
 \TgenTREz_print:n {#2}
 \end{smallmatrix};#3\right]
}
\seq_new:N \l_TgenTREz_list_seq
\cs_new_protected:Npn \TgenTREz_print:n #1
{
  \seq_set_split:Nnn \l_TgenTREz_list_seq { , } { #1 }
  \seq_use:Nn \l_TgenTREz_list_seq { , & }
}
\ExplSyntaxOff

\ExplSyntaxOn
\NewDocumentCommand{\BZ}{m m}
{
 \zeta\left(\begin{smallmatrix}
 \BZ_print:n {#1} \\
 \BZ_print:n {#2}
 \end{smallmatrix}\right)
}
\seq_new:N \l_BZ_list_seq
\cs_new_protected:Npn \BZ_print:n #1
{
  \seq_set_split:Nnn \l_BZ_list_seq { , } { #1 }
  \seq_use:Nn \l_BZ_list_seq { , & }
}
\ExplSyntaxOff

\usepackage[pdfencoding=auto,bookmarks=true,hyperfigures=true]{hyperref}
\usepackage{graphicx}
\usepackage{float}
\usepackage[nosort]{cite}
\usepackage{lmodern}
\usepackage{amsbsy}
\usepackage[utf8]{inputenc}
\usepackage[font=small,labelfont=bf]{caption}
\usepackage{diagbox}
\usepackage{array}

\makeatletter
\g@addto@macro\bfseries{\boldmath}
\makeatother
\usepackage[usenames,dvipsnames]{xcolor}
\definecolor{dgreen}{rgb}{0,0.70,0.30}
\definecolor{gold}{rgb}{0.85,.66,0}
\definecolor{purple}{rgb}{1.0,0.3,0.6}

\usepackage{tikz}
\usetikzlibrary{calc} 
\usetikzlibrary{patterns} 
\usetikzlibrary{decorations.pathreplacing} 
\usetikzlibrary{decorations.markings} 
\usetikzlibrary{decorations.pathmorphing} 
\usetikzlibrary{positioning}
\usetikzlibrary{arrows.meta}

\makeatletter
\newsavebox{\apb@box}\newlength{\apb@width}
\newcommand{\autoparbox}[2][c]{\sbox{\apb@box}{#2}%
 \settowidth{\apb@width}{\usebox{\apb@box}}%
 \parbox[#1]{\apb@width}{\usebox{\apb@box}}}

\makeatother

\numberwithin{equation}{section}

\newcommand{\eqn}[1]{eq.~\eqref{#1}}
\newcommand{\Eqn}[1]{Equation~\eqref{#1}}
\newcommand{\eqns}[2]{eqs.~\eqref{#1} and~\eqref{#2}}

\providecommand{\href}[2]{#2}

\makeatletter
\def\mr@ignsp#1 {\ifx\:#1\@empty\else #1\expandafter\mr@ignsp\fi}%
\newcommand{\multiref}[1]{\begingroup
\xdef\mr@no@sparg{\expandafter\mr@ignsp#1 \: }%
\def\mr@comma{}%
\@for\mr@refs:=\mr@no@sparg\do{\mr@comma\def\mr@comma{,}\ref{\mr@refs}}%
\endgroup}
\renewcommand{\eqref}[1]{(\multiref{#1})}
\makeatother

\makeatletter
\newcommand{\namedref}[2]{\hyperref[#2]{#1~\ref*{#2}}}
\newcommand{\secref}{\@ifstar{\namedref{Section}}{\namedref{section}}}
\newcommand{\subsecref}{\@ifstar{\namedref{Subsection}}{\namedref{subsection}}}
\newcommand{\appref}{\@ifstar{\namedref{Appendix}}{\namedref{appendix}}}
\newcommand{\tabref}{\@ifstar{\namedref{Table}}{\namedref{table}}}
\newcommand{\figref}{\@ifstar{\namedref{Figure}}{\namedref{figure}}}
\makeatother

\newcommand{\rcite}[1]{ref.~\cite{#1}}
\newcommand{\rcites}[1]{refs.~\cite{#1}}

\providecommand{\hypersetup}[1]{}

\hypersetup{plainpages=false}
\hypersetup{pdfpagemode=UseNone}
\hypersetup{bookmarksnumbered=true}
\hypersetup{pdfstartview=FitH}
\hypersetup{colorlinks=false}
\hypersetup{citebordercolor={.5 1 .5}}
\hypersetup{urlbordercolor={.5 1 1}}
\hypersetup{linkbordercolor={1 .7 .7}}

\makeatletter
\let\@keywords\@empty
\let\@subject\@empty
\providecommand{\keywords}[1]{\gdef\@keywords{#1}}
\providecommand{\subject}[1]{\gdef\@subject{#1}}
\def\thetitle{\@title}
\def\theauthor{\@author}
\def\thesubject{\@subject}
\def\thedate{\@date}
\def\thekeywords{\@keywords}
\makeatother
\AtBeginDocument{
\hypersetup{pdftitle={\thetitle}}%
\hypersetup{pdfauthor={\theauthor}}%
\hypersetup{pdfsubject={\thesubject}}%
\hypersetup{pdfkeywords={\thekeywords}}%
}

\RequirePackage{verbatim}

\newif\ifnote 
\notetrue

\allowdisplaybreaks

\newcommand{\ad} {\mathrm{ad}}
\def\Re{{\rm Re\,}}
\def\Im{{\rm Im\,}}

\newcommand{\pd}{\partial}

\newcommand{\ve}{\varepsilon}

\newcommand{\om}{\omega}

\newcommand{\SL}{\mathrm{SL}}

\newcommand{\te}{\textrm}

\newcommand{\co}{\ , \ \ \ \ \ \ }
\newcommand{\dd}{\mathrm{d}}

\newcommand{\ap}{\alpha'}

\newcommand{\nbeta}{b}

\newcommand{\ZR}{\mathbb R}
\newcommand{\ZC}{\mathbb C}
\newcommand{\ZH}{\mathbb H}
\newcommand{\ZN}{\mathbb N}
\newcommand{\ZZ}{\mathbb Z}
\newcommand{\ZQ}{\mathbb Q}

\newcommand{\nnl}{\nonumber\\}

\DeclareMathOperator{\TL}{T}
\DeclareMathOperator{\TRE}{T^\textit{R}_\ve}
\DeclareMathOperator{\TTRE}{\tilde{T}^\textit{R}_\ve}

\DeclareMathOperator{\GL}{\Gamma}

\DeclareMathOperator{\GGs}{G}

\DeclareMathOperator{\zm}{\zeta}

\DeclareMathOperator{\omm}{\omega}
\DeclareMathOperator{\ommz}{\omm_0}

\newcommand{\GG}[1]{\GGs_{#1}}

\newcommand{\GLargz}[2]{\GL\left(\begin{smallmatrix}#1\\#2\end{smallmatrix};z\right)}
\newcommand{\GLarg}[3]{\GL\left(\begin{smallmatrix}#1\\#2\end{smallmatrix};#3\right)}

\usepackage{nicefrac}
\newcommand{\tauh}{\nicefrac{\tau}{2}}
\newcommand{\oneh}{\nicefrac{1}{2}}
\newcommand{\oned}{\nicefrac{1}{3}}
\newcommand{\onezd}{\nicefrac{2}{3}}
\newcommand{\taud}{\nicefrac{\tau}{3}}
\newcommand{\tauv}{\nicefrac{\tau}{4}}
\newcommand{\tauf}{\nicefrac{\tau}{5}}
\newcommand{\tauzf}{\nicefrac{2\tau}{5}}

\newcommand{\temzv}{teMZV}

\vspace*{1.8cm}
\title{\textbf{Twisted elliptic multiple zeta values \\ 
and non-planar one-loop open-string amplitudes}}
\author{Johannes Broedel$^{\te{a}}$, Nils Matthes$^{\te{b}}$, Gregor Richter$^{\te{a,c}}$,
Oliver Schlotterer$^{\te{c}}$}
\date{\today}

\begin{document}
\pdfbookmark[1]{Title Page}{title} \thispagestyle{empty}
\begin{flushright}
  \verb!HU-EP-17/11!\\
  \verb!HU-Mathematik-2017-2!
\end{flushright}
\vspace*{0.4cm}
\begin{center}%
  \begingroup\LARGE\bfseries\thetitle\par\endgroup
  \vspace{1.4cm}

\begingroup\large\theauthor\par\endgroup
\vspace{8mm}
\begingroup\itshape
$^{\te{a}}$Institut f\"ur Mathematik und Institut f\"ur Physik,\\
Humboldt-Universit\"at zu Berlin\\
IRIS Adlershof, Zum Gro\ss{}en Windkanal 6, 12489 Berlin, Germany
\par\endgroup
\vspace{4mm}
\begingroup\itshape
$^{\te{b}}$Max-Planck-Institut f\"ur Mathematik,\\
Vivatsgasse 7, 53111 Bonn, Germany
\par\endgroup
\vspace{4mm}
\begingroup\itshape
$^{\te{c}}$Max-Planck-Institut f\"ur Gravitationsphysik,\\
Albert-Einstein-Institut\\
Am M\"uhlenberg 1, 14476 Potsdam, Germany
\par\endgroup

\vspace{0.8cm}

\begingroup\ttfamily
jbroedel@physik.hu-berlin.de, 
nilsmath@mpim-bonn.mpg.de,\\
grichter@physik.hu-berlin.de, 
olivers@aei.mpg.de 
\par\endgroup

\vspace{1.2cm}

\bigskip

\textbf{Abstract}\vspace{5mm}

\begin{minipage}{13.4cm}
We consider a generalization of elliptic multiple zeta values, which we call
twisted elliptic multiple zeta values.  These arise as iterated integrals on an
elliptic curve from which a rational lattice has been removed.  At the cusp,
twisted elliptic multiple zeta values are shown to degenerate to cyclotomic
multiple zeta values in the same way as elliptic multiple zeta values
degenerate to classical multiple zeta values.  We investigate properties of
twisted elliptic multiple zeta values and utilize them in the
evaluation of the non-planar part of the four-point one-loop open-
superstring amplitude.
\end{minipage}

\vspace*{4cm}

\end{center}

\newpage

\setcounter{tocdepth}{2}
\tableofcontents


\section{Introduction}

Within the program of studying iterated integrals on Riemann surfaces of
various genera, the genus-zero case, which leads to multiple zeta values
(MZVs)~\cite{Hoffman:MHS,Zagier23,Goncharov:2001iea}, is the starting point and
takes the most prominent r\^{o}le.  During the last years, however, the
genus-one situation has received more attention: various iterated integrals on
the elliptic curve as well as associated periods and elliptic associators have
been investigated \cite{LevinRacinet,KZB,BrownLev,Enriquez:EllAss}.  

\begin{figure}[h]
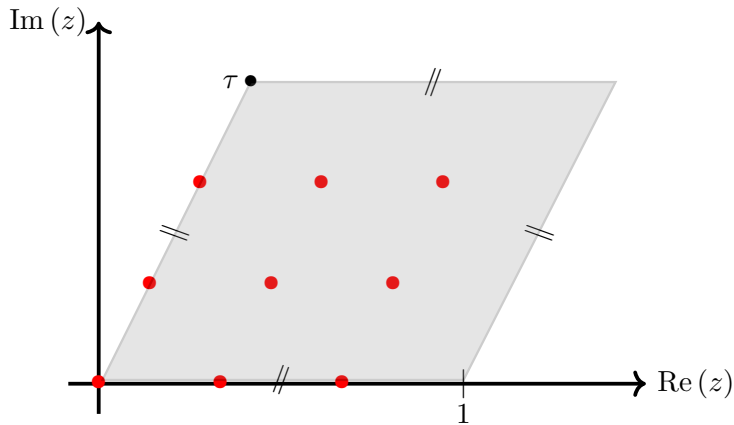

\begin{center}
\tikzpicture[line width=0.5mm]
\draw[->] (-0.4,0) -- (7.2,0)node[right]{$\Re(z)$};
\draw[->] (0,-0.4) -- (0,4.8)node[left]{$\Im(z)$};
\draw[red] (0,0.02)node{$\boldsymbol{\bullet}$} ;
\draw[red] (1.6,0.02)node{$\boldsymbol{\bullet}$} ;
\draw[red] (3.2,0.02)node{$\boldsymbol{\bullet}$} ;
\draw[red] (0.67,1.33)node{$\boldsymbol{\bullet}$} ;
\draw[red] (2.27,1.33)node{$\boldsymbol{\bullet}$} ;
\draw[red] (3.87,1.33)node{$\boldsymbol{\bullet}$} ;
\draw[red] (1.33,2.67)node{$\boldsymbol{\bullet}$} ;
\draw[red] (2.93,2.67)node{$\boldsymbol{\bullet}$} ;
\draw[red] (4.53,2.67)node{$\boldsymbol{\bullet}$} ;
\draw (4.8,0)node{$|$};
\draw (4.8,-0.4)node{$1$};
\draw (1,2) node{\rotatebox[origin=c]{-110}{$| \hspace{-1pt} |$}};
\draw (5.8,2) node{\rotatebox[origin=c]{-110}{$| \hspace{-1pt} |$}};
\draw (2.4,0.05) node{\rotatebox[origin=c]{-20}{$| \hspace{-1pt} |$}};
\draw (4.4,4) node{\rotatebox[origin=c]{-20}{$| \hspace{-1pt} |$}};
\draw(2,4)node{$\bullet$} node[left]{$\tau$};
\coordinate (AA) at (0.05,0.05);
\coordinate (BB) at (4.8,0.05);
\coordinate (CC) at (6.8,4);
\coordinate (DD) at (2,4);
\draw[fill=gray, opacity=0.2,line width=0.3mm] (AA) -- (BB) -- (CC) -- (DD) -- cycle;
\endtikzpicture
\caption{The shaded region represents the elliptic curve $\ZC / (\ZZ+\ZZ\tau)$,
  where edges marked by 
  \protect\rotatebox[origin=c]{-20}{$| \hspace{-1pt} |$} and
  \protect\rotatebox[origin=c]{-110}{$| \hspace{-1pt} |$} 
  are identified. In the setup of teMZVs, points from the lattice $\ZQ+\ZQ\tau$
are removed from the elliptic curve. Here we show the example
$\{0,\frac{1}{3},\frac{2}{3} \}+\{0,\frac{1}{3},\frac{2}{3} \}\tau$.} 
\label{fig:intro}
\end{center}
\end{figure}

The simplest genus-one generalizations of MZVs are elliptic multiple zeta values
(eMZVs), which arise from iterated integrals on the once-punctured elliptic
curve, that is the elliptic curve where the origin is removed
\cite{Enriquez:Emzv}. In this article, the notion of eMZVs is extended to
\textit{twisted} elliptic multiple zeta values (teMZVs), which are iterated
integrals on a multiply-punctured elliptic curve. While for eMZVs it is
sufficient to remove the origin, teMZVs arise when a lattice with rational
coordinates as visualized in \figref{fig:intro} is removed from the elliptic
curve. 

The iterated integrals to be considered are performed over a path parallel to
the real axis and are therefore a generalization of Enriquez' $A$-cycle eMZVs.
If no lattice is removed, we obtain A-cycle eMZVs by definition.  \footnote{It
  is worth noting that there is no structural problem in defining twisted
$B$-cycle eMZVs by considering integration paths parallel to the direction of
the modular parameter $\tau$ \cite{Calaque}.} A slight technical difficulty,
which was absent for eMZVs, is that the integrands giving rise to teMZVs might
have additional poles along the path of integration. We address the problem by
suggesting a rather natural regularization scheme, which essentially amounts to
integrating over an infinitesimal deformation of the real axis.

A crucial tool in the study of eMZVs was the existence of a certain first-order
linear differential equation, expressing eMZVs as special linear combinations
of iterated integrals of Eisenstein series and MZVs
\cite{Enriquez:EllAss,Enriquez:Emzv,Broedel:2015hia, Matthes:Thesis}. In
particular, this is an efficient way to compute their $q$-expansion, which is
instrumental for finding (as well as excluding) linear relations among eMZVs.
Even more so, since iterated integrals of Eisenstein series are linearly
independent \cite{LMS}, the differential equation amounts to a decomposition of
eMZVs into basic constituents, which reduces the study of relations among eMZVs
to solving linear systems of equations. One of the main results of this article
is the generalization of this differential equation to teMZVs. We find that the
occurrence of the classical Eisenstein series (for $\SL_2(\ZZ)$) in the
differential equation for eMZVs naturally generalizes to an occurrence of
certain weighting functions $f^{(n)}(s+r\tau,\tau)$, where $r,s \in \ZQ$, and
the latter are also known to be modular forms for congruence subgroups of
$\SL_2(\ZZ)$. Likewise, one can again identify a procedure delivering the
boundary data for teMZVs at the cusp $i\infty$ of the modular parameter $\tau$,
and relating them to integrals over genus-zero Riemann surfaces in a natural
way.  While for eMZVs this procedure leads to MZVs, in the case of teMZVs we
obtain \textit{cyclotomic} MZVs
\cite{Goncharov:2001iea,Racinet:Doubles,Zhao:2008,Glanois:Thesis}.  A further
parallel to eMZVs is the existence of shuffle and Fay relations
\cite{Broedel:2015hia,Matthes:edzv}. 

Scattering amplitudes in open-superstring theories have been recently noticed
as a rewarding setup where iterated integrals on Riemann surfaces appear
naturally. Generalizing the ubiquity of MZVs in tree-level
amplitudes\footnote{See \cite{Aomoto, Terasoma, Brown:0606} for a discussion of
  the contributing iterated integrals on a genus-zero surface in the
  mathematics literature and \cite{Broedel:2013tta, Mafra:2016mcc} for a
  treatment via polylogarithms in a physics context.  Moreover, the expansion
  of $n$-point disk integrals has been addressed via motivic MZVs
  \cite{Schlotterer:2012ny} and the Drinfeld associator \cite{Broedel:2013aza}
  (also see \cite{Drummond:2013vz}).
As a complementary approach, the relation of disk integrals to hypergeometric
functions has been used to obtain $(n\leq5)$-point expansions
\cite{Boels:2013jua, Puhlfuerst:2015gta} and certain ranges of low-energy
orders at $n\leq 7$ points, see e.g.\ \cite{Oprisa:2005wu, Stieberger:2006te,
Stieberger:2007jv}.}, one-loop scattering amplitudes (corresponding to
genus-one surfaces) provide a natural testing ground for eMZVs
\cite{Broedel:2014vla}. However, the analysis in \rcite{Broedel:2014vla} was
focused on the \textit{planar} sector of the one-loop amplitude where the
integrations are performed over a single boundary of a genus-one surface with
cylinder topology.

In this article, teMZVs will be identified as the suitable language for the
calculation of the \textit{non-planar} part of the open-string one-loop
amplitude: the extension of the iterated integrals to both boundaries of the
cylinder leads to the class of teMZVs with twist $\tauh$.  We will employ these
teMZVs to calculate non-planar contributions to the low-energy
expansion\footnote{The low-energy expansion of string amplitudes refers to an
expansion in the inverse string tension $\ap$.} of the four-point one-loop
open-string amplitude. Explicit results will be given up to the third
subleading low-energy order which are checked to match the expressions
available in the literature at the first subleading low-energy order
\cite{Hohenegger:2017kqy} and at the cusp \cite{Green:1981ya}. As in the
planar case, their efficient computation crucially relies on the differential
equation satisfied by teMZVs.

Interestingly, our final result for the non-planar part of string scattering
amplitudes in the cases considered can be expressed in terms of eMZVs alone,
i.e.~teMZVs with twist $\tauh$ cancel out. This observation relies on providing
explicit formulas for certain linear combinations of teMZVs with twist $\tauh$
in terms of eMZVs, which can in turn be checked using their differential
equation. We will provide physics arguments bolstering the conjecture that this
feature will persist to all orders in the low-energy expansion, and it would be
very interesting to find a mathematical explanation for this effect.

Finally, we expect the teMZVs defined here to be closely related to the
monodromy of the universal twisted elliptic KZB equation to be studied in upcoming work
of Calaque and Gonzalez \cite{Calaque}. A particularly important aspect of
their work is the definition of a twisted version of the derivation algebra,
the untwisted version of which \cite{Tsunogai,KZB,Pollack,HM} already appeared
in the study of eMZVs
\cite{Broedel:2015hia,Matthes:Decomposition,Matthes:Thesis}. Similar to the
situation for eMZVs, this twisted derivation algebra might be capable of encoding
the number of indecomposable teMZVs of a given weight and length. 

In \secref{sec:npe} we introduce teMZVs, and discuss the expansion of their
constituents with respect to the modular parameter of the elliptic curve.
Thereby we set the stage for \secref{sec:new}, where a differential equation
for teMZVs w.r.t.~$\tau$ as well as a procedure to extract their $\tau
\rightarrow i \infty$ limit is presented. In \secref{sec:oneloop}, the
formalism is applied to the calculation of the non-planar contribution to the
open-string one-loop scattering amplitude and the r\^{o}le of teMZVs therein is
discussed.  After concluding and pointing out a couple of open problems in
\secref{sec:conclusion} we provide various appendices containing collections of
definitions for the numerous objects appearing as well as several detailed
calculations omitted in the main text. 


\section{From elliptic to twisted elliptic multiple zeta values}
\label{sec:npe}
Elliptic multiple zeta values can be represented as iterated integrals on the
multiply punctured elliptic curve $
\ZC / (\ZZ+\ZZ\tau) \setminus \lbrace b_1 , \dots , b_\ell \rbrace$ with
parameter $\tau$ in the upper half plane $\ZH$, where we denote $q = \exp(2 \pi
i \tau)$. 
Starting from $\GL(;z) = 1$, elliptic iterated integrals are defined
recursively via 
\begin{equation}
  \GLargz{n_1 &n_2 &\ldots &n_\ell}{b_1 &b_2 &\ldots &b_\ell} =
  \int^z_0 \dd t \, f^{(n_1)}(t-b_1) \,
  \GLarg{n_2 &\ldots &n_\ell}{b_2 &\ldots &b_\ell}{t},
  \qquad z \in [0,1]\,,
  \label{eqn:defGell}
\end{equation}
where the interval of integration is $[0,z]$. As will be discussed in
\subsecref{ssec:regteMZVs}, regularization prescriptions have to be
specified, if $n_i=1$ and $b_i\in[0,z]$.

The weighting functions $f^{(n)}(z,\tau)$ arise as expansion coefficients of
the doubly-periodic completion of the Eisenstein--Kronecker series, starting with
\begin{equation}
 f^{(0)}(z,\tau)=1 \, ,
\qquad
  f^{(1)}(z,\tau)= \frac{ \theta_1'(z,\tau)}{\theta_1(z,\tau)} + 2\pi i \frac{ \Im(z) }{\Im (\tau)}
  \, ,
  \label{eqn:expl}
\end{equation}
see \appref{app:weighting} for details and conventions.  They are doubly-periodic
functions of alternating parity
\begin{equation}
 f^{(n)}(z+1,\tau)=
 f^{(n)}(z+\tau,\tau)=
 f^{(n)}(z,\tau)\, ,
\qquad
  f^{(n)}(-z,\tau)=(-1)^{n}f^{(n)}(z,\tau) \, ,
  \label{eqn:fparity1}
\end{equation}
and the function $f^{(1)}(z-b_i,\tau)$ in \eqn{eqn:expl} acquires a pole at
$z=b_i$ which requires regularization of \eqn{eqn:defGell}. Throughout the
article, we will frequently omit noting the $\tau$-dependence of both weighting
functions $f^{(n)}$ and elliptic iterated integrals \eqn{eqn:defGell}. 

In \rcites{Broedel:2014vla,Broedel:2015hia}, the main focus was on elliptic
multiple zeta values, whose shifting parameters~$b_i$ -- referred to as
\textit{twists} -- have been limited to $b_i=0$.
Correspondingly, the elliptic curve in question has a single puncture only:
$E_\tau^\times = \ZC / (\ZZ+\ZZ\tau) \setminus \{0\}$.  Evaluating this
subclass of elliptic iterated integrals at $z=1$ leads to the definition of
Enriquez' $A$-cycle elliptic multiple zeta values or eMZVs for short:
\begin{align}
  \label{eqn:defeMZV}
\omm(n_1,n_2,\ldots,n_\ell) &= \! \! \! \! \! \! \int \limits_{0 \leq z_i \leq  z_{i+1} \leq 1}  \! \! \! \! \! \!
f^{(n_1)}(z_1) \dd z_1\, f^{(n_2)}(z_2)\dd z_2\, \ldots f^{(n_\ell)}(z_\ell) \dd z_\ell \\
&= 
\GLarg{n_\ell &\ldots & n_2 & n_1}{0 &\ldots & 0 & 0}{1} = \GL(n_\ell  ,\ldots ,n_2 ,n_1;1) \, .\notag
\end{align}
The quantities $w=\sum_{i=1}^\ell n_i$, and the number $\ell$ of integrations in
\eqns{eqn:defGell}{eqn:defeMZV} are referred to as \textit{weight} and \textit{length} of the
elliptic iterated integral and of the corresponding eMZV, respectively.  

Allowing for rational values $s_i$ and $r_i$ in $b_i=s_i+r_i\tau$, leads to
\textit{twisted} elliptic multiple zeta values or teMZVs: 
\begin{align}
\label{eqn:defteMZV}
  	\omwb{n_1,n_2,\ldots,n_\ell}{\nbeta_1,\nbeta_2,\ldots,\nbeta_\ell}  &= \! \! \! \! \! \! \int \limits_{0\leq z_i \leq z_{i+1} \leq 1}  \! \! \! \! \! \!
   	f^{(n_1)}(z_1-\nbeta_1) \dd z_1\, f^{(n_2)}(z_2-\nbeta_2)\dd z_2\, \ldots f^{(n_\ell)}(z_\ell-\nbeta_\ell) \dd z_\ell \nnl[3pt]
	&=\quad\GLarg{n_\ell & n_{\ell-1} & \ldots & n_1}{\nbeta_\ell & \nbeta_{\ell-1} & \ldots & \nbeta_1}{1} \,,
\end{align}
where the notions of \textit{weight} and \textit{length} carry over from
\eqn{eqn:defeMZV} directly.  Taking the double-periodicity \eqref{eqn:fparity1}
of the weighting functions $f^{(n)}$ into account, one can limit the attention
to the fundamental domain of the
elliptic curve with $r_i,s_i \in [0,1)$, i.e.\ the shaded region in
  \figref{fig:twistandshout}.

In this article we are going to limit our attention to twists $\ZQ+\ZQ\tau$,
that is $r_i,s_i\in\ZQ$. In order to classify those, let us introduce
\begin{equation}
\Lambda_N= \Big\{0, \frac 1N, \frac 2N,\ldots, \frac{N-1}{N} \Big\} \ , \ \ \ \ \ \ 
\Lambda_N^{\times} = \Lambda_N \setminus \{0\}
\ .
\label{deflambda}
\end{equation}
If $b_i\in\Lambda_N^\times$, the twist is referred to as \textit{proper
rational}. Correspondingly, all other twists -- that is those with
$b_i\in(\Lambda_N+\Lambda_N\tau)\setminus \Lambda_N^\times$ as visualized
in \figref{fig:twistandshout} -- are called
\textit{generic} twists. While in the latter situation divergences occur at
endpoints only and can be addressed using the methods in
\rcite{Enriquez:Emzv}, the presence of a proper rational twist requires more
work as discussed in \subsecref{ssec:regteMZVs}.

\begin{figure}
\begin{center}
\tikzpicture[line width=0.5mm]
\draw[->] (-0.4,0) -- (7.2,0)node[right]{$\Re(z)$};
\draw[->] (0,-0.4) -- (0,4.8)node[left]{$\Im(z)$};
%
%
\draw[blue] (0,0.02)node{$\boldsymbol{\bullet}$} ;
\draw[red] (1.6,0.02)node{$\boldsymbol{\bullet}$} ;
\draw[red] (3.2,0.02)node{$\boldsymbol{\bullet}$} ;
\draw[blue] (0.67,1.33)node{$\boldsymbol{\bullet}$} ;
\draw[blue] (2.27,1.33)node{$\boldsymbol{\bullet}$} ;
\draw[blue] (3.87,1.33)node{$\boldsymbol{\bullet}$} ;
\draw[blue] (1.33,2.67)node{$\boldsymbol{\bullet}$} ;
\draw[blue] (2.93,2.67)node{$\boldsymbol{\bullet}$} ;
\draw[blue] (4.53,2.67)node{$\boldsymbol{\bullet}$} ;
%
%
\draw (4.8,0)node{$|$};
\draw (4.8,-0.3)node{$1$};
%
%
\draw (1,2) node{\rotatebox[origin=c]{-110}{$| \hspace{-1pt} |$}};
\draw (5.8,2) node{\rotatebox[origin=c]{-110}{$| \hspace{-1pt} |$}};
\draw (2.4,0.05) node{\rotatebox[origin=c]{-20}{$| \hspace{-1pt} |$}};
\draw (4.4,4) node{\rotatebox[origin=c]{-20}{$| \hspace{-1pt} |$}};
\draw(2,4)node{$\bullet$} node[left]{$\tau$};
\coordinate (AA) at (0.05,0.05);
\coordinate (BB) at (4.8,0.05);
\coordinate (CC) at (6.8,4);
\coordinate (DD) at (2,4);
\draw[fill=gray, opacity=0.2,line width=0.3mm] (AA) -- (BB) -- (CC) -- (DD) -- cycle;
\endtikzpicture
\caption{Example of the lattice $\Lambda_N+\Lambda_N\tau$ at $N=3$: Proper
  rational twists and generic twists are marked in red and blue, respectively.
  Edges marked by \protect\rotatebox[origin=c]{-20}{$| \hspace{-1pt} |$} and
  \protect\rotatebox[origin=c]{-110}{$| \hspace{-1pt} |$} are identified in
$\ZC / (\ZZ+\ZZ\tau) $, respectively.} 
\label{fig:twistandshout}
\end{center}
\end{figure}
Twisted eMZVs based on proper rational twists do not make an appearance in the
open-string one-loop amplitude. However, they are interesting from a
number-theoretic point of view because their constant terms give rise to
cyclotomic generalizations of MZVs or ``cyclotomic MZVs'' for short
\cite{Goncharov:2001iea,Racinet:Doubles,Zhao:2008,Glanois:Thesis}.  The set of
(generic) twists $b_i\in\lbrace 0,\tau/2\rbrace$ turns out to lead to teMZVs
relevant for the non-planar open-string amplitude, which we are going to discuss
in \secref{sec:oneloop}.

\subsection{Regularization}
\label{ssec:regteMZVs}

\begin{figure}[h]
\begin{center}
\begin{tikzpicture}[line width=0.30mm]

\draw (0,0) -- (5,0);
\draw (0.75,1.75) -- (5.75,1.75);
\draw (0,0) -- (0.75,1.75);
\draw (5,0) -- (5.75,1.75);
\draw[below] (0,0) node{$0$};
\draw[below] (5,0) node{$1$};
\draw[above] (0.75,1.75) node{$\tau$};
\draw[above] (5.75,1.75) node{$\tau+1$};

\draw[color=red] (0.05,0.1) -- (3.15,0.1);
\draw[color=red] (3.65,0.1) -- (5,0.1);
\draw[red, arrows={-Stealth[width=1.8mm, length=2.1mm]}](2.8,0.1)--(2.81,0.1);

\draw[color=blue] (0.75,0) node{$\bullet$};
\draw[below,color=blue] (0.75,0) node{\small$\frac{1}{N}$};
\draw[color=blue] (1.5,0) node{$\bullet$};
\draw[below,color=blue] (1.5,0) node{\small$\frac{2}{N}$};
\draw[color=blue] (2.25,0) node{$\bullet$};
\draw[below,color=blue] (2.25,0) node{\small$\frac{3}{N}$};
\draw[below,color=blue] (3.375,0) node{$\ldots$};
\draw[above,color=red] (3.375,0) node{$\ldots$};
\draw[color=blue] (4.25,0) node{$\bullet$};
\draw[below,color=blue] (4.25,0) node{\small$\frac{N-1}{N}$};

\begin{scope}[xshift=9cm]

\draw (0,0) -- (5,0);
\draw (0.75,1.75) -- (5.75,1.75);
\draw (0,0) -- (0.75,1.75);
\draw (5,0) -- (5.75,1.75);
\draw[below] (0,0) node{$0$};
\draw[below] (5,0) node{$1$};
\draw[above] (0.75,1.75) node{$\tau$};
\draw[above] (5.75,1.75) node{$\tau+1$};

\draw[below] (0,0) node{$0$};
\draw[below] (5,0) node{$1$};
\draw[color=blue] (0.75,0) node{$\bullet$};
\draw[below,color=blue] (0.75,0) node{\small$\frac{1}{N}$};
\draw[color=blue] (1.5,0) node{$\bullet$};
\draw[below,color=blue] (1.5,0) node{\small$\frac{2}{N}$};
\draw[color=blue] (2.25,0) node{$\bullet$};
\draw[below,color=blue] (2.25,0) node{\small$\frac{3}{N}$};
\draw[above,color=red] (3.375,0) node{$\ldots$};
\draw[color=blue] (4.25,0) node{$\bullet$};
\draw[below,color=blue] (4.25,0) node{\small$\frac{N-1}{N}$};

\draw[color=red] (0.05,0.1) -- (0.5,0.1);
\draw[color=red] (0.5,0.1) arc (180:0:0.25);
\draw[color=red] (1,0.1) -- (1.25,0.1);
\draw[color=red] (1.25,0.1) arc (180:0:0.25);
\draw[color=red] (1.75,0.1) -- (2,0.1);
\draw[color=red] (2.0,0.1) arc (180:0:0.25);
\draw[color=red] (2.5,0.1) -- (3.15,0.1);
\draw[below,color=blue] (3.375,0) node{$\ldots$};
\draw[color=red] (3.65,0.1) -- (4,0.1);
\draw[color=red] (4,0.1) arc (180:0:0.25);
\draw[color=red] (4.5,0.1) -- (5,0.1);
\draw[red, arrows={-Stealth[width=1.8mm, length=2.1mm]}](2.8,0.1)--(2.81,0.1);

\end{scope}

\end{tikzpicture}
\end{center}
\caption{Deformation of the straight-line path $[0,1]$ to the path
$[0,1]_{\varepsilon}$, avoiding the possible singularities of $f^{(1)}$.}
\label{figDeformation}
\end{figure}
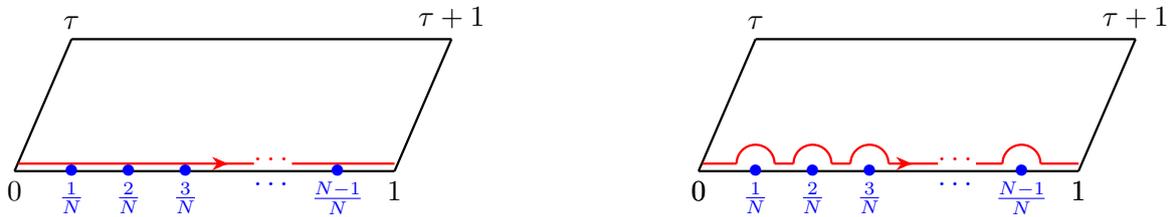

In order to regularize the divergences in \eqn{eqn:defteMZV} caused by twists
$b_1,\ldots,b_\ell \in \Lambda^\times_N$, we propose to replace the straight
line $[0,1]$ by the domain of integration $[0,1]_{\varepsilon}$ in the right
panel of \figref{figDeformation}.

\noindent Here, $\varepsilon>0$ is an additional real
parameter, which determines the radii of the semicircles around the proper rational
twists in figure \ref{figDeformation}. One then defines regularized values of
teMZVs
\begin{equation} 
  \label{eqn:rigorous}
  \omwb{n_1,n_2,\ldots,n_\ell}{b_1,b_2,\ldots,b_\ell} =
  \lim_{\varepsilon \to 0} \int\limits_{[0,1]_{\varepsilon}}f^{(n_1)}(z_1-b_1)\,\dd z_1\,f^{(n_2)}(z_2-b_2)\,\dd z_2\ldots f^{(n_\ell)}(z_\ell-b_\ell)\,\dd z_\ell \ ,
\end{equation}
which agree with \eqn{eqn:defteMZV} if all twists are generic.
The existence of the limit in \eqn{eqn:rigorous} requires some explanation
because $f^{(1)}(z-b_i)$ has a pole at $z=b_i$. For a single proper rational twist $b \in
\Lambda^\times_N$, the path $[0,1]_{\varepsilon}$ can be written as the
composition of a straight line from $0$ to $b-\varepsilon$, followed by a
semicircle from $b-\varepsilon$ to $b+\varepsilon$ \textit{above} $b$, and then
followed by a straight line from $b+\varepsilon$ to $1$:
\begin{align} 
  \label{eqn:om1}
  \omwb{1}{b}=\lim_{\varepsilon \to 0}\int_{[0,b-\varepsilon]}\dd z \ f^{(1)}(z-b)+\int_{[b-\varepsilon,b+\varepsilon]}\dd z \ f^{(1)}(z-b)+\int_{[b+\varepsilon,1]}\dd z \ f^{(1)}(z-b) \ .
\end{align}
Clearly, the contribution to \eqn{eqn:om1} coming from the non-holomorphic part
$2\pi i\frac{\Im(z)}{\Im(\tau)}$ of $f^{(1)}$ in \eqn{eqn:expl} vanishes in the limit $\varepsilon \to 0$.
The contribution coming from the closed one-form $\frac{\theta_1'(z,\tau)}{\theta_1(z,\tau)}\dd z$ in 
turn is independent of $\varepsilon$ by Stokes'
theorem, since the paths $[0,1]_{\varepsilon}$ belong to the same homotopy
class. Computing the right hand side of \eqn{eqn:om1}, we find that
\begin{equation}
 \omwb{1}{b}=\log\left(\frac{\theta_1(1-b,\tau)}{\theta_1(b,\tau)}\right)-i \pi=- i \pi \ ,
\end{equation}
where the first term comes from the first and third integral in \eqn{eqn:om1}
and vanishes by reflection and periodicity of the $\theta_1$ function. The
contribution of $-i\pi$ is due to the second integral in \eqn{eqn:om1}, by the residue theorem.

The higher-length case is handled similarly.  First we note that on the
semicircles we have additional contributions from the non-holomorphic parts of
the weighting functions $f^{(n_i)}(z_i - b_i)$ (cf.\ eq.\ \eqref{comp} below), given 
by powers of $2\pi i\frac{\Im(z_i-b_i)}{\Im(\tau)}$. These additional contributions on 
the semicircle are bounded by $\Im(z_i)\leq \varepsilon$, and the accompanying 
meromorphic functions have at most a simple pole at $b_i$. Hence,
the overall integrand on the semicircle is finite as $\varepsilon \to 0$.
Subsequently, we may use the composition of paths formula for iterated
integrals (cf.\ \eqn{eqn:comppath}) to check that the contributions from the
non-holomorphic parts on the semicircles are in fact of order
$\mathcal{O}(\varepsilon)$ and therefore do not contribute in the limit
$\varepsilon \rightarrow 0$.  Thus we are left with integrals over meromorphic
functions of the $z_i$, which do not depend on $\ve$ by homotopy of all paths
$[0,1]_{\ve}$. 

The upshot is then that, up to terms which vanish in the limit $\varepsilon \to 0$, 
the right hand side of \eqn{eqn:rigorous} is independent of $\varepsilon$, thus convergent. 
An example at length two can be found in \appref{app:example}.


\subsection{General properties of elliptic iterated integrals and (t)eMZVs}
In general, iterated integrals of the form \eqref{eqn:defGell} satisfy shuffle
relations. In terms of  combined letters $B_i = \begin{smallmatrix} n_i
\\ b_i \end{smallmatrix}$, the shuffle relation for elliptic iterated
integrals reads
\begin{equation}
  \GL(B_1,B_2,\ldots,B_{\ell_B};z) \GL(C_{1},\ldots,C_{\ell_C};z)  = \GL\big(( B_1,B_2,\ldots,B_{\ell_B}) 
  \shuffle (C_{1},\ldots,C_{\ell_C});z\big)\,,
\label{gl2}
\end{equation}
where $\shuffle $ denotes the shuffle product \cite{reutenauer}.
Naturally, the shuffle relation \eqn{gl2} straightforwardly
carries over to eMZVs,
\begin{equation}
  \omm(n_1,n_2,\ldots,n_{\ell_n}) \omm(m_1,m_2,\ldots,m_{\ell_m}) = \omega\big( (n_1,n_2,\ldots,n_{\ell_n}) \shuffle (m_1,m_2,\ldots,m_{\ell_m}) 
\big)\,,
\label{l2.7sh}
\end{equation}
and teMZVs
\begin{equation}
  \omm(B_1,B_2,\ldots,B_{\ell_B}) \omm(C_{1},\ldots,C_{\ell_C})  = \omm\big(( B_1,B_2,\ldots,B_{\ell_B}) 
  \shuffle (C_{1},\ldots,C_{\ell_C})\big)\,,
	\label{eqn:np_shuffle}
\end{equation}
where the $B_i,C_i$ are combined letters as defined above.

Taking into account the parity property \eqref{eqn:fparity1} of the weighting
functions $f^{(n_i)}$ and the definition of elliptic iterated integrals, one
finds the reflection identity
\begin{equation}
      \GLargz{n_1 &n_2 &\ldots &n_\ell}{b_1 &b_2 &\ldots &b_\ell}
      =(-1)^{n_1+n_2+\ldots+n_\ell}
        \GLargz{n_\ell &\ldots &n_2 &n_1}{z-b_\ell &\ldots &z-b_2 &z-b_1}\,,
     \label{eqn:reflection}
\end{equation}
which is, however, valid only if the combined letter $B_i = \begin{smallmatrix} 1
\\ b_i \end{smallmatrix}$ with $b_i$ a proper rational twist does not occur. 
The need to exclude such instances of $B_i$ stems from the regularization
of \secref{ssec:regteMZVs} which does not preserve the reflection property.

Again, as in the case of the shuffle relation, there is an echo of the
reflection identity for eMZVs and teMZVs:
\begin{align}
  \label{eqn:reflectioneMZV}
  \omm(n_1,n_{2} ,\ldots ,n_{\ell-1} ,n_\ell)&=
     (-1)^{n_1+n_2+\ldots+n_\ell}\omm(n_\ell,n_{\ell-1} ,\ldots ,n_2 ,n_1)\nnl
  	\omwb{n_1,n_2,\ldots,n_\ell}{\nbeta_1,\nbeta_2,\ldots,\nbeta_\ell}
		   &=    (-1)^{n_1+n_2+\ldots+n_\ell} \omwb{n_\ell,n_{\ell-1},\ldots,n_1}{\tilde\nbeta_\ell,\tilde\nbeta_{\ell-1},\ldots,\tilde\nbeta_1}\,,
\end{align}
where 
$\tilde{b}_i$ is the representative of $-b_i$ in the fundamental domain of the
elliptic curve
and letters $B_i =
\begin{smallmatrix} 1 \\ b_i \end{smallmatrix}$ with $b_i$ a proper rational
  twist are again excluded.


\subsection{$q$-expansion of teMZVs}
\label{taudep}

In contrast to usual MZVs, which are just
numbers, eMZVs and teMZVs are functions of
the modular parameter $\tau$ and can be expanded in its exponentiated
cousin $q= e^{2\pi i \tau}$.
The $q$-expansions of eMZVs and teMZVs rely on the available $q$-expansions of
the weighting functions $f^{(n)}$. The discussion below will simplify
considerably, if we consider in addition a class of meromorphic weighting
functions $g^{(n)}$. 

While the weighting functions $f^{(n)}$ appear as expansion coefficients of the
doubly-periodic completion $\Omega(z,\alpha,\tau)$ of the Eisenstein--Kronecker
series $F(z,\alpha,\tau)$ (cf.~\eqn{alt5}) \cite{BrownLev}
\begin{equation}
  \label{alt51}
  \Omega(z,\alpha,\tau)= 
  \exp \bigg( 2\pi i \alpha\, 
  \frac{\Im(z)}{\Im(\tau)} \bigg) F(z,\alpha,\tau)=
  \sum_{n=0}^{\infty}f^{(n)}(z,\tau)\alpha^{n-1} \, ,
\end{equation}
the functions $g^{(n)}$ are the expansion coefficients of the
Eisenstein--Kronecker series \cite{Weil:1976,ZagierF}
\begin{align}
    F(z,\alpha,\tau) &= \sum_{n =0}^{\infty} g^{(n)}(z,\tau) \alpha^{n-1}\,.
    \label{eqn:exp_F}
\end{align}
The set of meromorphic functions $g^{(n)}(z,\tau)$ starts with $g^{(0)}=1$ and
$g^{(1)}(z,\tau)=\frac{ \theta_1'(z,\tau)}{\theta_1(z,\tau)}$ and can be
related to their doubly-periodic but non-holomorphic\footnote{Note that by
Liouville's theorem, every meromorphic, doubly-periodic function, which has at
most a simple pole at zero must be constant. Therefore, one either has to
include non-holomorphic factors (as we do here, following \cite{BrownLev}) or
allow poles of order $\geq 2$ (as in \cite{LevinRacinet}, \S 5.1.2).}
completions via \eqn{alt51}:
\begin{equation}
  f^{(n)}(z,\tau) = \sum_{j=0}^{n} \frac{1}{(n-j)!} \left( 2 \pi i  \frac{\Im(z)}{\Im(\tau)} \right)^{n-j} g^{(j)}(z,\tau) \; .
  \label{comp}
\end{equation}
Quasi-periodicity and the reflection property of $F(z,\alpha,\tau)$ (see
\eqns{eqn:period_F}{eqn:period_F1}) imply the following properties of the
$g^{(n)}(z,\tau)$,
\begin{equation}
  g^{(n)}(z ) =  g^{(n)}(z + 1)  \,, \quad  
  g^{(n)}(z + \tau) = \sum_{j=0}^n \frac{(-2 \pi i)^j}{j!} g^{(n-j)}(z)  \,, \quad  
  g^{(n)}(-z) = (-1)^n g^{(n)}(z)\,,
\label{eqn:props_gn}
\end{equation}
and their Fourier expansions are given by \cite{ZagierF, BrownLev, Broedel:2014vla} 
\begin{align}
    g^{(1)}(z,\tau)    &= \pi \, \cot(\pi z) - 2 i (2 \pi i) \sum_{n,m=1}^\infty \sin(2 \pi m z) \, q^{m n} \notag \\
    g^{(2k)}(z,\tau)   &= -2 \zeta_{2k} -2 \frac{(2\pi i)^{2k}}{(2k-1)!} \sum_{n,m=1}^\infty \cos(2 \pi m z) \,  n^{2k-1} q^{m n}  \, , &k>0 \, \phantom{.} \label{eqn:gn_explicit} \\
    g^{(2k+1)}(z,\tau) &= -2i \frac{(2\pi i)^{2k+1}}{(2k)!} \sum_{n,m=1}^\infty \sin(2 \pi m z) \,  n^{2k} q^{m n} \, ,&k>0 \, .
  \notag
\end{align}
For real values of $z$ one finds from \eqns{alt51}{comp} that
$f^{(n)}(z)=g^{(n)}(z)$ and their $q$-expansions agree. In particular, they can
be employed to find $q$-expansions for eMZVs
\begin{equation}
  \omm(n_1,n_2,\ldots,n_\ell)=\om_0(n_1,n_2,\ldots,n_\ell)+\sum_{k=1}^{\infty}c_k(n_1,n_2,\ldots,n_\ell) q^k \ .
  \label{eqn:qexp}
\end{equation}
The $q$-independent quantity $\omega_0$ in \eqn{eqn:qexp} is called the
\textit{constant term} of the eMZV $\omega$ and is known to be a $\ZQ[(2\pi
i)^{-1}]$-linear combination of MZVs (see refs.~\cite{Broedel:2014vla,
Broedel:2015hia, Matthes:Thesis}). 

In order to describe the $q$-dependence of teMZVs in a
similar manner, we consider the twist~$b=s+r\tau \in \Lambda_N+\Lambda_N \tau$ in
  the weighting function $f^{(n)}(z-b)$ \eqn{comp} for real values of $z$:
\begin{equation}
  f^{(n)}(z-s-r\tau,\tau) = \sum_{j=0}^{n} \frac{ ( - 2 \pi i r)^{n-j} }{(n-j)!} g^{(j)}(z-s-r\tau,\tau) \; , \quad  z \in \mathbb R\; .
  \label{eqn:f_via_g}
\end{equation}
Employing eqns.~\eqref{eqn:gn_explicit}, the functions $g^{(j)}(z-b,\tau)$, can be
expanded in non-negative rational powers of $q$,
\begin{align}
  &g^{(2k+1)}(z - s - r \tau ,\tau) = \delta_{k,0} \pi \, \cot(\pi(z-s-r\tau)) + \frac{ (2\pi i)^{2k} }{(2k)!} \! \sum_{m,n=1}^{\infty} n^{2k} q^{mn}  \notag \\
  &\ \ \ \times \Big\{ \cos(2\pi m(z-s)) (q^{mr}-q^{-mr}) - i \sin(2\pi m(z-s)) (q^{mr}+q^{-mr}) \Big\} 
   \,, &k \geq 0 \, \phantom{.}
\notag \\
  & g^{(2k)}(z - s - r \tau ,\tau) = -2 \zeta_{2k} - \frac{ (2\pi i)^{2k-1} }{(2k-1)!} \! \sum_{m,n=1}^{\infty}  n^{2k-1} q^{mn}     \label{eqn:gnq} \\
  &\ \ \ \times  \Big\{ \cos(2\pi m(z-s)) (q^{mr}+q^{-mr}) - i \sin(2\pi m(z-s)) (q^{mr}-q^{-mr}) \Big\}  \,, &k > 0
   \,.  \notag
\end{align}
If $0<r<1$, i.e.~if $b$ is a generic twist, then the cotangent term in
$g^{(1)}$ may be rewritten as
\begin{equation}
  \pi \cot(\pi(z-s-r\tau)) = i \pi (1+q^r e^{2\pi i(s-z)}) \sum_{n=0}^{\infty}(q^r e^{2\pi i (s-z)})^n \,.
  \label{eqn:cot}
\end{equation}
On these grounds, $f^{(n)}(z-s-r\tau)$ can be expanded in powers of $q^r$ and $q^{1-r}$ 
such that every teMZV admits an expansion in $q^{p}$,
\begin{equation}
  \begin{split}
  	\omwb{n_1,n_2,\ldots,n_\ell}{\nbeta_1,\nbeta_2,\ldots,\nbeta_\ell} 
  	= \omwbc{n_1,n_2,\ldots,n_\ell}{\nbeta_1,\nbeta_2,\ldots,\nbeta_\ell} +
  	\sum_{k=1}^{\infty} c_{k} \left(\begin{smallmatrix} n_1,n_2,\ldots,n_\ell \\ \nbeta_1,\nbeta_2,\ldots,\nbeta_\ell \end{smallmatrix}\right) (q^{p})^k \ ,
  \end{split}
  \label{eqn:np_qexp}
\end{equation}
where $1/p\in \ZQ$ is the least common denominator of all
occurring $r_i$. The $q$-independent quantity $\omega_0$ in \eqn{eqn:np_qexp}
is called the \textit{constant term} of the teMZV, which we are going to study
in \secref{sec:new}. Depending on the set of twists $b_i$, different classes of
objects appear as constant terms: while MZVs cover constant terms for generic
twists, proper rational twists lead to cyclotomic MZVs
\cite{Goncharov:2001iea,Racinet:Doubles,Zhao:2008, Glanois:Thesis}.  We will
refer to teMZVs for which $c_{k} \left(\begin{smallmatrix}
n_1,n_2,\ldots,n_\ell \\ \nbeta_1 ,\nbeta_2,\ldots, \nbeta_\ell
\end{smallmatrix}\right)=0$ for all $k \in \ZN^+$ as \textit{constant}.


\section{$q$-expansion for twisted elliptic
multiple zeta values}
\label{sec:new}

The goal of this section is to set up an initial value problem for teMZVs
\eqn{eqn:defteMZV} and to obtain their $q$-expansion without performing any
integral over their trigonometric constituents in \eqn{eqn:gnq}.  The
differential equation to be derived in this section will prove to be the main
tool to allow the efficient computation and comparison of teMZVs in a
convenient representation as iterated integrals. In particular, this
representation will prove useful in the context of calculating non-planar
contributions to the one-loop amplitude in \secref{sec:oneloop}. 

Following the strategy for computing the usual eMZV's $q$-expansion in
\cite{Enriquez:Emzv, Broedel:2015hia}, in a first step we derive a first-order
differential equation in $\tau$ for teMZVs. In the second step, a boundary
value at the cusp $\tau \rightarrow i \infty$ will be determined to identify a
unique solution to the differential equation. Since the action of $\pd_\tau$
reduces the length of teMZVs, one can derive the $q$-expansion for teMZVs
recursively.

For eMZVs, classical Eisenstein series and MZVs are the building blocks for the
$\tau$-derivative and constant term respectively\cite{Enriquez:Emzv,
Broedel:2015hia,Matthes:Thesis}. Similarly, we will show that the weighting
functions $f^{(n)}(b,\tau)$ evaluated at lattice points
$b\in\Lambda_N+\Lambda_N\tau$ and cyclotomic MZVs are suitable generalizations
thereof for teMZVs.

After deriving the differential equation in \subsecref{ssec:differential},
the constant term will be elaborated on in \subsecref{ssec:constant} for
generic twists and modifications when including proper rational twists are
discussed in \subsecref{ssec:realboundary}.

\subsection{Differential equation}
\label{ssec:differential}

We begin by defining a generating series for \temzv{}s of length $\ell$,
\begin{align}
    \Tgen{\alpha_1, \alpha_2, \dots, \alpha_\ell}{\nbeta_1, \nbeta_2, \dots, \nbeta_\ell}  & =  
    \! \! \! \! \! \! \int\limits_{0\leq z_i \leq z_{i+1} \leq 1}  \! \! \! \! \! \!
    \Omega(z_1 - \nbeta_1,\alpha_1,\tau)\,\dd z_1\, \Omega(z_2 - \nbeta_2,\alpha_2,\tau)\,\dd z_2  \ldots \Omega(z_\ell - \nbeta_\ell,\alpha_\ell,\tau)\,\dd z_\ell  \nnl
    & = \sum_{n_1,n_2,\ldots,n_\ell=0 }^{\infty} \alpha_1^{n_1-1} \alpha_2^{n_2-1} \dots \alpha_\ell^{n_\ell-1} \omwb{n_1, n_2, \dots, n_\ell}{\nbeta_1, \nbeta_2, \dots, \nbeta_\ell} 
    \; ,
  \label{eqn:generating_T}
\end{align}
generalizing a construction of \rcite{Enriquez:Emzv}. For simplicity, we will
assume in this subsection that the twists $b_i=s_i+r_i\tau$ are generic, i.e
$r_i \in (0,1)$. The case of proper rational twists is discussed in
\appref{app:diff_eq_rat}.

First, since the domain of integration in \eqn{eqn:generating_T} is the
interval $[0,1] \subset \ZR$, it is natural to restrict the function $z\mapsto
\Omega(z-b,\alpha,\tau)$ to real arguments of $z$. With this restriction, the
following differential equation is then a consequence of the mixed heat
equation \eqref{mixedheat} 
\begin{equation}
  \begin{split}
    \pd_\tau \Omega(z - s - r \tau,\alpha,\tau) & = 
    \exp(-2 \pi i r \alpha) \pd_\tau F(z - s - r \tau,\alpha,\tau) \\
    & = \exp(- 2 \pi i r \alpha)
    \Big(- r \pd_{z} + \frac{1}{2 \pi i} \pd_{z} \pd_{\alpha} \Big)
    F(z - s - r \tau,\alpha,\tau) \\
    & = \frac{1}{2 \pi i}  \pd_{z}  \pd_{\alpha}   \Omega(z - s - r \tau,\alpha,\tau)\,.
  \end{split}
  \label{eqn:mixed_heat_type_eq}
\end{equation}
Here the partial derivative $\pd_\tau$ is understood to act on all occurrences
of the variable $\tau$. Furthermore, we have used that $\pd_z r=\pd_\tau r=0$,
since the twist $b=s+r\tau$ is fixed, and therefore neither depends on $z$ nor
$\tau$. Note that in going from the first to the second line in
\eqn{eqn:mixed_heat_type_eq}, the term $r \partial_z$ appears by taking the
occurrence of $\tau$ in the first argument of the Kronecker series into
account.  However, this additional term gets neatly absorbed when returning to
the doubly-periodic completion $\Omega(z-s-r\tau,\alpha,\tau)$ in the last
line.

The $\tau$-derivative of the generating function in
\eqn{eqn:generating_T} reads
\begin{equation}
  \begin{split}
  \label{eqn:taugen}
   2 \pi i \frac{\pd}{\pd \tau} \Tgen{\alpha_1, \alpha_2, \dots, \alpha_\ell}{\nbeta_1, \nbeta_2, \dots, \nbeta_\ell}
    &=  
\! \! \! \! \! \! \int \limits_{0\leq z_i \leq z_{i+1} \leq 1}  \! \! \! \! \! \!    
    \dd z_1 \, \dd z_2 \, \ldots \, \dd z_\ell
      \sum_{i=1}^\ell  \pd_{z_i} \pd_{\alpha_i} \Omega(z_i - \nbeta_i,\alpha_i) \prod_{j \neq i}^{\ell} \Omega(z_j - \nbeta_j,\alpha_j) \\
    &= \pd_{\alpha_\ell} \Omega(-\nbeta_\ell,\alpha_\ell) \Tgen{\alpha_1, \dots, \alpha_{\ell-1}}{\nbeta_1, \dots, \nbeta_{\ell-1}} -
      \pd_{\alpha_1} \Omega(-\nbeta_1,\alpha_1) \Tgen{\alpha_2, \dots, \alpha_\ell}{\nbeta_2, \dots, \nbeta_\ell} \\
    & \quad + \sum_{i=2}^{\ell} \Big(
    \Tgen{\alpha_1, \dots ,\alpha_{i-2}, \alpha_{i-1} + \alpha_i , \, \alpha_{i+1}, \dots ,\alpha_\ell}{\nbeta_1, \dots ,\nbeta_{i-2}, \nbeta_{i}, \, \nbeta_{i+1}, \dots ,\nbeta_\ell} 
    \pd_{\alpha_{i-1}} \Omega(\nbeta_i-\nbeta_{i-1},\alpha_{i-1}) \\
    & \qquad\qquad - \Tgen{\alpha_1, \dots, \alpha_{i-2}, \alpha_{i-1} + \alpha_i, \, \alpha_{i+1}, \dots, \alpha_\ell}{\nbeta_1, \dots ,\nbeta_{i-2} , \nbeta_{i-1} , \, \nbeta_{i+1}, \dots ,\nbeta_\ell} 
    \pd_{\alpha_i} \Omega(\nbeta_{i-1}-\nbeta_i,\alpha_i)
    \Big) \,,
    \end{split}
\end{equation}
where we used \eqn{eqn:mixed_heat_type_eq} in the first line and suppressed the
dependence of $\TL$ and $\Omega$ on the modular parameter $\tau$. In the
second equality, the number of integrations is reduced by evaluating $\int \dd
z_i \, \pd_{z_i} \pd_{\alpha_i} \Omega(z_i - \nbeta_i,\alpha_i) $ via boundary
terms $\pd_{\alpha_i} \Omega(z_i - \nbeta_i,\alpha_i)|_{z_{i-1}}^{z_{i+1}}$
with $z_0=0$ and $z_{\ell+1}=1$. The resulting products of the form
$\Omega(z_i-b_{i-1},\alpha_{i-1}) \Omega(z_i-b_i,\alpha_i)$ are rewritten using
the Fay identity \eqn{fayfay_Om} such that each integration variable $z_i$
appears in at most one factor of $\Omega$.  The details of the computation can
be found in \appref{app:diff_eq_der}.

Upon expanding $\Omega$ and $\TL$  in \eqn{eqn:taugen} in $\alpha_i$, one
can compare the coefficients of the monomials
$\alpha_1^{m_1} \ldots \alpha_\ell^{m_\ell}$. The coefficient of each monomial is a
linear combination of some $f^{(n)}$ multiplied by \temzv{}s of length $\ell-1$. Working 
out the details yields the following
differential equation for \temzv{}s ($\ell \geq2$),
and using the shorthand
\begin{equation}
  h^{(n)}(u,\tau) = (n-1) f^{(n)} (u,\tau)\,.
  \label{eqn:defh}
\end{equation}
for $u\in\ZC/(\ZZ+\ZZ\tau)$, we find
\begin{align}
  \label{eqn:diff_eq_nemzv}
    & 2 \pi i \pd_\tau \omwb{n_1, \dots , n_\ell}{\nbeta_1, \dots , \nbeta_\ell} 
    = h^{(n_\ell+1)}(-\nbeta_\ell)
    \omwb{n_1, \dots , n_{\ell-1}}{\nbeta_1, \dots , \nbeta_{\ell-1}} - h^{(n_1+1)}(-\nbeta_1)\omwb{n_2, \dots , n_{\ell}}{\nbeta_2, \dots , \nbeta_{\ell}} \nnl
    & \quad + \sum_{i=2}^\ell \Bigg[
   \theta_{n_{i} \geq 1} \sum_{k=0}^{n_{i-1}+1} \binom{n_i + k - 1}{k} 
   h^{(n_{i-1}-k+1)}(\nbeta_i - \nbeta_{i-1} )  \omwb{n_1, \dots, n_{i-2}, n_i+k ,\, n_{i+1}, \dots, n_\ell}{\nbeta_1, \dots, \nbeta_{i-2}, \nbeta_{i} , \,\nbeta_{i+1}, \dots, \nbeta_\ell}
      \nnl
    & \quad\qquad - \theta_{n_{i-1} \geq 1} \sum_{k=0}^{n_{i}+1} \binom{n_{i-1} + k -1}{k} 
  h^{(n_{i}-k+1)}(\nbeta_{i-1} - \nbeta_{i} )  \omwb{n_1, \dots, n_{i-2}, n_{i-1}+k ,\, n_{i+1}, \dots, n_\ell}{\nbeta_1, \dots, \nbeta_{i-2}, \nbeta_{i-1} ,\, \nbeta_{i+1}, \dots, \nbeta_\ell}  \nnl
    & \quad\qquad + (-1)^{n_{i}+1}  \theta_{n_{i-1} \geq 1} \theta_{n_i \geq 1} h^{(n_{i-1} + n_i + 1)}(\nbeta_{i}-\nbeta_{i-1})
    \omwb{n_1, \dots ,n_{i-2}, 0, n_{i+1}, \dots ,n_\ell}{\nbeta_1, \dots, \nbeta_{i-2},  0, \nbeta_{i+1}, \dots, \nbeta_\ell} 
    \Bigg]\,,
\end{align}
where we have introduced $\theta_{n \geq 1} = 1- \delta_{n,0}$ for non-negative
$n$ to indicate that some of the contributions in the last three lines vanish
for $n_i=0$. For vanishing twists $\nbeta_i=0$, \eqn{eqn:diff_eq_nemzv} reduces to the
differential equation for eMZVs stated in eq.\ (2.47) of
\rcite{Broedel:2015hia} since the weighting functions $f^{(n)}$ are related to
holomorphic Eisenstein series (with $ \GG{0}(\tau)=-1$) via 
\begin{equation}
-\lim_{z \to 0}f^{(k)}(z,\tau)=
 \GG{k}(\tau) = \left\{ \begin{array}{rl}
 \displaystyle 2\zm_k+\frac{2 (2\pi i)^k}{(k-1)!} \sum_{m,n=1}^{\infty} m^{k-1} q^{mn} &: \ k \geq 2 \ \te{even} \\
 -1 &: \ k=0 \  \\
  0 &: \ k \neq 1 \ \te{odd}
  \end{array} \right.\, ,
  \label{eqn:h0Eisen}
\end{equation}
where the limit is understood to be taken along the real axis. In other words,
the functions $h^{(n)}(b,\tau)$ occurring in the differential equation
\eqref{eqn:diff_eq_nemzv} (which are modular forms for congruence subgroups of
$\SL_2(\ZZ)$) take the r\^{o}le of Eisenstein series in the differential
equation for eMZVs. Also, note that the exceptional case $f^{(1)}(z,\tau)$
(which has a pole at $z=0$) does not appear in \eqn{eqn:taugen} since it is
accompanied by $\alpha^0=1$ in the generating series $\Omega(z,\alpha,\tau)$
and is therefore annihilated upon application of the $\alpha$-derivative.

\subsection{Constant terms for generic twists}
\label{ssec:constant}

In this subsection we are going to extend the constant-term procedure for
eMZVs studied in \cite{Broedel:2015hia,Matthes:Thesis} to a procedure
delivering the constant terms for teMZVs. 
Calculating the constant term for \temzv{}s amounts to the computation of the
limit $\tau \to i\infty$ of \eqn{eqn:defteMZV}. This limit will figure as the
initial value for the differential equation (\ref{eqn:diff_eq_nemzv}) 
discussed in the previous subsection. 

In order to make the bookkeeping more efficient, it is convenient to consider
a suitable generating series of teMZVs, which is a generalization of the $A$-part
of Enriquez' elliptic KZB associator \cite{Enriquez:EllAss} to the realm of
teMZVs: More precisely, for every $N\geq 1$ we will consider a formal power series in nested commutators
\begin{equation}
\ad^n_x(y)= \underbrace{[x, \ldots [x,[x,}_{n \ {\rm times}} y]] \ldots ]\,, \quad n \geq 0\, ,
\label{defcommut}
\end{equation}
of the non-commutative variables $y, \{ x_{b_i}\}_{b_i \in (\Lambda_N + \Lambda_N \tau) \setminus \Lambda_N^{\times}}$ 
as follows:
\begin{align}
A^\textrm{twist}_{(\Lambda_N+\Lambda_N\tau)\setminus\Lambda_N^\times}(\tau)&= 
\sum_{\ell \geq 0}(-1)^\ell
\hspace{-20pt}
  \sum\limits_{
   \begin{minipage}{100pt}\vspace*{-0.8pt}\footnotesize
    \begin{centering}
    $n_1,n_2,\ldots,n_\ell\geq 0$ \\
    ${b_1,b_2,\ldots,b_\ell \in (\Lambda_N+\Lambda_N\tau) \setminus \Lambda_N^{\times}}$
  \end{centering}
  \end{minipage}}
\hspace{-20pt}
\omwb{n_1,n_2,\ldots,n_\ell}{\nbeta_1,\nbeta_2,\ldots,\nbeta_\ell}\ad^{n_\ell}_{x_{b_\ell}}(y)\ldots \ad^{n_2}_{x_{b_2}}(y) \, \ad^{n_1}_{x_{b_1}}(y) 
\notag\\[-4pt]
&=\tilde {\cal P} \, \exp 
\bigg(
- \int^1_0 \dd z 
\sum\limits_{k=0}^{\infty} 
\hspace{5pt}
\sum\limits_{
\begin{minipage}{80pt}\vspace*{-2pt}\footnotesize
    $b\in(\Lambda_N+\Lambda_N\tau) \setminus \Lambda_N^{\times}$
\end{minipage}
}
\hspace{5pt}
f^{(k)} (z-b,\tau) \, \te{ad}_{x_{b}}^{k}(y) 
\bigg)\,,
\label{eqn:genfunction}
\end{align}
where $\Lambda_N$ and $\Lambda_N^\times$ were defined in \eqn{deflambda}, and
$\tilde{\mathcal{P}} \,\exp(\ldots)$ denotes the path-ordered exponential with
reverted order of multiplication for the non-commutative variables in
comparison to the order of the integration variables\footnote{Although
  composed of several non-commutative variables $x_{b_i}$ and $y$, each nested
commutator $\ad^{n_i}_{x_{b_i}}(y)$ is treated as a single letter when
reversing the order of multiplication.} $z$. We note that there is no loss of
generality in studying the lattice $\Lambda_N+\Lambda_N\tau$ rather than
$\Lambda_N+\Lambda_M\tau$ with $M\neq N$: the latter can be embedded into the
lattice $\Lambda_{N'}+\Lambda_{N'}\tau$ with $N'$ the least common multiple of
$M$ and $N$. Also, proper rational twists $b \in \Lambda_N^{\times} $ have been
excluded from the summation range for $b$ in \eqn{eqn:genfunction} in order to
relegate a discussion of the additional ingredients required in these cases to
section \ref{ssec:realboundary}.

The series \eqref{eqn:genfunction} combines different instances of the
generating series $\Tgen{\alpha_1& \alpha_2& \dots& \alpha_\ell}{\nbeta_1&
\nbeta_2& \dots& \nbeta_\ell}$ in \eqn{eqn:generating_T},
\begin{equation}
  A^\textrm{twist}_{(\Lambda_N+\Lambda_N\tau)\setminus\Lambda_N^\times}(\tau) \longleftrightarrow 
  \sum_{\ell \geq 0}(-1)^\ell\sum_{ b_1,\ldots,b_\ell\in (\Lambda_N+\Lambda_N\tau) \setminus \Lambda_N^{\times}} 
  \Tgen{\alpha_\ell& \alpha_{\ell-1}& \dots& \alpha_1}{b_\ell& b_{\ell-1}& \dots & b_1}\,,
\end{equation}
summing over all values of the length $\ell \geq 0$ and the generic twists $b_i \in
(\Lambda_N+\Lambda_N\tau) \setminus \Lambda_N^{\times}$.  The non-commutative
product of $\ad_{x_{b_i}}^{k_i}(y)$ corresponds to commutative variables
$\alpha_i^{k_i-1}$ in \eqn{eqn:generating_T}, which accompany individual teMZVs
in the respective generating series. While the organization via
$\alpha_i^{k_i-1}$ is better suited for the study of the differential equation
of teMZVs, the non-commutative variables $\ad_{x_{b_i}}^{k_i}(y)$ in
\eqn{defcommut} are well adapted to the subsequent analysis of their constant
terms\footnote{The use of two, essentially equivalent, generating series of
teMZVs goes back to Enriquez' original work on eMZVs \cite{Enriquez:Emzv}.}.

\subsubsection{Degeneration of weighting functions}
In order to compute $\lim_{\tau \to
i\infty}A^\textrm{twist}_{(\Lambda_N+\Lambda_N\tau)\setminus\Lambda_N^\times}(\tau)$,
we need to study the degeneration of the weighting functions
$f^{(k)}(z-b,\tau)$ as $\tau \to i\infty$ (or equivalently $q\to0$). 
Conveniently, the limit is expressed in the variables
\begin{equation}
w = e^{2\pi i z} \co \dd z = \frac{1}{2\pi i } \, \frac{ \dd w }{w} \,.
\label{eqn:drin12}
\end{equation}
Using the $q$-expansions \eqns{eqn:gnq}{eqn:cot} together with
\eqn{eqn:f_via_g} we obtain, for generic twists and $k>1$,
\begin{align}
  \lim_{\tau \to i\infty} f^{(k)}(z-s-r\tau,\tau) \, \dd z &= \bigg(
  \frac{\pi i (-2 \pi i r)^{k-1} }{(k-1)!} 
  - 2 \sum_{m=0}^{ \lfloor \frac k2\rfloor } \frac{(-2 \pi i r)^{k-2m}}{(k-2m)!} \zeta_{2m}  \bigg) \frac{1}{2 \pi i} \frac{\dd w}{w}\nnl
  &= -\frac{\dd w}{w}\sum_{m=0}^k\frac{B_m(-2\pi i)^{m-1}}{m!}\frac{(-2\pi ir)^{k-m}}{(k-m)!}\,,\qquad \ k >1 \,.
  \label{eqn:drin11a}
\end{align}
Here, we have used $\zeta_{2m}=-\frac{B_{2m}(2\pi i)^{2m}}{2(2m)!}$, where
$B_k$ denotes the $k^{\rm th}$ Bernoulli number (such that $B_1=-\frac 12$). While
$f^{(0)}=1$, the case of $f^{(1)}(z-b)$ is special and we find 
\begin{equation}
\lim_{\tau \to i\infty}f^{(1)}(z-s-r\tau,\tau) \, \dd z = \left\{
\begin{array}{cl}\displaystyle
\left(\frac 12-r\right) \, 
\frac{ \dd w }{w} &: \ r \neq 0 \\\\
\displaystyle-\frac{1}{2} \, 
\frac{ \dd w }{w}  + \frac{ \dd w }{w-1} &: \ r=s=0
\end{array}
\right.\,.
\label{eqn:drin11b}
\end{equation}
Combining \eqns{eqn:drin11a}{eqn:drin11b} allows to rewrite the exponent of
\eqn{eqn:genfunction} as follows: 
\begin{equation}
  \lim_{\tau \to i\infty}-  \dd z \ \sum_{k=0}^{\infty} 
  \bigg(\sum_{b \in (\Lambda_N+\Lambda_N\tau) \setminus \Lambda_N^\times} f^{(k)}(z-b,\tau) \, \te{ad}_{x_b}^{k}(y) \bigg)
   = \tilde{y}_N \, \frac{ \dd w}{w} + t \, \frac{ \dd w }{w-1} \ ,
  \label{eqn:drin23}
\end{equation}
where
\begin{equation}
\tilde{y}_N = -\frac{\ad_{x_0}}{e^{2\pi i\,\ad_{x_0}}-1}(y)+\sum_{b \in (\Lambda_N+\Lambda_N\tau) \setminus \Lambda_N}\frac{\ad_{x_b}e^{-2\pi ir\,\ad_{x_b}}}{e^{-2\pi i\,\ad_{x_b}}-1}(y)  \ , \ \ \ \ \ \
t = [y,x_0] \ .
\label{eqn:drin24}
\end{equation}
These definitions of $\tilde{y}_N$ and $t$ are tailored to track the appearance
of the forms $\frac{ \dd w }{w}$ and $\frac{ \dd w }{w-1}$ in the degeneration
limits \eqns{eqn:drin11a}{eqn:drin11b} of $f^{(k)}(z-b,\tau)$. In absence of
twists, for instance, the first contribution to $\tilde{y}_N$ in
\eqn{eqn:drin24} stems from setting $r=0$ in \eqn{eqn:drin11a} and identifying
the generating series of $\frac{B_k(-2\pi i)^{k-1}}{k!}\ad^k_{x_0}(y)$:
\begin{align}
  \lim_{\tau \to i\infty}-  \dd z \ \sum_{k=0}^{\infty} 
\bigg( f^{(k)}(z,\tau) \, \te{ad}_{x_0}^{k}(y) \bigg)&=-\frac{\dd w}{w}\sum_{k=0}^{\infty}\frac{B_k(2\pi i)^{k-1}}{k!}\ad^k_{x_0}(y)-\frac{\dd w}{w-1}\ad_{x_0}(y) \notag \\
&=-\frac{\ad_{x_0}}{e^{2\pi i\ad_{x_0}}-1}(y)\frac{\dd w}{w}-\frac{\dd w}{w-1}\ad_{x_0}(y) \, .
\label{explain}
\end{align}
In the generalization to non-zero twists $r \neq 0$, the second contribution to
$\tilde{y}_N$ in \eqn{eqn:drin24} arises as the generating series of
$\sum_{m=0}^k\frac{B_m(-2\pi i)^{m-1}}{m!}\frac{(-2\pi ir)^{k-m}}{(k-m)!}
\ad^k_{x_b}(y)$.  Note in particular that the dependence on $r$ gives rise to
the factor of $e^{-2\pi ir\ad_{x_b}}$ in the numerator.

By comparing the degeneration behaviour in \eqn{eqn:drin23} with
\eqn{eqn:genfunction}, we deduce that
\begin{equation} \label{eqn:drin16}
\lim_{\tau \to i\infty}A^\textrm{twist}_{(\Lambda_N+\Lambda_N\tau)\setminus\Lambda_N^\times}(\tau)=\tilde {\cal P} \, \exp \bigg(
\int \limits_{C_0^{2\pi}(1)} \bigg[   \frac{ \tilde{y}_N }{w} + \frac{  t  }{w-1}  \bigg] \, \dd w \bigg) \, ,
\end{equation}
where the unit circle $w \in C_0^{2\pi}(1)$ arises from the path of integration $[0,1]\subset \ZC$ under the change of variables \eqn{eqn:drin12}.
Strictly speaking, \eqn{eqn:drin16} requires regularization, due to
divergences at $w=1$. They are treated in analogy to eMZVs as described in refs.
\cite{Enriquez:Emzv,Matthes:Thesis,Broedel:2014vla} and cause modifications to
be pointed out in the subsequent discussion.

\subsubsection{Deforming the integration contour}

We have expressed $\lim_{\tau \to
i\infty}A^\textrm{twist}_{(\Lambda_N+\Lambda_N\tau)\setminus\Lambda_N^\times}(\tau)$
as a generating series of iterated integrals of explicit differential forms
along the unit circle $C_0^{2\pi}(1)$, see the left panel of \figref{figNils}
below. Although the functions $\lim_{\tau \to i\infty}f^{(k)}(z-b,\tau)$ in
\eqns{eqn:drin11a}{eqn:drin11b} were restricted to real values of $z$, they can
be extended to the doubly-punctured complex plane $z \in \ZC \setminus \{0,1\}$
(with $r$ kept constant). In this way, the integrand in \eqn{eqn:drin23} is
holomorphic on $\ZC \setminus \{0,1\}$, and the resulting path-ordered
exponential \eqn{eqn:drin16} is homotopy-invariant. Therefore one can replace
the unit circle by a contour homotopic to it, visualized in the right panel of
\figref{figNils}.

This deformed contour can in turn be viewed as the composition of straight
paths $P_1,P_1^{-1}$ connecting the points $w=0,1$ along with a circle
$C_0^{2\pi}(\varepsilon)$ of infinitesimal radius around the origin, as shown
in \figref{figNils}. Moreover, the regularization alluded to above manifests
itself in \figref{figNils}: both paths $C_0^{2\pi}(1)$ as well as the
composition $P_1^{-1} C_0^{2\pi}(\varepsilon) P_1 \hat C_{\pi}^0(\varepsilon)$
have to leave $w=1$ with velocity $-1$ and arrive back at $w=1$ with the same
velocity. More precisely, both paths, which really are smooth functions $[0,1]
\rightarrow \ZC^{\times}$ must have a derivative equal to $-\frac{\pd}{\pd w}
\in T_1(\ZC^{\times})$, where $T_1$ denotes the tangent space at $1$. This is
also the reason for the semicircle $\hat{C}_{\pi}^0(\varepsilon)$.

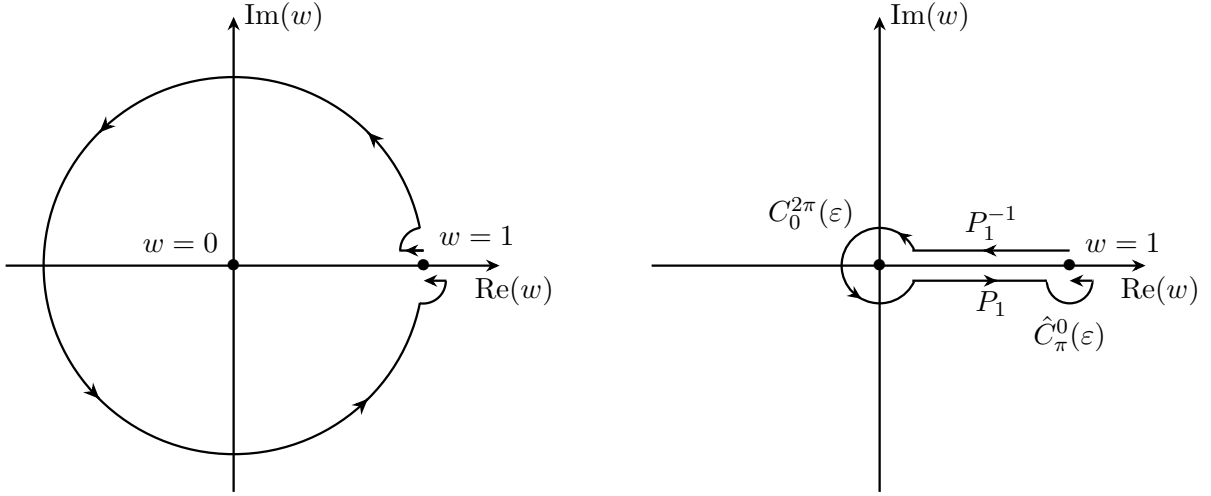
\begin{figure}[h]
	\begin{center}
\begin{tikzpicture}[line width=0.30mm]
\draw (2.5,0) arc (0:360:2.5cm);
\draw[white, fill=white] (2.6,0) circle (0.5cm);

\draw (2.5-0.05,-0.5) arc (-98:0:0.3cm);
\draw[arrows={-Stealth[width=1.6mm, length=1.8mm]}](2.8,-0.2)--(2.5,-0.2);

\draw (2.5-0.05, 0.5) arc (98:180:0.3cm);
\draw[arrows={-Stealth[width=1.6mm, length=1.8mm]}](2.5,0.2)--(2.25,0.2);
\draw(2.5,0.2)--(2.18,0.2);
\draw[arrows={-Stealth[width=1.8mm, length=2.1mm]}](1.767,1.765)--(1.766,1.766);
\draw[arrows={-Stealth[width=1.8mm, length=2.1mm]}](-1.767,-1.765)--(-1.766,-1.766);
\draw[arrows={-Stealth[width=1.8mm, length=2.1mm]}](-1.765,1.767)--(-1.766,1.766);
\draw[arrows={-Stealth[width=1.8mm, length=2.1mm]}](1.765,-1.767)--(1.766,-1.766);
%
\draw (0,0)node{$\bullet$} ;
\draw(-0.7,0.3)node{$w=0$};
\draw (2.5,0)node{$\bullet$} ;
\draw(3.2,0.4)node{$w=1$};
\draw [arrows={-Stealth[width=1.6mm, length=1.8mm]}] (-3,0) -- (3.5,0) node[below]{$\quad\te{Re}(w)$};
\draw [arrows={-Stealth[width=1.6mm, length=1.8mm]}] (0,-3) -- (0,3.3) node[right]{$\te{Im}(w)$};
\begin{scope}[xshift=8.5cm]
  \draw[fill=white] (0,0) circle (0.5cm);
  \draw[white, fill=white] (0.5,0) circle (0.2cm);
  \draw[fill=white] (2.5,-0.2) circle (0.3cm);
  \draw[white, fill=white] (2.5,0) circle (0.35cm);
  \draw(2.5,-0.9)node{$\hat C_{\pi}^0(\varepsilon)$};
  \draw(2.5,0.2) -- (0.43,0.2);
  \draw(2.2,-0.2) -- (0.43,-0.2);
  \draw(1.47,0.2) node[above]{$P_1^{-1}$};
  \draw[arrows={-Stealth[width=1.6mm, length=1.8mm]}](1.37,0.2)--(1.36,0.2);
  \draw(1.47,-0.2) node[below]{$P_1$};
  \draw[arrows={-Stealth[width=1.6mm, length=1.8mm]}](1.54,-0.2)--(1.55,-0.2);
  %
  \draw[arrows={-Stealth[width=1.6mm, length=1.8mm]}](2.8,-0.2)--(2.5,-0.2);
  \draw[arrows={-Stealth[width=1.6mm, length=1.8mm]}](0.25,0.433)--(0.24,0.44);
  \draw[arrows={-Stealth[width=1.6mm, length=1.8mm]}](-0.25,-0.433)--(-0.24,-0.44);
  %
  \draw (-0.9,0.7)node{$C_0^{2\pi}(\varepsilon)$};
  \draw (0,0)node{$\bullet$} ;
  %
  \draw (2.5,0)node{$\bullet$} ;
  \draw(3.2,0.3)node{$w=1$};
  \draw [arrows={-Stealth[width=1.6mm, length=1.8mm]}] (-3,0) -- (3.5,0) node[below]{$\quad\te{Re}(w)$};
  \draw [arrows={-Stealth[width=1.6mm, length=1.8mm]}] (0,-3) -- (0,3.3) node[right]{$\te{Im}(w)$};
  \end{scope}
\end{tikzpicture}
\end{center}
\caption{Deformation of the unit circle $C_0^{2\pi}(1)$ to the path composition
$P_1^{-1} C_0^{2\pi}(\varepsilon) P_1 \hat C_{\pi}^0(\varepsilon)$.}
\label{figNils}
\end{figure}
The virtue of deforming the path of integration is that \eqn{eqn:drin16} can
now be computed rather explicitly. First, in view of the reversal operation
contained in the definition of $\tilde {\cal P}$, the composition of paths
$\alpha$ and $\beta$ translates into a concatenation of the non-commutative
series with reversed order,
\begin{equation}
  \tilde{\cal P} \exp\left( \int_{\alpha\beta} \omega \right) = 
  \tilde{\cal P} \exp\left( \int_{\beta} \omega \right) \tilde{\cal P} \exp\left( \int_{\alpha} \omega \right)
\end{equation}
regardless of the differential form $\omega$. Hence, the equality of homotopy
classes of paths (relative to the tangent vector $-\frac{\pd}{\pd w}$ at $1$) 
$[C_0^{2\pi}(1)]=[P_1^{-1} C_0^{2\pi}(\varepsilon)P_1 \hat
C_{\pi}^0(\varepsilon)]$ allows to rewrite \eqn{eqn:drin16} as
\begin{equation}
\lim_{\tau \rightarrow i \infty} A^\textrm{twist}_{(\Lambda_N+\Lambda_N\tau)\setminus\Lambda_N^\times}(\tau) 
=e^{i \pi t}\Phi(\tilde{y}_N,t)e^{2\pi i\tilde{y}_N}\Phi(\tilde{y}_N,t)^{-1},
\label{eqn:drin17}
\end{equation}
where $\Phi(\tilde{y}_N,t)$ denotes the Drinfeld associator \cite{Drinfeld2}. In deducing
\eqn{eqn:drin17}, we have used the identities
\begin{align}
\Phi(\tilde{y}_N,t) &=  \tilde{\cal P} \exp\bigg( \, \int \limits_{P_1} \Big[  \frac{ \tilde{y}_N }{w}+   \frac{ t }{w-1} \Big] \,  \dd w \bigg)
\,,\\
e^{2\pi i \tilde{y}_N } &=  \tilde{\cal P} \exp\bigg( \, \int \limits_{C_0^{2\pi}(\varepsilon)} \Big[  \frac{ \tilde{y}_N }{w}+   \frac{ t }{w-1} \Big] \,  \dd w \bigg) \,,\\
\label{drin18}
e^{i \pi t }  &=  \tilde{\cal P} \exp\bigg( \, \int \limits_{\hat C_{\pi}^0(\varepsilon)} \Big[  \frac{ \tilde{y}_N }{w}+   \frac{ t }{w-1} \Big] \,  \dd w  \bigg) \,.
\end{align}
Since the coefficients of $\Phi$ are $\ZQ$-linear combinations of MZVs
\cite{LeMura}, an implementation of \eqn{eqn:drin17} using a standard computer
algebra system can be used to explicitly write the constant terms of teMZVs for
generic twists as $\ZQ[(2\pi i)^{-1}]$-linear combinations of MZVs.
\Eqn{eqn:drin17} is a generalization of a similar formalism for eMZVs, which
has been established in \rcites{Enriquez:EllAss} (see also \cite{Broedel:2015hia}).  Examples for
constant terms of teMZVs computed via  \eqn{eqn:drin17} are gathered in
\appref{app:emzvA}.


\subsection{Constant terms for all twists}
\label{ssec:realboundary}

So far, we have only considered the constant terms of teMZVs with generic twists. The presence of proper rational
twists $b \in \Lambda^\times_N$ requires a separate discussion due to the additional features of 
the corresponding weighting function $f^{(1)}(z-b,\tau)$: 
\begin{itemize}
\item its simple pole in the \textit{interior} of the domain of integration requires the regularization 
procedure of \subsecref{ssec:regteMZVs}
\item its $\tau \to i\infty$ limit introduces singularities $\frac{\dd w}{w-\zeta}$ with $\zeta$ denoting
a root of unity
\end{itemize}
In order to facilitate the computation of the constant terms of teMZVs including proper rational twists, 
we again introduce a generating series for bookkeeping purposes 
\begin{align}
A^\textrm{twist}_{\Lambda_N+\Lambda_N\tau}(\tau)
&=\sum_{\ell \geq 0}(-1)^\ell 
\hspace{-20pt}
  \sum\limits_{
   \begin{minipage}{100pt}\vspace*{-0.8pt}\footnotesize
    \begin{centering}
    $n_1,n_2,\ldots,n_\ell\geq 0$ \\
    ${b_1,b_2,\ldots,b_\ell \in \Lambda_N+\Lambda_N\tau}$
  \end{centering}
  \end{minipage}}
\hspace{-20pt}
\omwb{n_1,n_2,\ldots,n_\ell}{\nbeta_1,\nbeta_2,\ldots,\nbeta_\ell}\ad^{n_\ell}_{x_{b_\ell}}(y)\ldots  \ad^{n_2}_{x_{b_2}}(y) \, \ad^{n_1}_{x_{b_1}}(y) \notag\\
&=\lim_{\varepsilon \to 0}\tilde {\cal P} \, \exp \bigg(
- \int_{[0,1]_{\varepsilon}} \dd z \sum_{k=0}^{\infty} \hspace{5pt}\sum_{b\in\Lambda_N+\Lambda_N\tau} f^{(k)} (z-b,\tau) \, \te{ad}_{x_{b}}^{k}(y) 
\bigg) \ ,
\label{eqn:genfunction2}
\end{align}
which generalizes \eqn{eqn:genfunction} to the case of arbitrary twists.

\subsubsection{Degeneration of weighting functions}

As in the above situation we need to determine the degeneration limit of the
weighting functions $f^{(k)}(z-b,\tau)$ as $\tau \to i\infty$. The only case,
where this degeneration limit differs from the results of the previous
subsection (cf.~\eqns{eqn:drin11a}{eqn:drin11b}) is $k=1$ and
$r=0$:
\begin{equation}
  \lim_{\tau \to i\infty}
 f^{(1)}(z-s,\tau) \, \dd z = -\frac{1}{2} \frac{ \dd w}{w} + \frac{ \dd w}{w-e^{2\pi is}} \ .
\label{eqn:drin20}
\end{equation}
Note the occurrence of the root of unity $e^{2\pi is}$. Denoting
the set of $N^{\rm th}$ roots of unity by
\begin{equation}
\mu_N =
 \big\lbrace e^{2\pi is} \, \vert \, s \in \Lambda_N \big\rbrace \,,
\end{equation}
the generalization of \eqn{eqn:drin23} to twists $b \in \Lambda_N+\Lambda_N\tau$ reads
\begin{equation}
\lim_{\tau \to i\infty}-  \dd z 
\sum_{k=0}^{\infty} 
\;
\sum_{
  \begin{minipage}{60pt}\vspace*{-0.9pt}\footnotesize
  $b \in \Lambda_N{+}\Lambda_N\tau$
\end{minipage}} 
f^{(k)}(z-b,\tau) \, \te{ad}_{x_b}^{k}(y) = \tilde{y}_N \, \frac{ \dd w}{w} + \sum_{\zeta \in \mu_N}t_{\zeta} \, \frac{ \dd w }{w-\zeta} \ ,
\label{eqn:drin23b}
\end{equation}
where\footnote{Note that the definition of $\tilde{y}_N$ for all twists
is different from \eqn{eqn:drin24} in the previous subsection which is
valid for generic twists only.}
\begin{align}
\tilde{y}_N &= -\sum_{b \in \Lambda_N}\frac{\ad_{x_b}}{e^{2\pi i\ad_{x_b}}-1}(y)+
  \hspace{-2pt}
\sum_{\begin{minipage}{80pt}\vspace{-1pt}\footnotesize
  $b \in (\Lambda_N+\Lambda_N\tau) \setminus \Lambda_N$
\end{minipage}}
\frac{\ad_{x_b}e^{-2\pi ir\ad_{x_b}}}{e^{-2\pi i\ad_{x_b}}-1}(y)  \notag \\
 t_{\zeta}  &=  [y,x_{s}] \ , \ \ \ \ \mbox{for} \ \zeta = e^{2\pi i s} \in \mu_N \ .
\label{eqn:drin24b}
\end{align}
\Eqn{eqn:drin24b} is the generalization of \eqn{eqn:drin24} to
arbitrary twists in the lattice $\Lambda_N+\Lambda_N\tau$, and can be proved
along the lines of the previous subsection. In particular, the expression for
$\tilde{y}_N$ follows by repeating the steps which have been detailed around \eqn{explain}.

\subsubsection{Deforming the integration contour}

Now the image of the integration contour $[0,1]_{\varepsilon}$ under the transformation $z \mapsto w=e^{2\pi iz}$ is 
the unit circle around $0$, which is dented at the roots of unity $e^{2\pi is}  \in \mu_N$ as pictured in \figref{figdent} below.
However, the point $w=1$ is special and will be taken care of by the regularization of 
\subsecref{ssec:constant}. Similar to the situation above, the dented unit circle is homotopic to $[P_1^{-1}
  C_0^{2\pi}(\varepsilon)P_1 \hat C_{\pi}^0(\varepsilon)]$ as depicted in \figref{figdent} for twists in $\Lambda_3$.
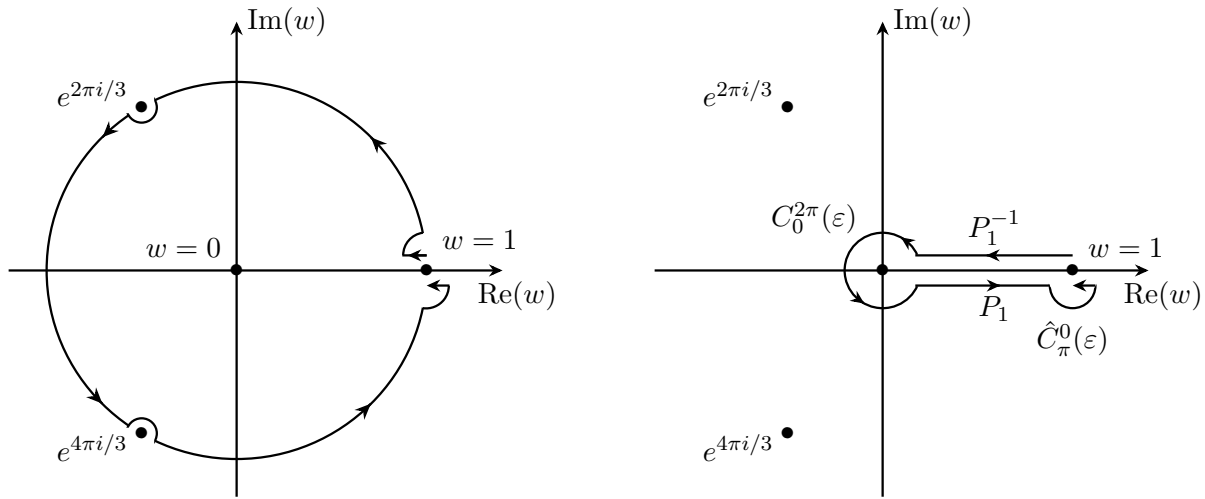
\begin{figure}[H]
\begin{center}
\begin{tikzpicture}[line width=0.30mm]
\draw (2.5,0) arc (0:360:2.5cm);
\draw[white, fill=white] (2.6,0) circle (0.5cm);

\draw (2.5-0.05,-0.5) arc (-98:0:0.3cm);
\draw[arrows={-Stealth[width=1.6mm, length=1.8mm]}](2.8,-0.2)--(2.5,-0.2);

\draw (2.5-0.05, 0.5) arc (98:180:0.3cm);
\draw[arrows={-Stealth[width=1.6mm, length=1.8mm]}](2.5,0.2)--(2.25,0.2);
\draw(2.5,0.2)--(2.18,0.2);

\draw[fill=white] (-1.25,2.16) circle (0.2cm);
\draw[fill=white] (-1.25,-2.16) circle (0.2cm);
\draw[white, fill=white] (-1.1,2.10) rectangle (-1.7,2.5);
\draw[white, fill=white] (-1.1,-2.10) rectangle (-1.7,-2.5);
%
\draw (-1.25,2.16) node{$\bullet$} ;
\draw (-1.25,-2.16) node{$\bullet$} ;
\draw(-1.9,2.35)node{$e^{2\pi i /3}$};
\draw(-1.9,-2.35)node{$e^{4\pi i /3}$};
\draw[arrows={-Stealth[width=1.8mm, length=2.1mm]}](1.767,1.765)--(1.766,1.766);
\draw[arrows={-Stealth[width=1.8mm, length=2.1mm]}](-1.767,-1.765)--(-1.766,-1.766);
\draw[arrows={-Stealth[width=1.8mm, length=2.1mm]}](-1.765,1.767)--(-1.766,1.766);
\draw[arrows={-Stealth[width=1.8mm, length=2.1mm]}](1.765,-1.767)--(1.766,-1.766);
%
%
\draw (0,0)node{$\bullet$} ;
\draw(-0.7,0.3)node{$w=0$};
\draw (2.5,0)node{$\bullet$} ;
\draw(3.2,0.4)node{$w=1$};
\draw [arrows={-Stealth[width=1.6mm, length=1.8mm]}] (-3,0) -- (3.5,0) node[below]{$\quad\te{Re}(w)$};
\draw [arrows={-Stealth[width=1.6mm, length=1.8mm]}] (0,-3) -- (0,3.3) node[right]{$\te{Im}(w)$};
	\begin{scope}[xshift=8.5cm]
		\draw[fill=white] (0,0) circle (0.5cm);
		\draw[white, fill=white] (0.5,0) circle (0.2cm);
		\draw[fill=white] (2.5,-0.2) circle (0.3cm);
		\draw[white, fill=white] (2.5,0) circle (0.35cm);
		\draw(2.5,-0.9)node{$\hat C_{\pi}^0(\varepsilon)$};
                \draw (-1.25,2.16) node{$\bullet$} ;
                \draw (-1.25,-2.16) node{$\bullet$} ;
                \draw(-1.9,2.35)node{$e^{2\pi i /3}$};
                \draw(-1.9,-2.35)node{$e^{4\pi i /3}$};
		\draw(2.5,0.2) -- (0.43,0.2);
		\draw(2.2,-0.2) -- (0.43,-0.2);
		\draw(1.47,0.2) node[above]{$P_1^{-1}$};
                \draw[arrows={-Stealth[width=1.6mm, length=1.8mm]}](1.37,0.2)--(1.36,0.2);
		\draw(1.47,-0.2) node[below]{$P_1$};
                \draw[arrows={-Stealth[width=1.6mm, length=1.8mm]}](1.54,-0.2)--(1.55,-0.2);
		%
                \draw[arrows={-Stealth[width=1.6mm, length=1.8mm]}](2.8,-0.2)--(2.5,-0.2);
                \draw[arrows={-Stealth[width=1.6mm, length=1.8mm]}](0.25,0.433)--(0.24,0.44);
                \draw[arrows={-Stealth[width=1.6mm, length=1.8mm]}](-0.25,-0.433)--(-0.24,-0.44);
		%
		\draw (-0.9,0.7)node{$C_0^{2\pi}(\varepsilon)$};
		\draw (0,0)node{$\bullet$} ;
		%
		\draw (2.5,0)node{$\bullet$} ;
		\draw(3.2,0.3)node{$w=1$};
		\draw [arrows={-Stealth[width=1.6mm, length=1.8mm]}] (-3,0) -- (3.5,0) node[below]{$\quad\te{Re}(w)$};
		\draw [arrows={-Stealth[width=1.6mm, length=1.8mm]}] (0,-3) -- (0,3.3) node[right]{$\te{Im}(w)$};
		\end{scope}
\end{tikzpicture}
\end{center}
\caption{Deformation of the dented unit circle to the path composition $P_1^{-1} C_0^{2\pi}(\varepsilon) P_1 \hat C_{\pi}^0(\varepsilon)$.
Proper rational twists $s=0,\frac{1}{N},\ldots, \frac{N-1}{N}$ are mapped to unit roots $\zeta = e^{2\pi is} \in \mu_N$.}
\label{figdent}
\end{figure}
\noindent
Hence, \eqn{eqn:drin17} can be generalized to
\begin{align}
  \lim_{\tau \rightarrow i \infty} A^\textrm{twist}_{\Lambda_N+\Lambda_N\tau}(\tau) 
&=e^{i \pi t_1}\Phi_N(\tilde{y}_N,(t_{\zeta})_{\zeta \in \mu_N})e^{2\pi i\tilde{y}_N}
\Phi_N(\tilde{y}_N,(t_{\zeta})_{\zeta \in \mu_N})^{-1} \ ,  \label{eqn:drin29}
\end{align}
where $\Phi_N(e_0,(e_{\zeta})_{\zeta \in \mu_N})$ is the cyclotomic version of
the Drinfeld associator \cite{Enriquez:Cyclotomic}, defined by
\begin{equation}
\Phi_N(e_0,(e_{\zeta})_{\zeta \in \mu_N}) = \tilde {\cal P} \, \exp \bigg(
 \int^1_0 \bigg[ \frac{e_0}{w}+\sum_{\zeta \in \mu_N}\frac{e_{\zeta}}{w-\zeta} \bigg] \,  \dd w  
\bigg) \ .
\end{equation}
Since $\Phi_N$ is the generating series of $N$-cyclotomic MZVs
\cite{Goncharov:2001iea,Racinet:Doubles,Zhao:2008,Glanois:Thesis}, the constant terms of
teMZVs for arbitrary twists in the lattice $\Lambda_N+\Lambda_N\tau$ are
$\ZQ[(2\pi i)^{-1}]$-linear combinations of cyclotomic MZVs. Definitions and
properties of cyclotomic MZVs are collected in \appref{app:cyclotomic}, and
examples for constant terms of teMZVs
with proper rational twists can be found in \appref{app:emzvB}.

As exemplified by $ \omwbc{1}{\oneh} = - i \pi $, it is the
regularization of divergences occurring for proper rational twists, which
spoils the validity of the reflection property \eqn{eqn:reflectioneMZV} for
letters $B = \begin{smallmatrix} 1 \\ b \end{smallmatrix}$ with
  $b\in\Lambda_N^\times$. It would be
interesting to identify an alternative regularization scheme where
\eqn{eqn:reflectioneMZV} is preserved.

\section{One-loop open-string amplitude}
\label{sec:oneloop}

This section is devoted to the discussion of the appearance of teMZVs in a
physics context -- in the low-energy expansion of scattering amplitudes in
string theory\cite{Green:1987sp, Green:1987mn, Polchinski:1998rq,
Polchinski:1998rr, Blumenhagen:2013fgp}. In general, string amplitudes at lower 
loop orders\footnote{In the Ramond--Neveu--Schwarz approach to superstring theory, ($g\geq 5$)-loop 
amplitudes cannot be derived from the moduli space of ordinary Riemann surfaces
since the moduli space of the required super Riemann surfaces of genus $g\geq 5$ 
is not split \cite{Donagi:2013dua}.} $g\leq 2$, possibly also at $g=3,4$, can be represented by 
integrals over the moduli space of punctured Riemann surfaces of genus $g$.
For one-loop scattering of open strings, the Riemann surfaces or
\textit{worldsheets} of interest are the cylinder and the M\oe{}bius strip. The
punctures -- the insertion points of vertex operators for external states --
are then integrated over the boundary components of the worldsheets. These
boundary integrals are weighted by traces over Lie-algebra generators $t^a$
associated with the gauge degrees of freedom of the open-string states: Each
boundary component contributes a separate trace factor, in each of which the
order of multiplication matches the ordering of the associated punctures. 

A convenient parametrization of the one-loop open-string topologies -- the
cylinder and the M\oe{}bius strip -- can be obtained starting from the torus
 by restricting the modular parameter to $\tau_C = i t$ and to
$\tau_M = i t+\frac{1}{2}$, respectively, where $t \in \ZR_+$.
\begin{figure}
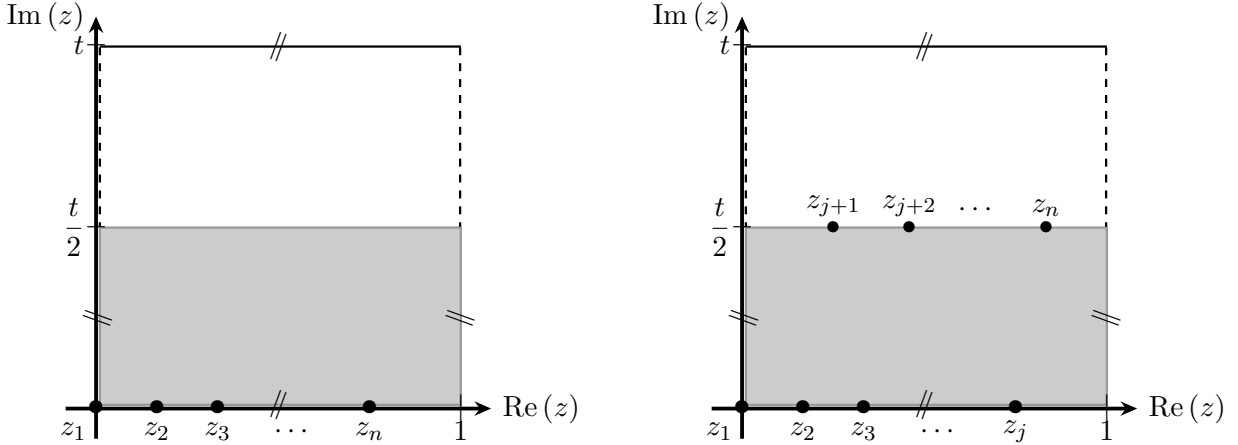

\begin{center}
\tikzpicture[line width=0.5mm]
\draw[arrows={-Stealth[width=1.9mm, length=2.2mm]}] (-0.4,0) -- (5.2,0)node[right]{$\Re(z)$};
\draw[arrows={-Stealth[width=1.9mm, length=2.2mm]}] (0,-0.4) -- (0,5.2)node[left]{$\Im(z)$};
\draw[dashed, line width=0.3mm] (0.05,4.79) -- (0.05,2.41);
\draw[dashed, line width=0.3mm] (4.79,4.79) -- (4.79,2.41);
\draw[line width=0.3mm](0.05,4.8) -- (4.8,4.8);
\draw[fill=gray, opacity=0.4,line width=0.3mm] (0.05,0.05) rectangle (4.8,2.4);
\draw (0,0.02)node{$\boldsymbol{\bullet}$} ;
\draw (0.8,0.02)node{$\boldsymbol{\bullet}$} ;
\draw (1.6,0.02)node{$\boldsymbol{\bullet}$} ;
\draw (3.6,0.02)node{$\boldsymbol{\bullet}$} ;
\draw (-0.3,-0.3)node{$z_1$} ;
\draw (0.8,-0.3)node{$z_2$} ;
\draw (1.6,-0.3)node{$z_3$} ;
\draw (3.6,-0.3)node{$z_n$} ;
\draw (2.6,-0.3)node{$\cdots$} ;
\draw (4.8,0)node{$|$};
\draw (4.8,-0.3)node{$1$};
\draw (0,2.41)node{$-$}node[left]{$\displaystyle \frac{t}{2}$};
\draw (0,4.82)node{$-$}node[left]{$\displaystyle t$};
\draw (0.02,1.2) node{\rotatebox[origin=c]{-110}{$| \hspace{-1pt} |$}};
\draw (4.79,1.2) node{\rotatebox[origin=c]{-110}{$| \hspace{-1pt} |$}};
\draw (2.4,0.05) node{\rotatebox[origin=c]{-20}{$| \hspace{-1pt} |$}};
\draw (2.4,4.8) node{\rotatebox[origin=c]{-20}{$| \hspace{-1pt} |$}};
\scope[xshift=8.5cm]
\draw[arrows={-Stealth[width=1.9mm, length=2.2mm]}] (-0.4,0) -- (5.2,0)node[right]{$\Re(z)$};
\draw[arrows={-Stealth[width=1.9mm, length=2.2mm]}] (0,-0.4) -- (0,5.2)node[left]{$\Im(z)$};
\draw[dashed, line width=0.3mm] (0.05,4.79) -- (0.05,2.41);
\draw[dashed, line width=0.3mm] (4.79,4.79) -- (4.79,2.41);
\draw[line width=0.3mm](0.05,4.8) -- (4.8,4.8);
\draw[fill=gray, opacity=0.4,line width=0.3mm] (0.05,0.05) rectangle (4.8,2.4);
\draw (0,0.02)node{$\boldsymbol{\bullet}$} ;
\draw (0.8,0.02)node{$\boldsymbol{\bullet}$} ;
\draw (1.6,0.02)node{$\boldsymbol{\bullet}$} ;
\draw (3.6,0.02)node{$\boldsymbol{\bullet}$} ;
\draw (-0.3,-0.3)node{$z_1$} ;
\draw (0.8,-0.3)node{$z_2$} ;
\draw (1.6,-0.3)node{$z_3$} ;
\draw (3.6,-0.3)node{$z_j$} ;
\draw (2.6,-0.3)node{$\cdots$} ;
\draw (1.2,2.4)node{$\bullet$} node[above]{$z_{j+1}$};
\draw (2.2,2.4)node{$\bullet$} node[above]{$z_{j+2}$};
\draw (4,2.4)node{$\bullet$} node[above]{$z_{n}$};
\draw (3.1,2.4)node{$ $} node[above]{$\cdots$};
\draw (4.8,0)node{$|$};
\draw (4.8,-0.3)node{$1$};
\draw (0,2.41)node{$-$}node[left]{$\displaystyle \frac{t}{2}$};
\draw (0,4.82)node{$-$}node[left]{$\displaystyle t$};
\draw (0.02,1.2) node{\rotatebox[origin=c]{-110}{$| \hspace{-1pt} |$}};
\draw (4.79,1.2) node{\rotatebox[origin=c]{-110}{$| \hspace{-1pt} |$}};
\draw (2.4,0.05) node{\rotatebox[origin=c]{-20}{$| \hspace{-1pt} |$}};
\draw (2.4,4.8) node{\rotatebox[origin=c]{-20}{$| \hspace{-1pt} |$}};
\endscope
\endtikzpicture
\caption{Worldsheets of cylinder topology are mapped to the shaded region, see
  the left and the right panel for the planar and the non-planar case,
  respectively. The punctures on the boundaries are taken to have coordinates
  with $\Re(z_j) \in [0,1]$ and either $\Im(z_j) =0$ or $\Im(z_j)
  =\frac{t}{2}$.  The identification of edges is marked by
  \protect\rotatebox[origin=c]{-110}{$| \hspace{-1pt} |$} and
  \protect\rotatebox[origin=c]{-20}{$| \hspace{-1pt} |$}, respectively, and
  inherited from a torus with modular parameter $\tau=it$. The M\oe{}bius
  topology is not drawn here, because all its contributions to the amplitude
  can be inferred from the planar cylinder topology. This can be seen by the
  change of variables described in \rcite{Green:1984ed}.}
\label{fig:cyl}
\end{center}
\end{figure}
In both cases, the boundary is parametrized via $z\in \mathbb C/(\mathbb Z
+ \mathbb Z \tau)$ with $\Re(z) \in [0,1]$ and $\Im(z)=0$ or $\Im(z)=\frac{t}{2}$ 
which is sometimes referred to as the closed-string channel.

In this setup, elliptic iterated integrals \eqn{eqn:defGell} appear naturally when
integrating over moduli spaces of cylinder- and M\oe{}bius-strip punctures. This is 
yet another example, where the iterated integrals on the 
boundary of open-string worldsheets yield special values of polylogarithms tailored
to the corresponding Riemann surfaces, generalizing the ubiquity of MZVs at genus zero.

Cylindrical worldsheets with all insertions on the same boundary are referred
to as \textit{planar} cylinders. For these contributions to the amplitude, all
integrals over the punctures were shown to boil down to eMZVs in
\cite{Broedel:2014vla}.  Moreover, the only difference between integrals over
punctures on the M\oe{}bius strip and those on the planar cylinder is the value
of the modular parameter $\tau$ \cite{Green:1984ed}: therefore one can
straightforwardly convert the contributions from the planar cylinder to those
of the M\oe{}bius strip by
\begin{itemize}
\item replacing $q_C  = e^{2\pi i \tau_C} = e^{-2\pi t}$ in the Fourier
  expansion of the eMZVs in the planar-cylinder contribution by $q_M = e^{2\pi
  i \tau_M} = e^{-2\pi t + i\pi} = -q_C$; this results in alternating relative
  signs between the coefficients in the $q_C$-expansion of the M\oe{}bius-strip
  contributions and the cylinder contributions.
\item inserting a factor of $\pm \frac{32}{N_G}$ for the M\oe{}bius strip to
  account for its single boundary of doubled length compared to individual
  boundary components of the cylinder \cite{Green:1984ed}. The ``$+$'' sign
  is for gauge groups $USp(N_G)$ and ``$-$'' for $SO(N_G)$. 
\end{itemize}
Hence, for a gauge group $SO(32)$, the constant term in the $q$-expansion
w.r.t.~$q_C$ which would give rise to a UV divergence upon integration over $t
\in \ZR_+$ cancels between the cylinder and the M\oe{}bius strip
\cite{Green:1984ed}.

The double-trace contributions, on the other hand, stem entirely from the cylinder topology
with punctures on two different boundaries -- \textit{non-planar} cylinder diagrams. 
We will see that the integrals over the punctures boil down to teMZVs with
purely imaginary modular parameter and twists $\nbeta \in
\{0,\frac{\tau_C}{2}\}=\{0,\frac{it}{2}\}$.

In the planar case mentioned above the link between eMZVs and the worldsheet
integral over the cylinder boundary was established as follows: Punctures on
the same boundary enter through the genus-one Green function at real arguments
which can be written as an integral over the weighting function $f^{(1)}(x)$ with 
real argument $x \in (0,1)$ \cite{Broedel:2014vla}.  Consequently, we will show
that the new ingredient in the case of the non-planar cylinder, i.e.~the
genus-one Green function for two insertions on different boundaries, is related
to integrals over $f^{(1)}(x - \tauh)$.  Hence, the non-planar contributions
may be expressed as iterated integrals on $E_\tau^\times \setminus \left\{
\tauh \right\}$ described above, which eventually lead to the teMZVs introduced
in \secref{sec:npe}.

In order to simplify the final formulas, we employ the differential equation of \secref{sec:new}
to build in all the relations among the teMZVs we encounter (see \appref{app:temzvcomputation}). 
In fact, as will be detailed in \secref{ssec:explicit}, these relations ultimately reduce all instances 
of teMZVs to eMZVs alone. Still, teMZVs are an essential tool for intermediate steps and to 
render the subsequent computations completely algorithmic.

\subsection{The four-point integrals}

We will illustrate the emergence of teMZVs through the non-planar contribution
to the four-point one-loop amplitude of the open superstring. Its dependence on
the external polarizations enters through a prefactor $K$ universal to all
worldsheet topologies \cite{Green:1982sw} and is irrelevant for the subsequent discussion. 
Then, setting $q_C=q$ and $q_M=-q$ as discussed above, the complete expression for 
the one-loop open-string four-point amplitude reads \cite{Green:1984ed}
\begin{align}
  {\cal A}_4^{\rm 1-loop} &= K\int \limits^1_{0} \frac{ \dd q }{q} \Big\{ {\rm Tr}(t^1 t^2 t^3 t^4) \,  \big[ N_G \, I_{1234}(q)  - 32  \,I_{1234}(-q)  \big] \notag \\
  & \qquad\qquad\quad+ {\rm Tr}(t^1 t^2 )\, {\rm Tr}(t^3 t^4)  I_{12|34}(q) + {\rm cyc}(2,3,4) \Big\} \; ,
  \label{eqn:singtrace}
\end{align}
where $t^a$ are traceless generators of the gauge group $SO(N_G)$ and the traces are taken in its fundamental representation. The accompanying integrals are given by\footnote{Note that
$I_{1234}(q)$ was denoted by $I_{\te{4pt}}(1,2,3,4)$ in
\rcite{Broedel:2014vla} and that $I_{12|34}(q) $ is defined with a factor of two in
comparison to the integral $h(s,u)$ in \rcite{Hohenegger:2017kqy} because we 
do not impose $x_3<x_4$ as done in the latter reference.}
\begin{align}
  I_{1234}(q) &= 
  \int^{1}_0 \dd x_4 \int^{x_4}_0 \dd x_3 \int^{x_3}_0 \dd x_2 \int^{x_2}_0 \dd x_1  \,\delta(x_1) \prod_{i<j}^4 \exp\bigg[ \frac{1}{2} s_{ij} G(x_{ij},\tau) \bigg] \; ,
  \label{eqn:itintA} \\
  I_{12|34}(q) &= 
  \int^{1}_0 \dd x_4 \int^{1}_0 \dd x_3 \int^{1}_0 \dd x_2 \int^{1}_0 \dd x_1  \,\delta(x_1)
  \label{eqn:itintB} \\
  & \ \ \ \times  \exp\bigg[ \frac{1}{2} s_{12} G(x_{12},\tau)+ \frac{1}{2} s_{34} G(x_{34},\tau)+\frac{1}{2} \sum_{i=1,2 \atop {j=3,4}}  s_{ij} G(x_{ij}-\tauh,\tau) \bigg] \ , \notag
\end{align}
where the insertion points $z_{1,2}= x_{1,2}$ and $z_{3,4}= x_{3,4} + \tauh$ of the vertex operators
are parametrized by real integration variables $x_i$ with $x_{ij} = x_i - x_j$, see \figref{fig:cyl}.
Translation invariance on a genus-one surface has been used to fix $x_1=0$
through the above delta function. The genus-one Green functions\footnote{
  As pointed out earlier, the cylinder and the M\oe{}bius-strip worldsheets for open-string one-loop amplitudes are
derived from a torus, and the restrictions of the modular parameters and the punctures can be understood in terms of
involutions. The open-string Green function \eqn{eqn:1l_prop} is constructed using the method
of images and therefore takes the same functional form as its closed-string counterpart adapted to the torus: This
follows from the localization of the open-string punctures on the boundaries of the cylinder and the Moebius-strip 
worldsheets which are in turn fixed points of the defining involutions \cite{Burgess:1986ah, Burgess:1986wt}.
}
\cite{Green:1981ya}
\begin{equation}
G(z,\tau) = \log \left| \frac{\theta_1(z,\tau)}{\theta_1'(0,\tau)}  \right|^2 - \frac{2 \pi}{\text{Im}(\tau)} \text{Im}(z)^2 
\label{eqn:1l_prop}
\end{equation}
depend on the differences of punctures $z_i$ and their second argument $\tau$ 
will often be suppressed. Considering the parametrization of the cylinder 
visualized in \figref{fig:cyl}, vertex insertions on different boundaries give rise to
arguments $x_{ij}  -\tauh$ as for
instance seen in the contribution $G(x_{13}-\tauh,\tau)$ to the exponent of
\eqn{eqn:itintB}.

Given the relations between the dimensionless
Mandelstam invariants\footnote{Mandelstam invariants are defined by $s_{ij} = \ap (k_i+k_j)^2$, 
where $k_{i}$ denote the momenta of the external open-string states with $i=1,2,3,4$ subject to 
momentum conservation $\sum_{i=1}^4 k_i=0$ and $k_i^2=0$.},
\begin{equation}
s_{34}=s_{12} \ , \ \ \ \ \ \ s_{14} = s_{23}  \ , \ \ \ \ \ \ s_{13}=s_{24}=-s_{12}-s_{23}  \, ,
\label{eqn:mand}
\end{equation}
the integrands of \eqns{eqn:itintA}{eqn:itintB} are unchanged if the Green
function \eqn{eqn:1l_prop} is shifted by a $z$-independent function. This
feature will be made use of in the following subsections. 

Configurations with three punctures on the same boundary lead to color factors
such as $ {\rm Tr}(t^1 t^2 t^3) {\rm Tr}( t^4)$ which vanish for traceless
$SO(N_G)$ generators considered in \eqn{eqn:singtrace}. Nevertheless, the
accompanying integral 
\begin{align}
  I_{123|4}(q) &= 
  \int^{1}_0 \dd x_4 \int^{1}_0 \dd x_3 \int^{x_3}_0 \dd x_2 \int_{0}^{x_2} \dd x_1  \,\delta(x_1)
  \label{eqn:itintC} \\
  & \ \ \ \times  \exp\bigg[
 \frac{1}{2}\sum_{1\leq i<j}^3 s_{ij} G(x_{ij},\tau)
 + \frac{1}{2}  \sum_{j=1}^3 s_{j4} G(x_{j4}-\tauh,\tau) \big]
   \bigg]  \notag
\end{align}
plays an important r\^{o}le for one-loop monodromy relations
\cite{Tourkine:2016bak, Hohenegger:2017kqy}.  It will be demonstrated in
\appref{app:3plus1} that \eqn{eqn:itintC} may be expanded in terms of teMZVs
using the same techniques as will be applied to the integral $I_{12|34}(q)$ in
\eqn{eqn:itintB} along with two punctures on each boundary. For both non-planar
integrals $I_{12|34}(q)$ and $I_{123|4}(q)$, our results up to and including
the order of $s_{ij}^{3}$ can be simplified to ultimately yield combinations of
eMZVs, i.e., as mentioned earlier, all of their twisted counterparts are found
to drop out at the orders considered.


\subsubsection{Analytic versus non-analytic momentum dependence}

The one-loop four-point amplitude is a non-analytic function of the Mandelstam
invariants \eqn{eqn:mand}: From the integration over $q$ in
\eqn{eqn:singtrace}, the region with $t\to\infty$ or \mbox{$q=e^{-2\pi
t}\to 0$} leads to branch cuts well-known from the Feynman integrals in
the field-theory limit \cite{Green:1982sw}. Moreover, the non-planar contribution 
$I_{12|34}(q)$ additionally integrates to kinematic poles 
in $s_{12}$, reflecting the exchange of
closed-string states between the cylinder boundaries
\cite{Green:1987mn}. Since both the poles and the branch cuts stem from the
integration over $q$, the integrals \eqns{eqn:itintA}{eqn:itintB} over the
punctures by themselves do not reflect the singularity structure of the
one-loop amplitude.

At fixed values of $q$ it is possible to separate the overall amplitude into
analytic and non-analytic parts. For the four-point closed-string one-loop
amplitude, a careful procedure to isolate the logarithmic dependence on
$s_{ij}$ has been developed in ref.\ \cite{Green:2008uj}: this method allows
for a focused study of the analytic sector where modular graph functions take
the r\^{o}le of eMZVs \cite{D'Hoker:2015foa, DHoker:2015wxz}. 

While the integrals in the expansion of \eqn{eqn:itintA} have been performed at
fixed value of $q$, integrating for instance \eqn{4ptja} below and its counterpart from the
M\oe{}bius strip over $q$ introduces divergences, also for the gauge group
$SO(32)$. The choice of regularization scheme for these divergences 
(see \cite{Hohenegger:2017kqy} for an example at the first subleading order in $\alpha'$) reflects a
particular way of splitting the analytic and non-analytic parts of the final
expression for the amplitude after integrating over $q$.


\subsubsection{The low-energy expansion in the single-trace sector}

In the following, we will study the analytic part of the non-planar one-loop
four-point amplitude by Taylor-expansion of \eqn{eqn:itintB} in $s_{ij}$ and
thereby in $\ap$, probing the low-energy behaviour. The analogous low-energy
expansion for the single-trace integral \eqn{eqn:itintA} has been performed in
\cite{Broedel:2014vla},
\begin{align}
&I_{1234}(q) = \omega(0,0,0) \, - \, 2\omega(0,1,0,0) \, (s_{12}+s_{23})  \, + \,  2 \omega(0,1,1,0,0)\,  \big( s_{12}^2  + s_{23}^2 \big)   \label{4ptja} \\
&\, - \, 2 \omega(0,1,0,1,0) \,s_{12}s_{23} \, + \, \beta_5 \, (s_{12}^3+2 s_{12}^2 s_{23} + 2s_{12} s_{23}^2+s_{23}^3)
\, + \, \beta_{2,3} \, s_{12} s_{23}(s_{12}+s_{23}) \, + \, {\cal O}(\ap^4)
\notag
\end{align}
with the following combinations of eMZVs at order $\ap^3$:
\begin{align}
\beta_5 &=  \frac{4}{3} \, \big[ \omm(0,0,1,0,0,2)+\omm(0,1,1,0,1,0) - \omm(2,0,1,0,0,0) - \zeta_2 \omm(0,1,0,0) \big]  \\
\beta_{2,3} &=\frac{\zeta_3}{12}+\frac{8\,\zeta_2}{3}\omm(0,1,0,0)-\frac{5}{18}\omm(0,3,0,0) \, .
\end{align}
It was explained in the reference that the dependence of the single-trace 
integral \eqn{4ptja} on $q$ is captured by
eMZVs at any order in $\ap$. Note that the contributions of the planar cylinder
and the M\oe{}bius strip to \eqn{eqn:singtrace} are obtained by integrating
\eqn{4ptja} over arguments $q \rightarrow e^{-2\pi t}$ and $q \rightarrow -
e^{-2\pi t}$, respectively, with $t \in \ZR_+$.

In analogy with \eqn{4ptja}, we will determine the $\ap$-expansion of the
non-planar integral \eqn{eqn:itintB} in the framework of teMZVs. Since the main
emphasis of this article is to exemplify the use of teMZVs in the
calculation of non-planar one-loop amplitudes, a detailed analysis of the
singularity structure after integration over $q$ is left for the future.


\subsection{The genus-one Green function as an elliptic iterated integral}

The link between open-string amplitudes and the framework of elliptic iterated
integrals is the holomorphic derivative\footnote{In contrast to
\secref{sec:new}, where $\partial_z$ denoted the derivative w.r.t.~the real
parameter $z$, $\partial_z$ denotes the holomorphic derivative in this section.
}
\begin{equation}
  \pd_z G(z,\tau) = f^{(1)}(z,\tau)  \ , \ \ \ \ \ \ z\in \mathbb C \setminus (\mathbb{Z} + \mathbb{Z} \tau ) \, ,
\label{eqn:del1l_prop}
\end{equation}
of the genus-one Green function \eqn{eqn:1l_prop} in the integrals
\eqns{eqn:itintA}{eqn:itintB}. It allows to rewrite the exponent in the
non-planar integral \eqn{eqn:itintB} into the form
\begin{align}
&\frac{1}{2} s_{12} G(x_{12},\tau)+ \frac{1}{2} s_{34} G(x_{34},\tau)+ \sum_{i=1,2 \atop {j=3,4}} \frac{1}{2} s_{ij} G \big(x_{ij}-\tauh,\tau \big)
\notag \\
&\ \ \ \ \ = s_{12} P(x_{12}) + s_{34} P(x_{34}) + \sum_{i=1,2 \atop {j=3,4}} s_{ij} Q(x_{ij})
 \; ,
  \label{eqn:ellipticKN}
\end{align}
where the entire dependence on the real parts\footnote{Restricting the first
  argument $G(x,\tau)$ to be real (as appropriate for our parametrization of
  the cylinder) leads to a relative factor of two between $\pd_x G(x,\tau) = 2
  f^{(1)}(x,\tau) , \ x\in \ZR$ and \eqn{eqn:del1l_prop}. This factor of two
has been neglected in early versions of \cite{Broedel:2014vla}, and it is not
altered by a complex shift $\pd_x G(x-\tauh,\tau) = 2 f^{(1)}(x-\tauh,\tau) , \
x\in \ZR$.}
$x_i \in \ZR$ 
of the punctures $z_{1,2}= x_{1,2}$ and $z_{3,4}= x_{3,4} + \tauh$ is captured 
via elliptic iterated integrals \eqn{eqn:defGell}: 
\begin{align}
P(x) &= \int^x_0 \dd y \ f^{(1)}(y) = \GLarg{1}{0 }{x}
  \label{eqn:defPP}  \\
Q(x) &= c(q)+ \int^x_0 \dd y \ f^{(1)} \left(y- \tauh \right) = c(q) + \GLarg{1}{ \tau/2 }{x}
 \; .
  \label{eqn:defQQ}
\end{align}
The appearance of the $x_{i}$-independent quantity
\begin{equation}
c(q) =  -\frac{i \pi }{2} - \frac{1}{8} \log(q) + 2 \sum_{m=1}^{\infty} \frac{1}{m} \Big[
\frac{ q^m}{1-q^m} - \frac{ q^{m/2} }{1-q^m}
\Big]  
\label{eqn:defcq}
\end{equation}
is special to the non-planar cylinder and caused by the different properties
of the Green functions $G(x_{ij})$ and $G(x_{ij}-\tauh)$ connecting
punctures on the same and different boundaries of the cylinder, respectively.

In passing to the right hand side of \eqn{eqn:ellipticKN} we have used momentum
conservation $\sum_{i<j}^4 s_{ij}=0$ to discard
\begin{align}
\frac{1}{2}G(x) -P(x) & 
= \frac{i\pi }{2} - \log(2 \pi) =  \frac{1}{2}G\big(x- \tauh \big) - Q(x) \ .
 \label{eqn:newlabel}
\end{align}
A detailed explanation of the relations above and the underlying regularization
will be given in the following two subsections.


\subsubsection{$G(x)$ versus $P(x)$}

In the case of punctures on the same boundary of the cylinder, the Green
functions with arguments $x\in [0,1]$ and $\tau \in i\ZR_+$ reduce to
$\frac{1}{2}G(x,\tau)=\log \frac{\theta_1(x,\tau)}{\theta_1'(0,\tau)} $.  Up to
an additive constant, this expression can be recovered by the following
integration with a regulator $\varepsilon>0$ in the lower limit:
\begin{align}
   \int_\varepsilon^{x} \dd y f^{(1)}(y) 
  	&= \log(\theta_1(x)) -  \log(\theta_1(\varepsilon)) \notag \\
  	&= \log(\theta_1(x)) -  \log( \theta_1'(0) ) -  \log( \varepsilon)  + \mathcal{O}(\varepsilon)  \label{eqn:Pbyint} \\
	&= \frac{1}{2}G(x) -  \log(-2\pi i \varepsilon) +  \log(-2 \pi i ) + \mathcal{O}(\varepsilon) \notag
	 \; .
\end{align}
The regularization scheme for the limit $\varepsilon\rightarrow 0$ has to be
chosen consistently with the treatment of divergent eMZVs: Following the
conventions of \cite{Broedel:2014vla}, the regularized value of an eMZV is
defined to be the constant term in an expansion\footnote{The expansion in $\log(-2\pi
i\varepsilon)$ will ensure that the constant terms of eMZVs are $\ZQ[(2\pi
i)^{-1}]$-linear combinations of MZVs, as opposed to $\ZQ[(2\pi
i)^{-1},\log(2\pi)]$-linear combinations of MZVs (as in \cite{Enriquez:Emzv},
Proposition 2.8).} w.r.t.\ $\log(-2 \pi i
\varepsilon)$, and we choose the principal branch of the logarithm, such that
$\log(-i)= -\frac{i\pi}{2}$.  Hence, $\log(-2 \pi i \varepsilon)$ is formally set to
zero in \eqn{eqn:Pbyint}, and we obtain
\begin{equation}
P(x) = \lim_{\varepsilon\rightarrow 0} {\rm Reg} \int_\varepsilon^{x} \dd y\,f^{(1)}(y) = \frac{1}{2}G(x) +  \log(-2 \pi i )
= \frac{1}{2}G(x) - \frac{i\pi }{2} + \log(2 \pi) \; ,
 \label{eqn:PbyintA} 
\end{equation}
reproducing the first equality in \eqn{eqn:newlabel}. 


\subsubsection{$G(x-\tauh)$ versus $Q(x)$}

Pairs of punctures on different boundaries of the cylinder lead to arguments
$x-\tauh$ of the Green function \eqn{eqn:1l_prop}, where $x  \in [0,1]$.
In these cases, the Green function may be related to the integral
\begin{align}
  \int_{0}^{x} \dd y \, f^{(1)}\left(y- \tauh \right) 
 &= \log \big(\theta_1\big(x - \tauh \big) \big) - i \pi x  -  \log \big(\theta_1 \big(-\tauh\big) \big)
  \notag \\
 &= \log \big| \theta_1 \big(x - \tauh \big) \big| - \frac{ i\pi }{ 2}  -  \log \big(-\theta_1 \big(\tauh \big) \big) \label{eqn:1l_prop_diffb} \\
 &= \frac{1}{2}G\big(x-\tauh \big) + \log \theta_1'(0) -\frac{1}{8} \log(q) - \frac{ i\pi }{ 2}  - \log \big(-i q^{-1/8}  \theta_4(0) \big) \notag 
 \\
&= \frac{1}{2} G \big(x-\tauh \big) 
	+ \log\left( \frac{ \theta_1'(0) }{ \theta_4(0) } \right) 
	= \frac{1}{2} G\big(x-\tauh \big)- \frac{1}{2} G\big(\tauh \big)  \; , \notag
\end{align}
without any need for regularization.  In passing to the second line, we have
chosen the principal branch of the logarithm to relate $\log(\theta_1(x -
\tau/2)) - i \pi x= \log| \theta_1(x - \tau/2) | - \frac{i \pi }{2 }$, see
\appref{app:thetas} for our conventions and several identities for the theta functions. In the 
next step, we have introduced the Green function via $ \log| \theta_1(x - \tau/2) |  = \frac{1}{2}G(x-\tau/2) + \log
\theta_1'(0) -\frac{1}{8} \log(q)$ and used the theta-function identity $
\theta_1( \tau/2) = i q^{-1/8}  \theta_4(0)$, cf.\ \eqn{nonameeq}.  

Using the infinite-product representations \eqn{eqn:conv_theta} of the theta
functions, the result of \eqn{eqn:1l_prop_diffb} can be rewritten as \cite{dlmf}
\begin{align}
 & \int\limits_{0}^{x} \dd y \, f^{(1)}\left(y- \tauh \right)  - \frac{1}{2} G(x- \tauh) =  \log\left( \frac{\theta_1'(0)}{ \theta_4(0)} \right) \notag \\
 & \ \ \ \  =  \log(2 \pi q^{1/8}) + 2 \sum_{n=1}^\infty \left[ \log( 1 - q^n ) - \log( 1 - q^{n-\oneh} ) \right] \notag \\
 & \ \ \ \  =  \log(2 \pi q^{1/8}) + 2 \sum_{m=1}^\infty \frac{1}{m}  \Big[
 \frac{ q^{m/2} }{1-q^m} - \frac{ q^m}{1-q^m} 
\Big] 
\label{moresteps} \\
  & \ \ \ \ = -\frac{i \pi }{2} + \log(2\pi) - c(q) \ .   \notag
\end{align}
In passing to the third line, we have rearranged the infinite
sums\footnote{Note the following useful intermediate expressions whose sum 
  is denoted by $-\frac{1}{2}Q_3$ in \rcite{Hohenegger:2017kqy}
\[
\sum_{n=1}^\infty \log( 1 - q^n ) = - \! \! \sum_{m,n=1}^\infty \frac{ q^{mn} }{m} = - \! \sum_{m=1}^\infty \frac{1}{m}  \frac{ q^{m} }{1-q^m}  \ , \ \ \ \ 
- \! \sum_{n=1}^\infty \log( 1 - q^{n-\frac{1}{2}} ) = \sum_{m,n=1}^\infty \frac{ q^{mn-m/2} }{m} =  \sum_{m=1}^\infty \frac{1}{m}  \frac{ q^{m/2} }{1-q^m} \; .
\]} to identify the quantity $c(q)$ in \eqn{eqn:defcq}.
In this way, one arrives at
\begin{equation}
Q(x) = c(q)+\int_0^{x} \dd y \,f^{(1)}\big(y-\tauh \big) = \frac{1}{2}G \big(x-\tauh \big)   - \frac{i\pi }{2} + \log(2 \pi) \; ,
 \label{gibmireinQ} 
\end{equation}
and the second equality in \eqn{eqn:newlabel} is confirmed.


\subsection{Non-planar contribution to the four-point amplitude}
\label{ssec:explicit}

Using the identities discussed in the previous subsections, the integral
\eqn{eqn:itintB} relevant to the non-planar cylinder can be written as
\begin{equation}
I_{12|34}(q) = \int_{12}^{34} \exp \bigg[ s_{12} P(x_{12}) + s_{34} P(x_{34}) + \sum_{i=1,2 \atop {j=3,4}} s_{ij} Q(x_{ij}) \bigg] \ ,
\label{eqn:newrep}
\end{equation}
where $P(x)$ and $Q(x)$ are given by the elliptic iterated integrals
\eqns{eqn:defPP}{eqn:defQQ} and we have introduced the following shorthand for
the integration measure:
\begin{equation}
\begin{split}
\int_{12}^{34} = \int_{0}^{1} \dd x_4 \int_{0}^{1} \dd x_3 \int_{0}^{1} \dd x_2 \int_{0}^{1} \dd x_1\, \delta(x_1) \; .
\end{split}
\label{eqn:int_bdy_np}
\end{equation}
We will now investigate the Taylor-expansion of \eqn{eqn:newrep} in the
dimensionless Mandelstam invariants \eqn{eqn:mand} and thus in $\ap$ by
expanding the exponentials $e^{s_{ij} P(x_{ij})}$ and $e^{s_{ij} Q(x_{ij})}$ in
the integrand:
\begin{equation}
I_{12|34}(q) = e^{-2s_{12}c(q)}  \int_{12}^{34} \sum_{n_{ij}=0}^{\infty} 
\frac{ (s_{12} \GLarg{1}{ 0 }{x_{12}} )^{n_{12}} ( s_{34} \GLarg{1}{ 0 }{x_{34}})^{n_{34}} }{n_{12}! \, n_{34}!}
\prod_{i=1,2 \atop {j=3,4}} \frac{ \left( s_{ij} \GLarg{1}{ \tau/2 }{x_{ij}} \right)^{n_{ij}}  }{n_{ij}!}\ .
\label{eqn:newnewrep}
\end{equation}
The punctures $x_i$ only enter via elliptic iterated integrals, and Fay
relations among the weighting functions $f^{(n)}$ guarantee that the individual
integrations over $x_i$
can always be performed in terms of further elliptic iterated integrals, see
\appref{app:deleven}. Hence, each order in $\ap$ can be expressed in terms of
teMZVs and the quantity $c(q)$ in \eqn{eqn:defcq}, where the latter will also
be related to teMZVs in \eqn{eqn:cversuslogq}. Moreover, the explicit results
up to the third order can in fact be expressed in terms of (untwisted) eMZVs only, without
the need to involve their twisted counterparts. Whether this behaviour persists
at any order in $\ap$ will be discussed in \subsecref{ssec:halfodd}.

\subsubsection{Structure of the leading orders $\ap^{\leq 3}$}

As a first step towards an expansion in terms of teMZVs, we classify the
inequivalent integrals w.r.t.\ the cycle structure of ${\rm Tr}(t^1 t^2 )\,
{\rm Tr}(t^3 t^4)$ which occur at the orders $\ap^{\leq 3}$ of
\eqn{eqn:newnewrep}: We will use the shorthand $P_{ij}=P(x_{ij})$ and
$Q_{ij}=Q(x_{ij})$ for the two integrals at order $\ap^{1}$,
\begin{equation}
d_1^1= \int_{12}^{34}  P_{12} \co d_2^1= \int_{12}^{34}  Q_{13} \ ,
 \label{npl9}
\end{equation}
the six integrals at order $\ap^{2}$,
\begin{align}
d_1^2 &= \frac{1}{2} \int_{12}^{34}  P_{12}^2 \co d^2_3 = \int_{12}^{34} P_{12} Q_{13} \co d^2_5 = \int_{12}^{34} Q_{13} Q_{14} \label{npl10} \\
d_2^2 &= \frac{1}{2} \int_{12}^{34}Q_{13}^2 \co d^2_4 = \int_{12}^{34} P_{12} P_{34} \co d^2_6 = \int_{12}^{34} Q_{13} Q_{24} \ , \notag
\end{align}
and the twelve integrals at order $\ap^{3}$:
\begin{align}
 d_1^3 &= \frac{1}{6} \int_{12}^{34} P_{12}^3 \co  \ \ \ \ \, \, d_5^3= \frac{1}{2} \int_{12}^{34} P_{12}^2 P_{34} \ , & d_9^3=   \int_{12}^{34} P_{12} Q_{13} Q_{24}  \ \phantom{.} \notag \\
  d_2^3 &= \frac{1}{6} \int_{12}^{34} Q_{13}^3 \co \ \ \ \ \, d_6^3= \frac{1}{2} \int_{12}^{34} Q_{13}^2 Q_{14} \ , & d_{10}^3=   \int_{12}^{34} P_{12} Q_{13} Q_{14}  \ \phantom{.} \label{npl11} \\
   d_3^3 &= \frac{1}{2} \int_{12}^{34} P_{12}^2 Q_{13} \co \hspace{0.1mm} d_7^3= \frac{1}{2} \int_{12}^{34} Q_{13}^2 Q_{24} \ ,& d_{11}^3=   \int_{12}^{34} P_{34} Q_{13} Q_{14} \ \phantom{.} \notag \\
    d_4^3 &= \frac{1}{2} \int_{12}^{34} P_{12} Q_{13}^2 \co \hspace{0.1mm}  d_8^3=  \int_{12}^{34} P_{12} P_{34} Q_{13} \ , & d_{12}^3=   \int_{12}^{34} Q_{13} Q_{14} Q_{23}  \ . \notag 
\end{align}
In fact, some of the above $d_i^j$ can be related via cyclicity and reflection
properties of the five-point open-string worldsheet setup where the integration measure \eqn{eqn:int_bdy_np}  is generalized to
\begin{equation}
\begin{split}
\int_{123}^{45} = \int_{0}^{1} \dd x_5 \int_{0}^{1} \dd x_4 \int_{0}^{1} \dd x_3  \int_{0}^{x_3} \dd x_2 \int_{0}^{x_2} \dd x_1 \,\delta(x_1)  \ .
\end{split}
\label{eqn:int_bdy_np5}
\end{equation}
Using $\pd_i Q_{ij}=-\pd_j Q_{ij}$ and the vanishing of $\int^1_0 \dd x_j \ \pd_j
Q_{ij}$ by double periodicity of the Green function, we find one relation
\begin{equation}
0 = \int_{123}^{45} Q_{25} \, \partial Q_{14} = \int_{123}^{45} Q_{25} \, (Q_{24}-Q_{34}) \ \ \ \Rightarrow \ \ \ d_5^2 = d_6^2
\label{npl22}
\end{equation}
among the integrals \eqn{npl10} at order $\ap^2$. The same methods 
yield the two relations
\begin{equation}
\left. \begin{array}{l} 0= \int_{123}^{45} Q_{25}^2 \pd Q_{14} 
  \\ 0 = \int_{123}^{45} P_{23} Q_{25} \pd Q_{14}
    \end{array} \right \} \ \ \ \Rightarrow \ \ \ \left\{ \begin{array}{l} 
    d_7^3 = d_6^3 \\ d_9^3 = d_{10}^3
    \end{array} \right. 
\label{npl0}
\end{equation}
among the integrals \eqn{npl11} at order $\ap^3$.

\subsubsection{teMZVs at orders $\ap^{\leq 3}$}

As a next step, we exploit the representations \eqns{eqn:defPP}{eqn:defQQ} of
$P_{ij}$ and $Q_{ij}$ to express the above $d^i_j$ in terms of teMZVs: The two
instances at order $\ap$ yield
\begin{align}
d_{1}^1	&= \int_0^1 \dd x_2 \int_0^{x_2} \dd y \; f^{(1)}(y) =  \omwb{1,0}{0,0}  
\label{eqn:d11} \\
d_{2}^1 	&=  c(q) +  \int_0^{1} \dd x_3 \int_0^{x_3} \dd y \ f^{(1)}\big(y - \tauh\big) \notag \\
&= c(q) +  \omwb{1,0}{\tauh,0} \; ,
\label{eqn:d12}
\end{align}
and we can similarly convert the five independent integrals at order $\ap^2$ to
\begin{align}
d^2_1 & = \omwb{1 , 1 , 0 }{0 , 0 , 0} 
\label{eqn:d21} \\[8pt]
d^2_2 & =  \frac{1}{2} c(q)^2 +  c(q) \omwb{1 , 0 }{ \tauh, 0 } +  \omwb{1 , 1 , 0 }{ \tauh,\tauh, 0} 
\label{eqn:d22} \\[8pt]
d^2_{3} &= d^1_1 d^1_2=  \omwb{1 , 0 }{0 , 0 } \left( c(q) +  \omwb{1 , 0 }{ \tauh,0 } \right)
\label{eqn:d23} \\
d^2_4 & =(d_1^1)^2= \omwb{1 ,  0 }{0 ,  0}^2 
\label{eqn:d24} \\[8pt]
d^2_5 & = (d_2^1)^2=  \left(  c(q) +  \omwb{1 , 0 }{ \tauh,0 } \right)^2 
\label{eqn:d25}  
\end{align}
and the ten independent integrals at order $\ap^3$ to
\begin{align}
d^{3}_{1} & =   \omwb{1 , 1 , 1 , 0 }{0 , 0 , 0, 0} 
\label{eqn:d31} \\[8pt]
d^{3}_{2} & =  \frac{1}{6} c(q)^3 + \frac{1}{2} c(q)^2 \omwb{1 , 0}{ \tauh, 0} +   c(q) \omwb{1 , 1 , 0}{ \tauh, \tauh, 0} +  \omwb{1 , 1 , 1 , 0}{ \tauh, \tauh, \tauh, 0} 
\label{eqn:d32} \\[8pt]
d^{3}_{3} &  = d^{2}_{1} d^{1}_{2} 
=  \omwb{1 , 1 , 0 }{0 , 0 , 0 } \left( c(q) +  \omwb{1 , 0}{ \tauh, 0} \right)
\label{eqn:d33} \\[8pt]
d^{3}_{4} &=  d^1_{1} d^2_{2}  =  \omwb{1 , 0}{0 , 0} \left( \frac{1}{2}  c(q)^2 +  c(q) \omwb{1 , 0}{ \tauh, 0} 
+  \omwb{1 , 1 , 0}{ \tauh, \tauh, 0} \right)
\label{eqn:d34} \\[8pt]
d^{3}_{5} & = d^1_{1} d^{2}_{1} 
=  \omwb{1 , 1 , 0 }{0 , 0 , 0} \omwb{1 , 0 }{0 , 0}
\label{eqn:d35} \\[8pt]
d^{3}_{7} &  = d^1_{2} d^2_{2}   = \left( c(q) +  \omwb{1,0}{\tauh,0} \right)  \left( \frac{1}{2}  c(q)^2 +  c(q) \omwb{1 , 0}{ \tauh, 0} 
+  \omwb{1 , 1 , 0}{ \tauh, \tauh, 0} \right)
\label{eqn:d37} \\[8pt]
d^3_{8} & = d^1_{1} d^2_{3} = (d^1_{1})^2 d^1_{2}  =  \omwb{1 , 0 }{0 , 0 } ^2 \left( c(q) +  \omwb{1 , 0 }{ \tauh, 0 } \right)
\label{eqn:d38} \\[8pt]
d^3_{9} & =  d^1_{2} d^2_{3} = d^1_{1} (d^1_{2})^2  =  \omwb{1 , 0 }{0 , 0 } \left( c(q) +  \omwb{1 , 0 }{ \tauh, 0 } \right)^2
\label{eqn:d39} \\[8pt]
d^3_{11} & =  \omwb{1 , 0 }{0 , 0 } \left( c(q) +  \omwb{1 , 0 }{ \tauh, 0 } \right)^2
+\frac{1}{3} \,\omwb{0 , 3 , 0 , 0 }{0 , 0 , 0 , 0}
\label{eqn:d311} \\[8pt]
d^3_{12} & = d^1_{2} d^2_{5} = (d^1_{2})^3   = \left( c(q) +  \omwb{1 , 0 }{ \tauh, 0 } \right)^3 \ .
\label{eqn:d312}
\end{align}
Note that $d_6^2,d_6^3$ and $d_{10}^3$ are determined by \eqns{npl22}{npl0},
and the derivation of the more involved integral $d_{11}^3$ is detailed in
\appref{app:deleven}.

\subsubsection{Assembling the orders $\ap^{\leq 3}$}

Once we apply momentum conservation \eqn{eqn:mand} to the integral
\eqn{eqn:newnewrep}, its leading orders in the $\ap$-expansion simplify to 
\begin{align}
I_{12|34}(q) = 
1\ &+\ 2 s_{12}\, (d_1^1 - d_2^1) \ +\ 
 s_{12}^2 \,(2 d_1^2 + 2 d_2^2 - 4 d_3^2 + d_4^2 + d_5^2) \ + \ s_{13}s_{23} \, (  2 d_5^2-4 d_2^2) \notag \\
 & + \ 
 2s_{12}^3\, (d_1^3 - d_2^3 - 2 d_3^3 + 2 d_4^3 + d_5^3 - d_7^3 - d_8^3 + d_9^3)  \label{npl12} \\
 & - \ 
 2s_{12}s_{13}s_{23}\,(d_{12}^3 - 3 d_2^3 + 4 d_4^3 - d_7^3 - 2 d_{11}^3) + \ {\cal O}(\ap^4) \ .
\notag
\end{align}
The $q$-dependence of the relevant teMZVs can be determined by solving the
initial-value problem set up in section \ref{sec:new}, yielding for instance
\begin{align}
\omwb{1,0}{0,0} &= 
    	- \frac{i \pi}{2} + 2\sum_{n,m=1}^\infty \frac{q^{m n}}{m}  \ , \ \ \ \ \ \ 
    	\omwb{1,0}{\tauh,0} =
	2 \sum_{n,m=1}^\infty \frac{q^{m(n-\oneh)}}{m} \; ,
  	\label{eqn:iti_ex_d11} 
\end{align}
and further examples can be found in \appref{app:emzvC}. By comparing the
$q$-expansion with the expression \eqn{eqn:defcq} for $c(q)$, we infer that
\begin{equation}
c(q) =  \omwb{1,0}{0,0}  - \omwb{1,0}{\tauh,0} - 
 \frac{1}{8} \log(q) \ ,
\label{eqn:cversuslogq}
\end{equation}
which identifies the prefactor $e^{-2s_{12}c(q)}$ in \eqn{eqn:newnewrep} as
$q^{s_{12}/4}$ multiplied by a series in teMZVs $\omwb{1,0}{0,0}$ and
$\omwb{1,0}{\tauh,0}$. Moreover, \eqn{eqn:cversuslogq} simplifies the order
$\ap^1$ of \eqn{npl12} to
\begin{equation}
I_{12|34}(q) \, \big|_{s_{12}} = 2(d_1^1 - d_2^1)  = 2 \Big\{  \omwb{1,0}{0,0}  - \omwb{1,0}{\tauh,0} - c(q) \Big\} = \frac{1}{4} \log(q) \ ,
\label{eqn:orderone}
\end{equation}
in agreement with \cite{Hohenegger:2017kqy}. Also at higher orders of \eqn{npl12}, we
convert any appearance of $c(q)$ into $-\frac{1}{8}\log(q)$ via
\eqn{eqn:cversuslogq} and obtain
\begin{align}
I_{12|34}(q) \, \big|_{s_{12}^2} &= \frac{ (\log(q)^2 ) }{32} + 2  \omwb{1,1 ,0}{0,0,0} -  \omwb{1,0}{0,0}^2 + 2  \omwb{1,1,0}{\tauh,\tauh ,0} -  \omwb{1,0}{\tauh,0}^2 \notag 
\\ 
&= \frac{ (\log(q)^2 ) }{32} + \frac{ 7\zeta_2 }{6} + 2\,\omega(0,0,2) 
 \\
I_{12|34}(q) \, \big|_{  s_{13} s_{23} }  & = 2  \omwb{1,0}{\tauh,0}^2 - 4  \omwb{1,1,0}{\tauh,\tauh,0}  =
-2 \,  \omega(0,0,2) -  \frac{2 \zeta_2 }{3} 
\end{align}
via teMZV relations \eqns{weight2Bnoq}{weight2Anoq} as well as
\begin{align}
I_{12|34}(q) \, \big|_{s_{12}^3} &=  \frac{1}{3!} \left( \frac{ \log(q) }{4} \right)^3 +  \frac{1 }{4} \log(q)  \,  \left( \frac{ 7\zeta_2 }{6} + 2\,\omega(0,0,2)  \right)
\notag \\
& \ \  + 
\frac{2}{3}   \omwb{1,0}{0,0}^3 - 2  \omwb{1,0}{0,0}  \omwb{1 ,1,0}{0,0,0}  + 2 \omwb{1,1,1,0}{0,0,0,0}  \\
& \ \ - \frac{2}{3}   \omwb{1,0}{\tauh,0}^3 + 2  \omwb{1,0}{\tauh,0}  \omwb{1 ,1,0}{\tauh,\tauh,0}  - 2 \omwb{1,1,1,0}{\tauh,\tauh,\tauh,0}
  \notag \\
&= \frac{1}{3!} \left( \frac{ \log(q) }{4} \right)^3 + \frac{1}{4} \log(q) \, \left( \frac{ 7\zeta_2 }{6} + 2\,\omega(0,0,2)  \right) - 4 \, \zeta_2  \, \omega(0,1,0,0) \notag \\
I_{12|34}(q) \, \big|_{s_{12}s_{23} s_{13}} &=  \frac{1 }{4} \log(q) \, 
\left( -2 \,  \omega(0,0,2) -  \frac{2 \zeta_2 }{3} \right)
+ \frac{4}{3} \, \omega(0,3,0,0) \notag \\
& \ \ +2   \omwb{1,0}{\tauh,0}^3 -6  \omwb{1,0}{\tauh,0}  \omwb{1 ,1,0}{\tauh,\tauh,0}  +6 \omwb{1,1,1,0}{\tauh,\tauh,\tauh,0} \\
&=  \frac{1 }{4} \log(q) \, 
\left( -2 \,  \omega(0,0,2) -  \frac{2 \zeta_2 }{3} \right) + 
 \frac{5}{3}\,\omega(0,3,0,0) + 4 \, \zeta_2 \, \omega(0,1,0,0)  - \frac{1}{2} \zeta_3
  \notag
\end{align}
via teMZV relations \eqns{weight2Dnoq}{weight2Cnoq}. As will be discussed shortly, the 
cancellation of teMZVs with nonzero twist manifests the absence of unphysical poles in
the string amplitude after integration over $q$.


\subsubsection{Summary of the orders $\ap^{\leq 3}$}

As is clear by comparing \eqn{eqn:newnewrep} with \eqn{eqn:cversuslogq}, any
appearance of $\log(q)$ can be traced back to the expansion of $q^{s_{12}/4}$.
Hence, the above orders $\ap^{\leq 3}$ can be summarized as
\begin{align}
I_{12|34}(q) & = q^{s_{12}/4} \bigg\{ 1+ s_{12}^2  \Big( \frac{ 7\zeta_2 }{6} + 2\,\omega(0,0,2)  \Big)
-2s_{13}s_{23} \Big( \frac{ \zeta_2 }{3} + \omega(0,0,2)  \Big)  \label{finalqtononzero} \\
 & \ \ - 4 \, \zeta_2 \, \omega(0,1,0,0) \, s_{12}^3 + s_{12}s_{13}s_{23} \Big(
 \frac{5}{3}\,\omega(0,3,0,0) + 4 \, \zeta_2 \, \omega(0,1,0,0)  - \frac{1}{2} \zeta_3
 \Big) + {\cal O}(\ap^4) \bigg\}\; ,
\notag
\end{align}
and the integral $I_{123|4}(q)$ in \eqn{eqn:itintC} admits a similar low-energy
expansion in terms of eMZVs only, see \eqn{31result}. While the $q$-expansions
of the above eMZVs are listed in \appref{app:emzvC}, their constant terms yield
\begin{align}
I_{12|34}(q) &= q^{s_{12}/4} \Big\{ 1+ \frac{1}{2} \zeta_2 s_{12}^2   - \frac{1 }{2} \zeta_3 s_{12}^3 +  {\cal O}(q,\ap^4)  \Big\}\; ,
\label{qtozeroA} 
\end{align}
in agreement with the all-order $\ap$ expression given in \cite{Hohenegger:2017kqy} 
\begin{equation}
I_{12|34}(q) =  \frac{ 2^{2s_{12}}  q^{s_{12}/4 }}{\pi}  \left\{ \left( \frac{ \Gamma(\tfrac{1}{2} + \tfrac{s_{12}}{2})}{\Gamma(1+ \tfrac{s_{12}}{2})}  \right)^2 
+  {\cal O}(q) \right\}\ .
\label{qtozeroB}
\end{equation}
With the above strategy and the techniques exemplified in \appref{app:deleven},
there is no limitation in obtaining higher orders of the $\ap$-expansion
\eqn{finalqtononzero}. Similarly, non-planar open-string amplitudes with five
and more external legs can be expanded along the same lines\footnote{See
  section 5.1 of \cite{Broedel:2014vla} for the analogous expansion of the
planar five-point one-loop amplitude in terms of eMZVs.} because their
integrands depend on the punctures and $\tau$ through products of $f^{(n)}$ and
Eisenstein series \cite{Tsuchiya:1988va, Stieberger:2002wk, Broedel:2014vla}.

\subsubsection{Higher orders in $\ap$ and eMZVs}
\label{ssec:halfodd}

The exclusive appearance of eMZVs at the leading orders of the $\ap$-expansion
\eqn{finalqtononzero} illustrates a general property of the non-planar integral
$I_{12|34}(q)$: apart from the prefactor $q^{s_{12}/4}$, its $q$-expansion
comprises integer powers only. This property is shared by eMZVs but not by
typical teMZVs involving twists $\tauh$. 

The absence of half-odd integer powers $q^{n+\oneh}$ with $n \in \ZN_0$ can be
explained from physical constraints on the pole structure of the open-string
amplitude: Performing the $q$-integration in the amplitude prescription
\eqn{eqn:singtrace} over expansions of the schematic form
\begin{equation}
I_{12|34}(q) =    q^{s_{12}/4 }  \sum_{n=0}^{\infty}(a_n q^n + c_n q^{n+\oneh})
 \label{justify1}
\end{equation}
yields kinematic poles at $s_{12} =-4n$ in case of integer powers of 
$q$ and at $s_{12} = -4n-2$ in case of half-odd integer powers, respectively, with $n\in\ZN_0$:
\begin{equation}
\int^1 \limits_0 \frac{ \dd q}{q} \; I_{12|34}(q) =    \sum_{n=0}^{\infty}\bigg\{ \frac{4a_n}{s_{12} + 4n} 
+ \frac{4c_n}{s_{12}+ 4n+2} \bigg\} \; .
 \label{justify2}
\end{equation}
The expansion coefficients $a_n$ and $c_n$ are understood to be formal
power series in the Mandelstam variables $s_{ij}$ accompanied by $\ZQ[(2\pi
i)^{-1}]$-linear combinations of MZVs.

The singular values of Mandelstam variables $s_{12}=\ap(k_1+k_2)^2$ in
scattering amplitudes with external momenta $k_i$ correspond to internal masses $-\ap m^2$.
As a general property of non-planar one-loop open-string string amplitudes,
the kinematic poles arising from the integration over $q$ reveal the appearance of 
closed-string modes among the internal states \cite{Green:1987mn}.
 In particular,
the poles in \eqn{justify2} with residues proportional to $a_n$ and $c_n$
signal the exchange of closed-string states with masses $m^2 = \frac{4n}{\ap}$
and $m^2 = \frac{4n+2}{\ap}$, respectively, with $n \in \ZN_0$.  However, the
closed-superstring spectrum only comprises masses $m^2 = \frac{4n}{\ap}$,
whereas states with $m^2 = \frac{4n+2}{\ap}$ cannot be found in GSO projected
string theories \cite{Green:1987sp, Green:1987mn, Polchinski:1998rr, Blumenhagen:2013fgp}. Hence,
the pole structure of $ \frac{4c_n}{s_{12}+ 4n+2}$ due to half-odd integer
powers of $q$ would signal the propagation of unphysical states and violate
unitarity if some of the $c_n$ were nonzero.

However, it is not immediately clear if an integer-power $q$-expansion
\eqn{justify1} with $c_n=0$ is necessarily expressible in terms of eMZVs. Given
that $q^{-s_{12}/4 }I_{12|34}(q)$ was argued to comprise teMZVs with twists
$\in \{0,\tauh\}$, this leads to the following, purely mathematical question:
If a linear combination of teMZVs with twists $\in \{0,\tauh\}$ is such that
its $q$-expansion has integer exponents only, can it be written as a linear
combination of eMZVs (i.e.\ teMZVs with vanishing twists) only?  

We expect that an answer to this question will necessitate a closer study of the
decomposition of teMZVs into iterated $\tau$-integrals of the functions
$h^{(n)}(b)$ in \eqn{eqn:defh}, with $b\in \{0,\tauh\}$. Given that $h^{(n)}(b)$ are 
modular forms for congruence subgroups of $\SL_2(\ZZ)$, this decomposition of teMZVs
generalizes the decomposition of eMZVs into linear combinations of iterated Eisenstein integrals
\cite{Broedel:2015hia,Matthes:Decomposition}. In particular, a natural first
step would be to prove linear independence of iterated $\tau$-integrals
comprised of the integrands $h^{(n)}(b)$ with $b \in \{0,\tauh\}$, which would
generalize results of \cite{LMS,Matthes:IterIntQMod}, and can presumably be
proved along similar lines.

In any case, based on the arguments presented in this subsection, we conjecture
that all orders in the $\ap$-expansion \eqn{finalqtononzero} are furnished by
eMZVs. The same is expected to hold for all non-planar $n$-point amplitudes 
at one loop prior to integration over $q$. It is conceivable that this can be derived from
the one-loop monodromy relations \cite{Tourkine:2016bak, Hohenegger:2017kqy}, 
and it would be interesting to work out a rigorous proof.


\section{Conclusions}
\label{sec:conclusion}

In this article, teMZVs have been introduced as iterated integrals on an
elliptic curve with multiple punctures on a lattice $\ZQ+ \ZQ \tau$. Our main
result is the identification of an initial-value problem satisfied by teMZVs, which
expresses them in terms of linear combinations of iterated $\tau$-integrals of
the weighting functions $f^{(n)}$ with coefficients given by cyclotomic MZVs. 

As an application of teMZVs in physics we have studied one-loop scattering
amplitudes of open-string states. In the non-planar sector of the four-point
amplitude, the low-energy expansion can be computed by integrals over the two
boundaries of a cylinder. A systematic procedure is established, which allows
to evaluate these integrals in terms of teMZVs. There is no conceptual
bottleneck in extending the procedure to one-loop amplitudes with an arbitrary
number of external states. Having calculated the non-planar amplitude up to the
third subleading low-energy order, we find that the results can ultimately be
simplified to eMZVs. 

The results of this article trigger a variety of questions: From a mathematical
perspective, the differential equation of teMZVs could serve as a starting
point to classify their relations and to understand the underlying algebraic
principles.  In the untwisted case, a crucial r\^{o}le was played by a certain
derivation algebra. We expect that a suitable twisted analogue of the
derivation algebra \cite{Calaque} will likewise control
the algebraic structure of teMZVs.

In a physics context, the methods of this article allow to compute higher
orders in the low-energy expansion of non-planar one-loop open-string
amplitudes and to investigate its structure. More interestingly, higher-loop
open-string amplitudes should require an extension of elliptic iterated
integrals to Riemann surfaces of higher genus and a suitable generalization of
teMZVs to accommodate multiple boundaries.


\subsection*{Acknowledgments}

We would like to thank Claire Glanois, Martin Gonzalez, David Meidinger and
Federico Zerbini for helpful discussions, and the Kolleg Mathematik und Physik
Berlin for support in various ways. JB and OS would like to thank Universit\"at
Hamburg for hospitality. The work of JB and GR is supported in part by the SFB
647 ``Raum--Zeit--Materie. Analytische und Geometrische Strukturen''.  GR is
furthermore supported by the International Max Planck Research School for
Mathematical and Physical Aspects of Gravitation, Cosmology and Quantum Field
Theory. This paper was written while NM was a Ph.D.\ student at Universit\"at
Hamburg under the supervision of Ulf K\"uhn.  The research of OS was
supported in part by Perimeter Institute for Theoretical Physics. Research at
Perimeter Institute is supported by the Government of Canada through the
Department of Innovation, Science and Economic Development and by the Province
of Ontario through the Ministry of Research, Innovation and Science.


\section*{Appendix}
\appendix


\section{Jacobi theta functions}
\label{app:thetas}

For later use we note as well the expansion of Jacobi theta functions (and
derivatives thereof) as an expansion in the parameter $q=e^{2\pi i \tau}$ \cite{dlmf}
\begin{equation}
  \begin{split}
    \theta_1(z,\tau) & = 2 q^{1/8} \text{ sin}(\pi z) \prod_{n=1}^{\infty} (1-q^n)(1 - 2q^n \text{ cos}(2\pi z) + q^{2n}) \\
    \theta_1'(0,\tau)& = 2 \pi q^{1/8} \prod_{n=1}^{\infty} (1-q^n)^3 \\
    \theta_4(z,\tau) & = \prod_{n=1}^{\infty} (1-q^n)(1 - 2q^{n-1/2} \text{ cos}(2\pi z) + q^{2n-1}) \; ,
  \end{split}
  \label{eqn:conv_theta}
\end{equation}
which all turn out to be positive given $z \in [ 0,1 ]$ and $q \in [ 0,1 ]$.
Furthermore, a periodicity property useful for calculating
\eqn{eqn:1l_prop_diffb} as well as a relation between $\theta_1$ and $\theta_4$
reads
\begin{align}
  \theta_1(z+m+n \tau,\tau) &= (-1)^{n+m} q^{-n^2/2} e^{-2\pi i n z}\,\theta_1(z,\tau)
\\
  \theta_4(z,\tau) &= -i e^{i\pi z}q^{1/8}\,\theta_1(z+\tauh,\tau)\,.
  \label{nonameeq}
\end{align}
%


\section{Weighting functions}
\label{app:weighting}
The weighting functions $f^{(n)}(z,\tau)$ arise as expansion coefficients of the
doubly-periodic completion $\Omega(z,\alpha,\tau)$ of the Eisenstein--Kronecker series $F(z,\alpha,\tau)$,
\begin{equation}
\Omega(z,\alpha,\tau)= 
\exp \bigg( 2\pi i \alpha\, 
\frac{\Im(z)}{\Im(\tau)} \bigg) F(z,\alpha,\tau)=
\sum_{n=0}^{\infty}f^{(n)}(z,\tau)\alpha^{n-1} \, ,
\label{alt5}
\end{equation}
which, in turn is given by \cite{Kronecker,BrownLev}
\begin{equation}
F(z,\alpha,\tau) = 
\frac{\theta_1'(0,\tau)\theta_1(z+\alpha,\tau)}{\theta_1(z,\tau)\theta_1(\alpha,\tau)} \ .
\label{alt1}
\end{equation}
The odd Jacobi theta function $\theta_1$ is defined in \appref{app:thetas}, and the derivative
with respect to the first argument is denoted by a tick. For 
real $z$, the expressions in \eqns{alt5}{alt1} agree, and the lowest-order examples
of $f^{(n)}$ are spelt out in \eqn{eqn:expl}. In fact, $f^{(1)}$ is the only weighting function with a 
simple pole on the lattice $\ZZ+\ZZ\tau$, while all $f^{(n)}$ with $n\neq
1$ are non-singular on the entire elliptic curve.

Both the Eisenstein--Kronecker series $F(z,\alpha,\tau) $ and its doubly
periodic completion $\Omega(z,\alpha,\tau) $ satisfy the Fay identity
\begin{equation}
  \begin{split}
    \Omega(z_1,\alpha_1,\tau) \Omega(z_2,\alpha_2,\tau) & = 
    \Omega(z_1,\alpha_1 + \alpha_2,\tau) \Omega(z_2 - z_1,\alpha_2,\tau)\\
    & \quad + \Omega(z_2,\alpha_1 + \alpha_2,\tau) \Omega(z_1 - z_2,\alpha_1,\tau) \,.
  \end{split}
  \label{fayfay_Om}
\end{equation}
Furthermore, the Eisenstein--Kronecker series satisfies the mixed heat equation
\begin{equation}
  2\pi i\,\pd_\tau F(z,\alpha,\tau) = \pd_\alpha \pd_z F(z,\alpha,\tau) \; .
\label{mixedheat}
\end{equation}
Both the Fay identity \eqn{fayfay_Om} and the mixed heat equation
\eqref{mixedheat} are relevant in the calculations of \secref{sec:new}.
Starting from the quasi-periodicity of the Eisenstein--Kronecker series,
\begin{equation}
  F(z+1,\alpha,\tau) = F(z,\alpha,\tau) \; , \qquad F(z+\tau,\alpha,\tau) = \exp(-2 \pi i \alpha) F(z,\alpha,\tau) \; ,
  \label{eqn:period_F}
\end{equation}
and its reflection property
\begin{equation}
F(-z,-\alpha,\tau)=-F(z,\alpha,\tau) \; ,
  \label{eqn:period_F1}
\end{equation}
it is straightforward to derive properties \eqn{eqn:fparity1} of the weighting functions $f^{(n)}$.


\section{A teMZV with proper rational twist $b=\oneh$} 
\label{app:example}

We illustrate the definition of teMZVs in the case of proper
rational twists $b \in \Lambda_N^{\times} =
\{\frac{1}{N},\ldots,\frac{N-1}{N}\}$, via an explicit computation of $\omwb{0
, 1}{0 , \oneh}$.  Our starting point is the definition \eqn{eqn:rigorous} of
teMZVs with twists $b_i \in \left\{ 0 , \oneh \right\}$ through the integral
\begin{align}
  \omwb{n_1 , n_2 , \dots , n_{\ell}}{b_1 , b_2 , \dots , b_\ell}
  =
  \lim_{\varepsilon \rightarrow 0}
  \int\limits_{\alpha_1 \delta_{\varepsilon} \alpha_2}
  f^{(n_1)}(z-b_1)\dd z_1 \, f^{(n_2)}(z-b_2)\dd z_2 \ldots f^{(n_\ell)}(z-b_\ell)\dd z_\ell \; .
  \label{eqn:def_real_twist}
\end{align}
We choose the parametrization of the individual path segments depicted in
\figref{figDeformation} as 
\begin{equation}
  \begin{split}
  	\alpha_1(t) 			&= \left(\frac{1}{2} - \varepsilon \right) t \\
  	\delta_{\varepsilon}(t) 	&=  \frac{1}{2} - \varepsilon \exp( - i \pi t ) \\
  	\alpha_2(t) 			&=  \frac{1}{2} + \varepsilon + \left(\frac{1}{2} - \varepsilon \right)t 
  \end{split}
  \label{eqn:path_para}
\end{equation}
with $t \in (0,1)$ in each case.
Then we may compute the iterated integral using the composition of paths
formula, for the smooth one-forms $\omega_i = f^{(n_i)}(z-b_i) \dd z$
(cf.~\cite{Hain:1985}, Proposition 2.9) 
\begin{align} \label{eqn:comppath}
  \int\limits_{\alpha\beta} \omega_1 \omega_2 \ldots\omega_\ell
  =&\sum_{k=0}^\ell \int\limits_{\alpha} \omega_1\omega_2\ldots\omega_k 
  \int\limits_{\beta} \omega_{k+1}\ldots\omega_\ell  \, ,
\end{align}
where the paths $\alpha, \beta$ are such that $\alpha(1)=\beta(0)$ and the
empty integral is defined to be one.

As the forms $\omega_i = f^{(n_i)}(z-b_i) \dd z$ admit an expansion in $q$ we
may treat the $q^0$ term separately from the rest, assuming the $q$-expansion
can be exchanged with the integration.  Then, as the coefficients of the
$j^{\rm th}$ power $q^j$ for $j \neq 0$ are well defined on the real line, we
may exchange the limit $\varepsilon \rightarrow 0$ with the integration and
compute this part of the integral over the much more mundane path $\gamma(t) =
t$.  Specifically, the $q$ dependent part is given by
\begin{align}
  I_{q} =
  -2 i (2 \pi i) \hspace{-0.4cm} \int\limits_{0 < t_1 < t_2 < 1} \hspace{-0.4cm} \dd t_1 \dd t_2 \sum_{n,m = 1}^{\infty} q^{m n} \sin\left(2 \pi m \left(t_2-\frac{1}{2} \right) \right) 
  =
  - 2 \sum_{n,m = 1}^{\infty} \frac{ (-1)^m q^{m n}}{m} \; .
  \label{eqn:qn_terms}
\end{align}
We note that this can be reproduced from the differential equation
(\ref{eqn:diff_eq_nemzv}) for real twists 
\begin{equation}
  \begin{split}
  	\omwb{0 , 1}{0 , \oneh} 
	&= 
	\omwbc{0 , 1}{0 , \oneh} + \int\limits_0^q \frac{\dd \log(q_1)}{-4 \pi^2} \left[ \omwb{2}{\oneh} - f^{(2)}\left(\frac{1}{2}, q_1\right) \right] \\
	&= 
	\omwbc{0 , 1}{0 , \oneh} + \int\limits_0^q \frac{\dd \log(q_1)}{-4 \pi^2} \left[ 8 \pi^2 \sum_{m,n=1}^\infty (-1)^m n \, q_1^{m n} \right] \; ,
  \end{split}
  \label{eqn:de_w010h}
\end{equation}
where the integration constant $\omwbc{0 , 1}{0 , \oneh}$ remains to be
determined.

Application of \eqn{eqn:comppath} for the constant term $f_0^{(n)}(z-b)$ of the
$q$ expansion of $f^{(n)}(z-b)$ yields, bearing in mind that $f^{(0)}(z)=1$,  
\begin{equation}
  \begin{split}
    	I_0 &=
        \int\limits_{\alpha_1 \delta_{\varepsilon} \alpha_2} \hspace{-0.2cm}
        \dd z_1 \, f^{(1)}_0\left(z_2 - \frac{1}{2} \right) \dd z_2 \\
  	&=
  	\int\limits_{\alpha_1} \dd z_1 \, f^{(1)}_0\left(z_2 - \frac{1}{2} \right)\dd z_2
  	+ \int\limits_{\delta_{\varepsilon}} \dd z_1 \, f^{(1)}_0\left(z_2 - \frac{1}{2} \right) \dd z_2
	+ \int\limits_{\alpha_2}\dd z_1 \, f^{(1)}_0\left(z_2 - \frac{1}{2} \right) \dd z_2  \\
  	& \quad
	+ \int\limits_{\alpha_1}  \dd w \int\limits_{\delta_{\varepsilon}} f^{(1)}_0\left(z - \frac{1}{2} \right) \dd z
  	+ \int\limits_{\alpha_1}  \dd w \int\limits_{\alpha_2} f^{(1)}_0\left(z - \frac{1}{2} \right) \dd z
  	+ \int\limits_{\delta_{\varepsilon}} \dd w \int\limits_{\alpha_2} f^{(1)}_0\left(z - \frac{1}{2} \right) \dd z \; .
  \end{split}
  \label{eqn:comppath_q0}
\end{equation}
The individual integrals are given by
\begin{align}
  	\int\limits_{\alpha_1} \dd z_1\, f^{(1)}_0\left(z_2-\frac{1}{2}\right) \dd z_2
	&=
	\int\limits_{0 < t_1 < t_2 < 1 } \hspace{-0.4cm}  \dd t_1 \dd t_2 \, \left(\frac{1}{2} - \varepsilon\right)^2 \pi \cot \left( \pi \left[ \left(\frac{1}{2}-\varepsilon \right)t_2 -\frac{1}{2} \right] \right)
	\notag \\
	&=\frac{\log(2)}{2} + \frac{ \log(\pi \varepsilon) }{2} + \mathcal{O}(\varepsilon)
	\label{eqn:i1}\\
  	\int\limits_{\delta_{\varepsilon}} \dd z_1 \, f^{(1)}_0 \left(z_2-\frac{1}{2} \right) \dd z_2
	&=
	\int\limits_{0 < t_1 < t_2 < 1 } \hspace{-0.4cm}  \dd t_1 \dd t_2 \, (i \pi)^2 \varepsilon^2 e^{- i \pi (t_1+t_2)} 
	\pi \cot( - \pi \varepsilon e^{- i \pi t_2})
	\notag \\
	&= \mathcal{O}( \varepsilon)
	\label{eqn:i2}\\
	\int\limits_{\alpha_2}  \dd z_1 \, f^{(1)}_0\left(z_2-\frac{1}{2} \right)  \dd z_2
	&=
	\int\limits_{0 < t_1 < t_2 < 1 } \hspace{-0.4cm}  \dd t_1 \dd t_2 \, \left(\frac{1}{2} - \varepsilon\right)^2 \pi \cot\left( \pi \left[  \left(\frac{1}{2}-\varepsilon \right)t_2 + \varepsilon \right] \right)
	\notag \\
	&= \frac{\log(2)}{2} + \varepsilon \log( \pi \varepsilon) + \mathcal{O}(\varepsilon)
	\label{eqn:i3}\\
	\int\limits_{\alpha_1} \dd w  \int\limits_{\delta_{\varepsilon}} f^{(1)}_0\left(z-\frac{1}{2} \right) \dd z
	&= i \pi \left(  \varepsilon - \frac{1}{2} \right) 
	\label{eqn:i4}\\
  	\int\limits_{\alpha_1} \dd w \int\limits_{\alpha_2} f^{(1)}_0\left(z-\frac{1}{2} \right) \dd z
	&= \left(  \varepsilon - \frac{1}{2} \right)  \log(\sin(\pi \varepsilon))
	\label{eqn:i5}\\
  	\int\limits_{\delta_{\varepsilon}} \dd w \int\limits_{\alpha_2} f^{(1)}_0\left(z-\frac{1}{2}\right)\dd z
	&= -2 \varepsilon  \log(\sin(\pi \varepsilon)) \, .
	\label{eqn:i6}
\end{align}
Note that due to $\lim_{\varepsilon \rightarrow 0} \varepsilon \log(\sin(\pi
\varepsilon)) = 0$ the only singular contributions stem from the integrals in
\eqns{eqn:i1}{eqn:i5}, which cancel in their sum. Then, we arrive at
\begin{align}
  \lim_{\varepsilon \rightarrow 0} I_0 = - \frac{i \pi}{2} + \log(2) \, ,
  \label{eqn:c_w010h}
\end{align}
as predicted by the constant-term procedure \eqn{eqn:drin29}, see
\eqn{eqn:real_twists}.  Finally, upon combination with the $q$-series in
\eqn{eqn:qn_terms}, the desired teMZV is given by
\begin{align}
  \omwb{0 , 1}{0 , \oneh}  
  = 
  \lim_{\varepsilon \rightarrow 0} ( I_0 + I_q )
  = 
  - \frac{i \pi}{2} + \log(2) 
  - 2 \sum_{n,m = 1}^{\infty} \frac{ (-1)^m q^{m n}}{m} \; .
  \label{eqn:f_w010h} 
\end{align}


\section{MZVs and cyclotomic MZVs}
\label{app:cyclotomic}

Cyclotomic MZVs (also called ``multiple polylogarithms at roots of unity'') are
generalizations of MZVs. While MZVs are represented by nested sums of the form
\begin{equation}
  \zm_{n_1,n_2,\ldots,n_r}=\zm(n_1,n_2,\ldots,n_r)=\sum^{\infty}_{0<k_1<k_2<\ldots<k_r}\frac{1}{k_1^{n_1}k_2^{n_2}\ldots k_r^{n_r}}, \quad n_1,\ldots,n_{r-1} \geq 1, \, n_r \geq 2,
\end{equation}
cyclotomic MZVs are represented by nested sums with additional ``coloring''
given by $N^{\rm th}$ roots of unity $\sigma_1,\ldots,\sigma_r \in \mu_N$:
\begin{equation}
 \BZ{n_1,n_2,\ldots,n_r}{\sigma_1,\sigma_2,\ldots,\sigma_r}
 = \sum_{0<k_1<k_2<\ldots<k_r}
\frac{\sigma_1^{k_1} \sigma_2^{k_2} \ldots \sigma_r^{k_r}}
{k_1^{n_1} k_2^{n_2} \ldots k_r^{n_r}}
\co n_1,\ldots,n_r \geq 1, \, n_r \geq 2 \ {\rm if} \ \sigma_r=1 \ .
\label{cyclo11}
\end{equation}
Likewise, the integral representation of MZVs 
\begin{equation}
\zeta(n_1,\ldots,n_r)=\int\limits_{0 \leq t_i \leq t_{i+1} \leq 1}\frac{\dd t_1}{1-t_1}\frac{\dd t_2}{t_2}\ldots \frac{\dd t_{n_1}}{t_{n_1}}\frac{\dd t_{n_1+1}}{1-t_{n_1+1}}\ldots \frac{\dd t_w}{t_w} \ , \quad w=n_1+\ldots +n_r \ ,
\end{equation} 
generalizes to an integral representation for cyclotomic MZVs 
\begin{equation}
\BZ{n_1,n_2,\ldots,n_r}{\sigma_1,\sigma_2,\ldots,\sigma_r}
 = \int\limits_{0 \leq t_i \leq t_{i+1} \leq 1}\frac{\dd t_1}{\eta_1-t_1}\frac{\dd t_2}{t_2}\ldots \frac{\dd t_{n_1}}{t_{n_1}}\frac{\dd t_{n_1+1}}{\eta_2-t_{n_1+1}}\ldots \frac{\dd t_w}{t_w} \ ,
\end{equation}
where $\eta_i=(\sigma_i \sigma_{i+1}\cdots\sigma_r)^{-1}$. The positive integer
$N$, implicit in the definition of cyclotomic MZVs, is sometimes considered an
additional datum, and one speaks of $N$-cyclotomic MZVs to emphasize the choice
of $N$.

Cyclotomic MZVs have first been considered by Goncharov
\cite{Goncharov:2001iea}. Suitable references for cyclotomic MZVs include
\cite{Racinet:Doubles,Zhao:2008,Glanois:Thesis}. For a detailed study of
$N$-cyclotomic MZVs where $N=2,3,4,6,8$, see \cite{Deligne:23468}. More
recently, the case $N=6$ has generated further interest
\cite{Ablinger:2011te,Broadhurst:2014jda,Henn:2015sem}.


\section{Details on the differential equation of teMZVs}
\label{app:diff_eq_der}

In this appendix, we give a detailed derivation of the differential equation
\eqref{eqn:diff_eq_nemzv} in the case $b_1,b_\ell \neq 0$. The case $b_1=0$ or
$b_\ell=0$ is technically more complicated, since the iterated integrals involved
need to be regularized according to \rcite{Enriquez:Emzv}. However, using
Proposition 3.1 of \cite{Enriquez:Emzv}, the arguments of this section go
through for $b_1=0$ or $b_\ell=0$ as well.

Using the mixed heat equation \eqref{eqn:mixed_heat_type_eq} for
$\Omega(z-b,\alpha)$, we may rewrite the $\tau$-derivative of the generating
function \eqn{eqn:generating_T} of length-$\ell$ teMZVs as  
\begin{align}
   2 &\pi i \frac{\pd}{\pd \tau} \Tgen{\alpha_1, \alpha_2, \dots, \alpha_\ell}{\nbeta_1, \nbeta_2, \dots, \nbeta_\ell}
    =     \! \! \! \! \! \! \int \limits_{0\leq z_p \leq z_{p+1} \leq 1}  \! \! \! \! \! \! \dd z_1 \, \dd z_2 \, \ldots \, \dd z_\ell 
      \sum_{i=1}^\ell \pd_{\alpha_i} \pd_{z_i} \Omega(z_i - \nbeta_i,\alpha_i) \prod^\ell_{j \neq i} \Omega(z_j - \nbeta_j,\alpha_j) \notag \\
%
    &=  \! \! \! \! \! \! \int \limits_{0\leq z_p \leq z_{p+1} \leq 1}  \! \! \! \! \! \! \dd z_2 \, \ldots \, \dd z_\ell  \, \pd_{\alpha_1} \Omega(z_2 - \nbeta_1,\alpha_1) \prod_{j =2}^\ell \Omega(z_j - \nbeta_j,\alpha_j) -
       \pd_{\alpha_1} \Omega(-\nbeta_1,\alpha_1) \Tgen{\alpha_2, \dots, \alpha_\ell}{\nbeta_2, \dots, \nbeta_\ell}  \notag\\
           & \quad + \pd_{\alpha_\ell} \Omega(-\nbeta_\ell,\alpha_\ell) \Tgen{\alpha_1, \dots, \alpha_{\ell-1}}{\nbeta_1, \dots, \nbeta_{\ell-1}} - \! \! \! \! \! \! \! \int \limits_{0\leq z_p \leq z_{p+1} \leq 1} \! \! \! \! \! \! \!\dd z_1 \,   \ldots \, \dd z_{\ell-1}\,  \pd_{\alpha_\ell} \Omega(z_{\ell-1} - \nbeta_\ell,\alpha_\ell) \prod_{j=1}^{\ell-1} \Omega(z_j - \nbeta_j,\alpha_j)  \notag \\
    & \quad + \sum_{i=2}^{\ell-1}     \int \limits_{0\leq z_p \leq z_{p+1} \leq 1}  \! \! \! \! \dd z_1 \, \ldots \, \dd z_{i-1} \,\dd z_{i+1} \, \ldots \, \dd z_{r} \,  \pd_{\alpha_i} \Omega(z_{i} - \nbeta_i,\alpha_i) \, \Big|^{z_i=z_{i+1}}_{z_i = z_{i-1}} \, \prod_{j \neq i}^\ell  \Omega(z_j - \nbeta_j,\alpha_j) \notag
    \\
%
    &= \pd_{\alpha_\ell} \Omega(-\nbeta_\ell,\alpha_\ell) \Tgen{\alpha_1, \dots, \alpha_{\ell-1}}{\nbeta_1, \dots, \nbeta_{\ell-1}} -
      \pd_{\alpha_1} \Omega(-\nbeta_1,\alpha_1) \Tgen{\alpha_2& \dots& \alpha_\ell}{\nbeta_2& \dots& \nbeta_\ell}    \label{eqn:diff_eq_gen_T}\\
    & \quad +  \! \! \! \! \! \!  \! \! \int \limits_{0\leq z_p \leq z_{p+1} \leq 1} \! \! \! \! \! \! \! \! \dd z_2 \,\dd z_3 \, \ldots \, \dd z_\ell \, \sum_{i=2}^{\ell}   \, \prod_{j \neq i,1}^\ell  \Omega(z_j - \nbeta_j,\alpha_j) \, ( \pd_{\alpha_{i-1}} - \pd_{\alpha_{i}})  \Omega(z_i - \nbeta_{i-1},\alpha_{i-1}) \Omega(z_i - \nbeta_{i},\alpha_{i}) \notag \\
 %
    &= \pd_{\alpha_\ell} \Omega(-\nbeta_\ell,\alpha_\ell) \Tgen{\alpha_1, \dots, \alpha_{\ell-1}}{\nbeta_1, \dots, \nbeta_{\ell-1}} -
      \pd_{\alpha_1} \Omega(-\nbeta_1,\alpha_1) \Tgen{\alpha_2, \dots, \alpha_\ell}{\nbeta_2, \dots, \nbeta_\ell} \notag \\
    & \quad + \sum_{i=2}^{\ell} \Big(
    \Tgen{\alpha_1, \dots, \alpha_{i-2}, \alpha_{i-1} + \alpha_i ,  \alpha_{i+1}, \dots, \alpha_\ell}{\nbeta_1, \dots, \nbeta_{i-2}, \nbeta_{i},  \nbeta_{i+1}, \dots, \nbeta_\ell} 
    \pd_{\alpha_{i-1}} \Omega(\nbeta_{i}-\nbeta_{i-1},\alpha_{i-1})  \notag \\
    & \qquad - \Tgen{\alpha_1, \dots, \alpha_{i-2}, \alpha_{i-1} + \alpha_i,  \alpha_{i+1}, \dots, \alpha_\ell}{\nbeta_1, \dots ,\nbeta_{i-2} , \nbeta_{i-1} , \nbeta_{i+1}, \dots, \nbeta_\ell} 
    \pd_{\alpha_i} \Omega(\nbeta_{i-1} - \nbeta_{i},\alpha_i)
    \Big) \; , \notag
\end{align}
where we mapped the integrals over $\pd_{z_i}(\ldots)$ to boundary terms in the
second equality and used the Fay identity (\ref{fayfay_Om}) in the last
equality to simplify
\begin{align}
&( \pd_{\alpha_{i-1}} - \pd_{\alpha_{i}})  \Omega(z_i - \nbeta_{i-1},\alpha_{i-1}) \Omega(z_i - \nbeta_{i},\alpha_{i})
 \\
& =
( \pd_{\alpha_{i-1}} - \pd_{\alpha_{i}})\Big[\Omega(z_i {-} \nbeta_{i},\alpha_{i-1}{+}\alpha_i) \Omega(\nbeta_i {-} \nbeta_{i-1},\alpha_{i-1}) + \Omega(z_i {-} \nbeta_{i-1},\alpha_{i-1}{+}\alpha_i) \pd_{\alpha_{i}} \Omega(\nbeta_{i-1} {-} \nbeta_{i},\alpha_{i})\Big]
\notag \\ 
&= \Omega(z_i - \nbeta_{i},\alpha_{i-1}+\alpha_i) \pd_{\alpha_{i-1}} \Omega(\nbeta_i - \nbeta_{i-1},\alpha_{i-1})
- \Omega(z_i - \nbeta_{i-1},\alpha_{i-1}+\alpha_i) \pd_{\alpha_{i}} \Omega(\nbeta_{i-1} - \nbeta_{i},\alpha_{i}) \, . \notag 
\end{align}
From the above differential equation for the generating function one may deduce
a differential equation for teMZVs,
\begin{equation}
  \begin{split}
    2 \pi i \pd_\tau \Tgen{\alpha_1, \alpha_2, \dots, \alpha_\ell}{\nbeta_1, \nbeta_2, \dots, \nbeta_\ell} &= 
    2 \pi i \hspace{-0.3cm} \sum_{n_1,n_2,\ldots,n_\ell=0}^{\infty} \hspace{-0.3cm} 
    \alpha_1^{n_1-1} \dots \alpha_\ell^{n_\ell-1}  \pd_\tau \omwb{n_1, \dots , n_\ell}{\nbeta_1, \dots , \nbeta_\ell}\\
    &= \pd_{\alpha_\ell} \Omega(-\nbeta_\ell,\alpha_\ell) 
    \Tgen{\alpha_1, \dots, \alpha_{\ell-1}}{\nbeta_1, \dots, \nbeta_{\ell-1}} -
      \pd_{\alpha_1} \Omega(-\nbeta_1,\alpha_1) 
      \Tgen{\alpha_2, \dots, \alpha_\ell}{\nbeta_2, \dots, \nbeta_\ell} \\
    & \quad + \sum_{i=2}^{\ell} \Big(
 \Tgen{\alpha_1, \dots, \alpha_{i-2}, \alpha_{i-1} + \alpha_i , \alpha_{i+1}, \dots, \alpha_\ell}{\nbeta_1, \dots, \nbeta_{i-2}, \nbeta_{i},  \nbeta_{i+1}, \dots, \nbeta_\ell} 
    \pd_{\alpha_{i-1}} \Omega(\nbeta_{i}-\nbeta_{i-1},\alpha_{i-1})  \\
    & \qquad -  
    \Tgen{\alpha_1, \dots, \alpha_{i-2}, \alpha_{i-1} + \alpha_i, \alpha_{i+1}, \dots, \alpha_\ell}{\nbeta_1, \dots ,\nbeta_{i-2} , \nbeta_{i-1} , \nbeta_{i+1}, \dots, \nbeta_\ell} 
    \pd_{\alpha_i} \Omega(\nbeta_{i-1} - \nbeta_{i},\alpha_i) 
    \Big) \ .
  \end{split}
  \label{eqn:diff_eq_again}
\end{equation}
Eventually we want to equate coefficients to extract the $\tau$-derivative of a
particular \temzv{}.  For this purpose let us consider the terms in
\eqn{eqn:diff_eq_again} separately.  Recalling the definition \eqn{eqn:defh} 
of $h^{(n)}$ we may rewrite the first term on the right hand side of \eqn{eqn:diff_eq_again} as 
\begin{equation}
  \begin{split}
    \pd_{\alpha_\ell} \Omega(-\nbeta_\ell,\alpha_\ell) 
    \Tgen{\alpha_1, \dots, \alpha_{\ell-1}}{\nbeta_1, \dots, \nbeta_{\ell-1}} & = 
    \sum_{n_\ell = 0}^\infty h^{(n_\ell)}(-\nbeta_\ell) \alpha_\ell^{n_\ell-2}
    \hspace{-0.65cm} \sum_{n_1,n_2,\ldots,n_{\ell-1}=0}^{\infty} \hspace{-0.65cm} 
    \alpha_1^{n_1 - 1} \dots \alpha_{\ell-1}^{n_{\ell-1} - 1} \omwb{n_1, \dots , n_{\ell-1}}{\nbeta_1, \dots , \nbeta_{\ell-1}} \\
    &= 
    \hspace{-0.6cm} \sum_{n_1,n_2,\ldots,n_\ell=0}^{\infty} \hspace{-0.6cm} 
    \alpha_1^{n_1-1} \dots \alpha_{\ell-1}^{n_{\ell-1}-1} \alpha_\ell^{n_\ell-1} h^{(n_\ell+1)}(-\nbeta_\ell)\omwb{n_1, \dots , n_{\ell-1}}{\nbeta_1, \dots , \nbeta_{\ell-1}} \\
    & \quad +  h^{(0)}(-\nbeta_\ell) \alpha_\ell^{-2} 
    \hspace{-0.65cm} \sum_{n_1,n_2,\ldots,n_{\ell-1}=0}^{\infty} \hspace{-0.65cm} 
    \alpha_1^{n_1-1} \dots \alpha_{\ell-1}^{n_{\ell-1}-1} \omwb{n_1, \dots , n_{\ell-1}}{\nbeta_1, \dots , \nbeta_{\ell-1}} \; ,
  \end{split}
  \label{eqn:diff_eq_bdy_term}
\end{equation}
and similarly for the second term. The sum $\sim \alpha_\ell^{-2}$ is canceled
by contributions from the last two lines of \eqn{eqn:diff_eq_again}, which we
will now turn to 
\begin{align}
  &\Tgen{\alpha_1, \dots, \alpha_{i-2}, \alpha_{i-1} + \alpha_i , \alpha_{i+1}, \dots, \alpha_\ell}{\nbeta_1, \dots, \nbeta_{i-2}, \nbeta_{i}, \nbeta_{i+1}, \dots, \nbeta_\ell} 
    \pd_{\alpha_{i-1}} \Omega(\nbeta_{i}-\nbeta_{i-1},\alpha_{i-1}) \nnl
    &\hspace{0.6cm}=  
    \hspace{-0.5cm}\sum_{n_1, \ldots, n_{i-2}, n_{i+1} , \ldots, n_{\ell} = 0}^{\infty} \hspace{-0.5cm} 
    \alpha_1^{n_1-1} \dots \alpha_{i-2}^{n_{i-2}-1} \alpha_{i+1}^{n_{i+1}-1} \dots \alpha_\ell^{n_\ell-1} \nnl
    & \qquad \times \,  \Big[
      \underbrace{(\alpha_{i-1} + \alpha_i)^{-1} \omwb{n_1, \dots, n_{i-2}, 0 , \, n_{i+1}, \dots, n_\ell}{\nbeta_1, \dots, \nbeta_{i-2}, \nbeta_{i}, \, \nbeta_{i+1}, \dots, \nbeta_\ell} 
      \sum_{j=0}^\infty h^{(j)}(\nbeta_{i}-\nbeta_{i-1}) \alpha_{i-1}^{j-2}}_{ = \ B_{i,+}} \nnl
      & \qquad\qquad+
      \underbrace{\sum_{j,k = 0}^\infty \sum_{p=0}^{k} \binom{k}{p} \alpha_{i-1}^{k-p+j-2} \alpha_i^{p} 
      \omwb{n_1, \dots, n_{i-2}, k+1 ,\, n_{i+1}, \dots, n_\ell}{\nbeta_1, \dots, \nbeta_{i-2}, \nbeta_{i}, \nbeta_{i+1}, \dots, \nbeta_\ell} h^{(j)}(\nbeta_{i} - \nbeta_{i-1} )}_{ = \ C_{i,+}}
    \Big]  
  \label{eqn:diff_eq_complicated_term}
\end{align}
with an analogous definition for $B_{i,-}$ and $C_{i,-}$ relevant to the last
term of \eqn{eqn:diff_eq_again}
\begin{align}
B_{i,-} &= 
(\alpha_{i-1} + \alpha_i)^{-1} \omwb{n_1, \dots ,n_{i-2}, 0 , \, n_{i+1}, \dots ,n_\ell}{\nbeta_1, \dots ,\nbeta_{i-2}, \nbeta_{i-1}, \, \nbeta_{i+1}, \dots ,\nbeta_\ell} 
      \sum_{j=0}^\infty h^{(j)}(\nbeta_{i-1}-\nbeta_{i}) \alpha_{i}^{j-2}
\\
C_{i,-} &= 
\sum_{j,k = 0}^\infty \sum_{p=0}^{k} \binom{k}{p} \alpha_{i}^{k-p+j-2} \alpha_{i-1}^{p} 
      \omwb{n_1, \dots ,n_{i-2}, k+1 , \, n_{i+1}, \dots ,n_\ell}{\nbeta_1, \dots ,\nbeta_{i-2}, \nbeta_{i-1}, \, \nbeta_{i+1}, \dots ,\nbeta_\ell} h^{(j)}(\nbeta_{i-1} - \nbeta_{i} ) \ .
\end{align}
In the following the manipulations only affect pairs $(\alpha_{i-1},\alpha_i)$,
hence we will suppress the summation over the other $\alpha$'s. Since
$h^{(0)}(\nbeta)=-1$ does not depend on $\nbeta$, we can set it to zero for the
iterated integral
\begin{equation}
  \begin{split}
  \omwb{n_1, \dots ,n_{i-2}, 0 , n_{i+1}, \dots ,n_\ell}{\nbeta_1, \dots, \nbeta_{i-2}, \nbeta_{i-1} , \nbeta_{i+1}, \dots ,\nbeta_\ell} =
  \omwb{n_1, \dots ,n_{i-2}, 0 , \, n_{i+1}, \dots ,n_\ell}{\nbeta_1, \dots, \nbeta_{i-2}, 0 , \nbeta_{i+1}, \dots, \nbeta_\ell} \; .
  \end{split}    
  \label{eqn:equality_n=0}
\end{equation}
Hence, for the terms not expressible by the binomial law $B_{i,\pm}$, we can use
$h^{(0)} = -1$, $h^{(1)}= 0$, $h^{(2)}(b) = h^{(2)}(-b)$  as well as
\begin{equation}
  \begin{split}
    (\alpha_{i-1}^j -(-1)^j \alpha_i^j) = (\alpha_{i-1} + \alpha_i) \sum_{a=0}^{j-1} (-1)^{j-1-a} \alpha_{i-1}^{a} \alpha_{i}^{j-1-a} \; , \;  j>0  
  \end{split}
  \label{eqn:random_sum_1}
\end{equation}
to obtain
\begin{align}
    & B_{i,+} - B_{i,-} \nnl
    &=(\alpha_{i-1} + \alpha_i)^{-1} \omwb{n_1, \dots ,n_{i-2}, 0 , \, n_{i+1}, \dots ,n_\ell}{\nbeta_1, \dots ,\nbeta_{i-2}, 0, \, \nbeta_{i+1}, \dots ,\nbeta_\ell} \sum_{j = 0}^\infty
  \left(  h^{(j)}(\nbeta_{i}-\nbeta_{i-1}) \alpha_{i-1}^{j-2} - 
 h^{(j)}(\nbeta_{i-1}-\nbeta_{i}) \alpha_{i}^{j-2} \right) \nnl
  & = 
  \omwb{n_1, \dots ,n_{i-2}, 0 , n_{i+1}, \dots ,n_\ell}{\nbeta_1, \dots ,\nbeta_{i-2}, 0   , \nbeta_{i+1}, \dots ,\nbeta_\ell} 
  \Big(   \frac{  \alpha_i^{-2} -  \alpha_{i-1}^{-2} }{ \alpha_{i-1}+\alpha_i }
   + \sum_{j = 0}^\infty h^{(j + 3)}(\nbeta_{i}-\nbeta_{i-1}) \sum_{a=0}^j (-1)^{j-a} \alpha_{i-1}^{a} \alpha_{i}^{j-a} 
  \Big) \; .
  \label{eqn:non_binomial_terms}
\end{align}
The singular term $\frac{  \alpha_i^{-2} -  \alpha_{i-1}^{-2} }{ \alpha_{i-1}+\alpha_i }
= \frac{1}{\alpha_{i-1} \alpha_i} ( \frac{1}{\alpha_i} - \frac{1}{\alpha_{i-1}} )$
will eventually cancel among different $B_{i,+}-B_{i,-}$.  The remaining
contribution may be rewritten into a form where we can easily read off the
coefficient of a given monomial in the $\alpha_i$,
\begin{equation}
  \begin{split}  
    & \omwb{n_1, \dots ,n_{i-2}, 0 , n_{i+1}, \dots ,n_\ell}{\nbeta_1, \dots ,\nbeta_{i-2}, 0   , \nbeta_{i+1}, \dots ,\nbeta_\ell} 
    \sum_{j=0}^\infty h^{(j + 3)}(\nbeta_{i}-\nbeta_{i-1}) \sum_{a=0}^j (-1)^{j-a} \alpha_{i-1}^{a} \alpha_{i}^{j-a} \\
    & \quad = \omwb{n_1, \dots ,n_{i-2}, 0 ,  n_{i+1}, \dots ,n_\ell}{\nbeta_1, \dots ,\nbeta_{i-2}, 0   , \nbeta_{i+1}, \dots ,\nbeta_\ell} 
    \sum_{m,n = 0}^\infty h^{(m + n + 3)}(\nbeta_{i}-\nbeta_{i-1}) (-1)^{n} \alpha_{i-1}^{m} \alpha_{i}^{n} \; ,
  \end{split}
  \label{eqn:non_binomial_part}
\end{equation}
which gives rise to the last line of \eqn{eqn:diff_eq_nemzv}.

For the contributions $C_{i,\pm}$ we have
\begin{equation}
  \begin{split} 
    C_{i,+} & = \sum_{j,k = 0}^\infty \sum_{p=0}^{k} \binom{k}{p} \alpha_{i-1}^{k+j-p-2} \alpha_i^{p} 
    \omwb{n_1, \dots ,n_{i-2}, k+1 ,  n_{i+1}, \dots ,n_\ell}{\nbeta_1, \dots ,\nbeta_{i-2}, \nbeta_{i} ,\, \nbeta_{i+1}, \dots ,\nbeta_\ell} 
    h^{(j)}(\nbeta_{i} - \nbeta_{i-1} ) \\
    &  = \sum_{j,k = 0}^\infty \sum_{p=0}^{k-1} \binom{k}{p} \alpha_{i-1}^{k+j-p-1} \alpha_i^p
    \omwb{n_1, \dots ,n_{i-2}, k+1 ,n_{i+1}, \dots ,n_\ell}{\nbeta_1, \dots ,\nbeta_{i-2}, \nbeta_{i} , \nbeta_{i+1}, \dots ,\nbeta_\ell} 
    h^{(j+1)}(\nbeta_{i} - \nbeta_{i-1} ) \\
    & \qquad + \sum_{k = 0}^\infty \sum_{p=0}^{k} \binom{k+1}{p} \alpha_{i-1}^{k-p-1} \alpha_i^p 
    \omwb{n_1, \dots ,n_{i-2}, k+2 , n_{i+1}, \dots ,n_\ell}{\nbeta_1, \dots ,\nbeta_{i-2}, \nbeta_{i}, \nbeta_{i+1}, \dots ,\nbeta_\ell} 
    h^{(0)}(\nbeta_{i} - \nbeta_{i-1} ) \\
    & \qquad + \alpha_{i-1}^{-2} \sum_{a = 0}^\infty h^{(0)}(\nbeta_{i}-\nbeta_{i-1}) \alpha_i^a
      \omwb{n_1, \dots ,n_{i-2}, a+1 , n_{i+1}, \dots ,n_\ell}{\nbeta_1, \dots ,\nbeta_{i-2}, \nbeta_{i}, \nbeta_{i+1}, \dots ,\nbeta_\ell} \ .
  \end{split}
  \label{eqn:binomial_part}
\end{equation}
The sums proportional to $\alpha_{i-1}^{-2}$ cancel among the corresponding
contributions from $C_{i,+}$ and $C_{i-1,-}$.  For the cases $i=2$ and $i=r$ it
is canceled by the corresponding sums in the last line of \eqn{eqn:diff_eq_bdy_term}. 
We then find for the remaining part of $C_{i,+}$
\begin{equation}
  \begin{split}
    & \sum_{j,k = 0}^\infty \sum_{p=0}^{k-1} \binom{k}{p} \alpha_{i-1}^{k+j-p-1} \alpha_i^p
    \omwb{n_1, \dots ,n_{i-2}, k+1 , n_{i+1}, \dots ,n_\ell}{\nbeta_1, \dots ,\nbeta_{i-2}, \nbeta_{i} , \nbeta_{i+1}, \dots ,\nbeta_\ell} h^{(j+1)}(\nbeta_{i-1} - \nbeta_{i} ) \\
    & \qquad + \sum_{k = 0}^\infty \sum_{p=0}^{k} \binom{k+1}{p} \alpha_{i-1}^{k-p-1} \alpha_i^p 
    \omwb{n_1, \dots ,n_{i-2}, k+2 , n_{i+1}, \dots ,n_\ell}{\nbeta_1, \dots ,\nbeta_{i-2}, \nbeta_{i} , \nbeta_{i+1}, \dots ,\nbeta_\ell} h^{(0)}(\nbeta_{i-1} - \nbeta_{i} ) \\
    & \quad = \sum_{m,n = 0}^\infty \alpha_{i-1}^{m-1} \alpha_i^n
    \sum_{j=0}^{m} \binom{m+n-j}{n} 
    \omwb{n_1, \dots ,n_{i-2}, m+n+1-j , n_{i+1}, \dots ,n_\ell}{\nbeta_1, \dots ,\nbeta_{i-2}, \nbeta_{i} , \nbeta_{i+1}, \dots ,\nbeta_\ell} h^{(j+1)}(\nbeta_{i-1} - \nbeta_{i} ) \\
    & \qquad + \sum_{m,n = 0}^\infty \alpha_{i-1}^{m-1} \alpha_i^n \binom{m+n+1}{n} 
    \omwb{n_1, \dots ,n_{i-2}, m+n+2 ,  n_{i+1}, \dots ,n_\ell}{\nbeta_1, \dots ,\nbeta_{i-2}, \nbeta_{i} ,  \nbeta_{i+1}, \dots ,\nbeta_\ell} h^{(0)}(\nbeta_{i-1} - \nbeta_{i} ) \\
    & \quad = \sum_{m,n = 0}^\infty \alpha_{i-1}^{m-1} \alpha_i^{n}
    \sum_{k=0}^{m+1} \binom{n+k}{k} 
    \omwb{n_1, \dots ,n_{i-2}, n+k+1 , n_{i+1}, \dots ,n_\ell}{\nbeta_1, \dots ,\nbeta_{i-2}, \nbeta_{i} , \, \nbeta_{i+1}, \dots ,\nbeta_\ell} h^{(m-k+1)}(\nbeta_{i-1} - \nbeta_{i} ) \; ,
  \end{split}
  \label{eqn:binomial_part_contd}
\end{equation}
which, in combination with an analogous contribution from $C_{i,-}$, gives rise to the second and third line of \eqn{eqn:diff_eq_nemzv}.


\section{Differential equation for proper rational twists} \label{app:diff_eq_rat}

The derivation of the differential equation given in \appref{app:diff_eq_der} 
was a priori only valid for generic twists. In this
appendix, we provide an argument why it also holds for proper rational twists $b_i \in \ZQ$.
The only difference is that in \eqn{eqn:diff_eq_gen_T} we now need to carefully
keep track of the effect of deforming the domain of integration $[0,1]$
infinitesimally to obtain $[0,1]_{\varepsilon}$ (cf.\ \figref{figDeformation}). The key 
point is that the integral along
$[0,1]_{\varepsilon}$ is the same as along $[0,1]$ up to terms which vanish
in the limit, and the differential equation \eqref{eqn:diff_eq_gen_T} goes
through without change.

As in the case of generic twists, it will be convenient to define a generating function
\begin{align}
\TgenR{\alpha_1,\dots,\alpha_\ell}{b_1,\dots, b_\ell} 
& = \lim_{\varepsilon \rightarrow 0}  \TgenRE{\alpha_1 , \dots , \alpha_\ell}{b_1 , \dots , b_\ell} \; , \\
\TgenRE{\alpha_1 , \dots , \alpha_\ell}{b_1 , \dots , b_\ell} 
& = \int\limits_{0\leq t_i \leq t_{i+1} \leq 1} \hspace{-0.7cm}
(\gamma_R^* \Omega(z_1 - b_1, \alpha_1,\tau) \dd z_1) \dots
(\gamma_R^* \Omega(z_\ell - b_\ell, \alpha_\ell,\tau) \dd z_\ell) \; ,
\label{eqn:tgenr_def1}
\end{align}
where $\gamma_R = [0,1]_\varepsilon$ and $\gamma_R^*$ denotes its pullback.
Here and throughout this appendix, we will denote the pullback of the
coordinate $z_i$ along $\gamma_R$ by $t_i$.  We note that in the case where all $b_i$
are generic twists, we may pass to the limit $\varepsilon \to 0$ immediately
and integrate along the line $\gamma_R(t) = t $, which leads to
\eqn{eqn:generating_T}. Pulling back $\Omega(z,\alpha,\tau) \dd z$ along
$\gamma_R$, we obtain
\begin{align}
\gamma_R^* \Omega(z_i - b_i, \alpha_i,\tau) \dd z_i
= \gamma_R^* \underbrace{e^{-2 \pi i r_i \alpha_i} F(z_i - b_i,\alpha_i,\tau) \dd z_i }_{= \ \tilde \Omega(z_i - b_i, \alpha_i, \tau) \dd z_i} 
+ \mathcal{O}(\varepsilon) \; ,
\label{eqn:better_omega}
\end{align}
since $\operatorname{Im}(z_i)$ is of order $\varepsilon$ on
$[0,1]_{\varepsilon}$.

The resulting form
$\tilde{\Omega}(z,\alpha,\tau) \dd z$ is now meromorphic; therefore the
integral over $\tilde{\Omega}(z,\alpha,\tau)\dd z$ along any path depends exclusively
on its homotopy class. In particular, since the paths $[0,1]_{\varepsilon}$ are
all homotopic for sufficiently small $\varepsilon$, for every such $\varepsilon$ we get\footnote{Recall
that the same argument has already been used in \subsecref{ssec:regteMZVs}
to show that our version of teMZVs for proper rational twists is well-defined
(more precisely that the limit in \eqn{eqn:rigorous} exists).}
\begin{align}
\TgenR{\alpha_1 , \dots , \alpha_\ell}{b_1 , \dots , b_\ell} 
& = \lim_{\varepsilon \to 0} \TgenTRE{\alpha_1 , \dots , \alpha_\ell}{b_1 , \dots , b_\ell}=\TgenTRE{\alpha_1 , \dots , \alpha_\ell}{b_1 , \dots , b_\ell} \; , 
\notag \\
\TgenTRE{\alpha_1 , \dots , \alpha_\ell}{b_1 , \dots , b_\ell}
& = \int\limits_{0\leq t_i \leq t_{i+1} \leq 1} \hspace{-0.7cm}
(\gamma_R^* \tilde \Omega(z_1 - b_1, \alpha_1,\tau) \dd z_1) \dots
(\gamma_R^* \tilde \Omega(z_\ell - b_\ell, \alpha_\ell,\tau) \dd z_\ell) \; .
\label{eqn:tgenr_def2}
\end{align}
Furthermore, we define the intermediate object 
\begin{align}
\TgenTREz{\alpha_1 , \dots , \alpha_i}{b_1 , \dots , b_i}{z_{i+1}}
& = \int\limits_{0\leq t_1 \leq \ldots \leq t_i \leq t_{i+1}} \hspace{-0.7cm}
(\gamma_R^* \tilde \Omega(z_1 - b_1, \alpha_1,\tau) \dd z_1) \dots
(\gamma_R^* \tilde \Omega(z_i - b_i, \alpha_i,\tau) \dd z_i)  \notag \\
& = \int\limits_0^{t_{i+1}} 
\gamma_R^* 
\left(
\tilde \Omega(z_i - b_i, \alpha_i,\tau)   	
\TgenTREz{\alpha_1 , \dots , \alpha_i}{b_1 , \dots , b_i}{z_i}
\dd z_i
\right)
\label{eqn:tgenr_inter_merda}
\end{align}
with $0<t_{i+1} < 1$. It satisfies
\begin{align}
\partial_{z_{i+1}} \TgenTREz{\alpha_1 , \dots , \alpha_i}{b_1 , \dots , b_i}{z_{i+1}}
=
\tilde \Omega(z_{i+1} - b_i, \alpha_i,\tau) \TgenTREz{\alpha_1 , \dots , \alpha_{i-1}}{b_1 , \dots , b_{i-1}}{z_{i+1}} \,.
\label{eqn:tgenr_inter_merda_2}
\end{align}
Using the above setup, we now show that all essential aspects of the computation
\eqref{eqn:diff_eq_gen_T} in \appref{app:diff_eq_der} remain unchanged.
Firstly, we compute the $\tau$-derivative of $\gamma_R^* \tilde \Omega$ and
obtain
\begin{align}
  2 \pi i \partial_\tau \gamma_R^* &\left( \tilde \Omega(z_i - b_i, \alpha_i,\tau) \dd z_i \right)\nnl
&\hspace{0.6cm}=
2 \pi i \partial_\tau \left( e^{-2 \pi i r_i \alpha_i } F(\gamma_R(t_i) - b_i, \alpha_i,\tau) \dd \gamma_R(t_i)  \right)  
\notag \\
&\hspace{0.6cm}= 
e^{-2 \pi i r_i \alpha_i }   
\left[
(- 2 \pi i r_i \partial_{\gamma_R(t_i)} +
\partial_{\gamma_R(t_i)} \partial_{\alpha_i}
)
F(\gamma_R(t_i) - b_i, \alpha_i,\tau) \right]
\dd \gamma_R(t_i)
\notag \\
&\hspace{0.6cm}=
\gamma_R^* \left( \partial_{z_i} \partial_{\alpha_i} \tilde \Omega(z_i - b_i, \alpha_i,\tau) \dd z_i  \right) \,,
\label{eqn:tau_diff_pb}
\end{align}
using the mixed heat equation \eqref{mixedheat} for $F(z,\alpha,\tau)$ in the
second step. In particular, $\gamma_R^*\tilde \Omega$ itself satisfies a
mixed-heat type equation.  Secondly, we may exchange the $\tau$-derivative with
the integration
\begin{align} 
2 \pi i \partial_\tau \TgenTRE{\alpha_1 , \dots , \alpha_\ell}{b_1 , \dots , b_\ell} 
= \sum_{i=1}^\ell \ 
& \int\limits_{ 0 \leq t_{j-1} \leq t_j \leq 1 } \  \prod^\ell_{j > i} (\gamma_R^* \tilde \Omega(z_j - b_j, \alpha_j,\tau) \dd z_j) 
\label{eqn:diff_eq_prt}
\\
& \! \!
\times
\int\limits_{0}^{t_{i+1}}
\gamma_R^*
\left(
(\partial_{z_i} \partial_{\alpha_i}
\tilde \Omega(z_i - b_i, \alpha_i,\tau))
\TgenTREz{\alpha_1 , \dots , \alpha_{i-1}}{b_1 , \dots , b_{i-1}}{z_i}
\dd z_i
\right) \,,
\notag
\end{align}
since the $\tau$-derivative of the integrand is bounded on the domain of
integration. Rewriting the $i^{\rm th}$ integration
using integration by parts yields
\begin{align}
& \int\limits_{0}^{t_{i+1}}
\gamma_R^*
\left(
(\partial_{z_i} \partial_{\alpha_i}
\tilde \Omega(z_i - b_i, \alpha_i,\tau))
\TgenTREz{\alpha_1 , \dots , \alpha_{i-1}}{b_1 , \dots , b_{i-1}}{z_i}
\dd z_i
\right) 
\notag \\
& =
\gamma_R^*
\left(
\partial_{\alpha_i}
\tilde \Omega(z_{i+1} - b_i, \alpha_i,\tau))
\TgenTREz{\alpha_1 , \dots , \alpha_{i-1}}{b_1 , \dots , b_{i-1}}{z_{i+1}}
\right)
\label{eqn:ith_int} \\
& \qquad - 
\int\limits_{0}^{t_{i+1}}
\gamma_R^*
\left(
(\partial_{\alpha_i}
\tilde \Omega(z_i - b_i, \alpha_i,\tau))
\tilde \Omega(z_{i-1} - b_{i-1}, \alpha_{i-1},\tau)
\TgenTREz{\alpha_1 , \dots , \alpha_{i-2}}{b_1 , \dots , b_{i-2}}{z_i}
\dd z_i
\right) \,,
\notag
\end{align}
as in the case of generic twists.  Therefore, we may proceed further as in the
computation of \eqn{eqn:diff_eq_gen_T} and arrive at the same result simply by
virtue of the replacements $\TL \rightarrow \TL^R_\varepsilon$ and $\Omega
\rightarrow \tilde \Omega$.


\section{Properties of teMZVs} 
\label{app:emzv}

The purpose of this appendix is to gather constant terms and $q$-expansions of teMZVs as well as selected relations
relevant to the one-loop open-string amplitude in \secref{sec:oneloop}.


\subsection{Constant terms for generic twists} 
\label{app:emzvA}

We start by listing a few simple examples for constant terms of teMZVs with
generic twists. These constant terms are obtained from \eqn{eqn:drin17} by
comparing the coefficients of words in the non-commutative variables ${\rm
ad}^n_{x_b}( y)$ on both sides, see \eqn{eqn:drin24} for the change of alphabet
between the two sides.  At length one, specializing
\eqns{eqn:drin11a}{eqn:drin11b} to $r=\frac{1}{2}$ yields
\begin{equation}
\omwbc{n}{\tauh} =
  	   \begin{cases} \displaystyle
    	    \frac{2^{n-1} -1}{2^{n-2}}\zeta_n 	&: \ n \ \text{even}\\
    	   \hspace{0.8cm} 0					&: \ n\ \text{odd} \ ,
  	   \end{cases}
	   \label{exlength1}
\end{equation}
see \eqn{eqn:np_qexp} for the $\omega_0(\ldots)$ notation. This immediately
implies that
\begin{equation}
  \omwbc{\ldots, 2k-1, \ldots}{\ldots, \tauh, \ldots}
  = 0 \ , \ \ \ \ \ \ k \in \ZN
\ ,
	   \label{exlengthn}
\end{equation}
regardless on the position of the combined letter $\begin{smallmatrix}  2k-1 \\
\tau/2\end{smallmatrix}$.  Similarly, higher-length examples include
\begin{align}
  \omwbc{n,0}{\tauh,0} &=
  	   \begin{cases}\displaystyle
    	    \frac{2^{n-1} -1}{2^{n-1}}\zeta_n 	&: \ n \ \text{even}\\
    	    \hspace{0.8cm}0					&: \ n\ \text{odd}
  	   \end{cases} \ , \ \ \ \ \ \ 
  \omwbc{1,n}{0,\tauh} =
  	   \begin{cases}\displaystyle
    	 ( -i\pi)  \frac{2^{n-1} -1}{2^{n-1}}\zeta_n 	&: \ n \ \text{even}\\
    	   \hspace{1.35cm} 0					&: \ n\ \text{odd}
  	   \end{cases}
	\notag
\\
\omwbc{2,0,0}{0,0,0}    &= - \frac{ \zeta_2}{3} \ , \ \ \ \ \ \
\omwbc{0,1,0,0}{0,0,0,0} =  \frac{3 \zeta_3}{4\pi^2} \ , \ \ \ \ \ \ \ \  \ \ \ \!
\omwbc{0,3,0,0}{0,0,0,0}=0
   \label{exlength3}
   \\
\omwbc{1,0,2}{0,0,\tauh} &= - \frac{ i \pi^3}{24} \ , \ \ \ \ \ \
\omwbc{0,2,2}{0,0,\tauh} = - \frac{ \pi^4}{108} \ , \ \ \ \ \ \
\omwbc{0,1,0,2}{0,0,0,\tauh} =  \frac{\zeta_3}{8}  \ ,
	\notag
\end{align}
and we obtain the following examples with more general twists $b \in
(\Lambda_N+\Lambda_N\tau) \setminus \Lambda^\times_N$:
\begin{align}
  \label{eqn:real_twists_firstline_2nd}
  	\omwbc{1}{\oneh + \taud} &= -\frac{i \pi}{3},    
	&&
	\;\omwbc{2, 1, 0, 1}{\tauh, \tauv, 0, 0}
	= - \frac{i \pi}{48} \zeta_3 + \frac{5}{8} \zeta_4,  \\  
	\omwbc{1,1}{\taud,\tauf} &= - \frac{3}{5} \zeta_2,
	&& 
	\omwbc{3,1,0,1}{\tauzf,0,0,\tauv} = 
	- \frac{9}{125} \zeta_2 \zeta_3\,.
  \label{eqn:real_twists_2nd}
\end{align}
Given a twist $b=s+r\tau$ with $r\neq 0$, the constant term
does not depend on $s$ (cf.~\eqns{eqn:drin11a}{eqn:drin11b}).

Up to weight five and length three (respectively weight three and length four), we have
checked the constant-term procedure for consistency with Fay relations among
teMZVs which can be derived along the lines of \cite{Broedel:2015hia}. As
already noted above, the constant-term procedure discussed in
\subsecref{ssec:constant} covers all eMZVs occurring in the one-loop
open-string amplitude at the orders considered in \secref{sec:oneloop}. The
examples presented here address all teMZVs relevant to the calculations in
\secref{ssec:explicit}.


\subsection{Constant terms for proper rational twists} 
\label{app:emzvB}

For the proper rational twist $b=\frac{1}{2}$ we arrive at the following
examples of constant terms
\begin{align}
  \label{eqn:real_twists_firstline}
 \omwbc{1}{\oneh} &= - i \pi,    
	&&\omwbc{2 , 0 , 1}{\oneh , 0 , \oneh} =  \frac{i \pi^3}{24} - \zeta_2 \log(2)\,, \\  
 \omwbc{0 , 1}{0 , \oneh} &=  - \frac{i \pi}{2} + \log(2) \ , \ \ \ \ \ \ 
	&&\;\;\omwbc{1 , 0 , 0}{\oneh , 0 , 0} = - \frac{i \pi}{8}  - \frac{\log(2)}{2}  \,,
  \label{eqn:real_twists}
\end{align}
while twists $b \in \Lambda^\times_3$ give rise to 
\begin{align}
  \label{eqn:real_twists_firstline_2nda}
 	\omwbc{1}{\oned} &= - i \pi,    
	&&\;\omwbc{1, 1}{\oned, \onezd}
	=  
	i \pi \left( \BZ{1}{e^{4\pi i /3}}
	-
	\BZ{1}{e^{2 \pi i /3}}  \right)  - 3 \zeta_2 , \\  
	\omwbc{1,1,2}{0,0,\oned} &= \frac{5}{2} \zeta_4,
	&& 
	\omwbc{1,1,0}{\oned,0,0}
	= 
	-\frac{i \pi}{2} \BZ{1}{e^{2\pi i /3}}
	+ \frac{1}{2} \BZ{2}{e^{2\pi i /3}}
	- \BZ{1,1}{e^{4\pi i /3},e^{2 \pi i /3}} ,
  \label{eqn:real_twists_2nda}
\end{align}
\begin{equation}
  \begin{split}
 	\omwbc{1, 1, 0, 1}{\oned, 0, 0, 0} &= 
	\frac{i \pi}{4} \zeta_2
	+ \zeta_2 \BZ{1}{e^{2\pi i/3}}
	+ \frac{i \pi}{4} \BZ{2}{e^{2\pi i/3}}
	+ \frac{1}{4} \BZ{3}{e^{2\pi i /3}} \\
	& \quad
	- \frac{i \pi}{2} \BZ{1,1}{e^{4\pi i /3},e^{2\pi i /3}} 
	- \frac{1}{2} \BZ{1,2}{e^{4\pi i /3},e^{2\pi i /3}} 
	- \frac{1}{2} \BZ{2,1}{e^{4\pi i /3},e^{2\pi i /3}} \,,
  \end{split}
  \label{eqn:awesome_dude}
\end{equation}
see \eqn{cyclo11} for the definition of cyclotomic MZVs $\BZ{n_1,n_2,\ldots,n_r}{\sigma_1,\sigma_2,\ldots,\sigma_r}$.

Again, consistency of the constant-term procedure with Fay relations among
teMZVs has been checked up to weights five and three at lengths three and four,
respectively.


\subsection{teMZV relations for the string amplitude}
\label{app:temzvcomputation}

The simplification of the string amplitude in \secref{ssec:explicit} requires
several relations among teMZVs and eMZVs.  The subsequent identities involving
teMZVs can be proven by comparing both the constant terms encoded in
\eqn{eqn:drin17} and the $\tau$-derivatives \eqn{eqn:diff_eq_nemzv} of both
sides. The relations among eMZVs follow from a combination of Fay and shuffle
identities \cite{Broedel:2015hia} and are listed at
\url{https://tools.aei.mpg.de/emzv/}. At the second order in $\ap$, we make use
of
\begin{align}
2\omwb{1 ,1 , 0}{0 ,0 , 0}  - \omwb{1 , 0}{0 , 0}^2 &=  \omwb{2 , 0, 0}{0 , 0, 0} + \frac{ 5 \zeta_2}{6}	\label{weight2Bnoq} \\
2\omwb{1 ,1 , 0}{\tauh,\tauh, 0}  - \omwb{1 , 0}{\tauh, 0}^2 &=   \omwb{2 , 0, 0}{0 , 0, 0} + \frac{  \zeta_2}{3}\; ,	\label{weight2Anoq} 
\end{align}
while the simplifications at the order $\ap^3$ are based on the relations
\begin{align}
& \omwb{1,0}{0,0}^3 - 3  \omwb{1,0}{0,0}  \omwb{1 ,1,0}{0,0,0}  + 3 \omwb{1,1,1,0}{0,0,0,0} \notag \\
&  \ \ \ = \frac{1}{6} \, \omwb{0,3,0,0}{0,0,0,0} - \frac{1}{4} \, \zeta_3 - 4 \, \zeta_2\, \omwb{0,1,0,0}{0,0,0,0}
 \label{weight2Dnoq} \\
& \omwb{1,0}{\tauh,0}^3 -3  \omwb{1,0}{\tauh,0}  \omwb{1 ,1,0}{\tauh,\tauh,0}  +3 \omwb{1,1,1,0}{\tauh,\tauh,\tauh,0}
 \notag \\
 & \ \ \ = \frac{1}{6} \, \omwb{0,3,0,0}{0,0,0,0} - \frac{1}{4} \, \zeta_3 +2 \, \zeta_2\, \omwb{0,1,0,0}{0,0,0,0} \; .
 \label{weight2Cnoq}
\end{align}


\subsection{$q$-expansions for the string amplitude} 
\label{app:emzvC}

The eMZVs seen in the final results \eqns{finalqtononzero}{31result} for the
integrals in the string amplitude have the following $q$-expansions
\begin{align}
\omega(0,0,2) &=- \frac{ \zeta_2 }{3}+ 2 \sum_{n,m=1}^{\infty} \frac{ m }{n^2} q^{mn} \notag \\
\omega(0,1,0,0)  &=\frac{ 3\zeta_3 }{4\pi^2} + \frac{3}{2\pi^2} \sum_{n,m=1}^{\infty} \frac{ 1 }{n^3} q^{mn} 
\label{qexps} \\
\omega(0,3,0,0) &= -3 \sum_{n,m=1}^{\infty} \frac{ m^2 }{n^3} q^{mn} \; .  \notag
\end{align}
These expressions follow from repeatedly integrating the Eisenstein series
\eqn{eqn:h0Eisen} in the $\tau$-derivatives \eqn{eqn:diff_eq_nemzv} of the
eMZVs in question. Analogous $q$-expansions for teMZVs with twists $b_i \in
\left\{0,\tauh\right\}$ can be determined based on
\begin{equation}
  f^{(k)}\left(\tauh\right) 
  = 
  \begin{cases} \displaystyle
    \ \ \frac{2^{k-1}-1}{2^{k-2}} \zeta_k - \frac{2 (2 \pi i)^k }{(k-1)!} \sum\limits_{n,m=1}^\infty \left(n-\frac{1}{2}\right)^{k-1} q^{m(n-\oneh)}	
    &: \ k  \text{ even } \\
  \hspace{4.5cm}  0																
    &: \ k  \text{ odd }
  \end{cases} \; .
  \label{eqn:fn_th}
\end{equation}


\section{The non-planar integral along with 
${\rm Tr}(t^1 t^2 t^3) {\rm Tr}( t^4)$}
\label{app:3plus1}

In this appendix, we investigate the low-energy expansion of the integral
\eqn{eqn:itintC} associated with the color factor $ {\rm Tr}(t^1 t^2 t^3) {\rm
Tr}( t^4)$ in the one-loop four-point open-string amplitude.  The representations
\eqns{eqn:defPP}{eqn:defQQ} of the Green functions allow to cast the integral
into the form
\begin{equation}
I_{123|4}(q) =  \int_{123}^{4} 
\prod_{1\leq i<j}^3 \sum_{n_{ij}=0}^{\infty}  \frac{ (s_{ij} \GLarg{1}{ 0 }{x_{ij}} )^{n_{ij}}  }{n_{ij}!}
\prod_{j=1}^3 \sum_{n_{j4}=0}^{\infty}   \frac{ \left(s_{j4} \GLarg{1}{ \tauh}{x_{j4} }  \right)^{n_{j4} } }{ n_{j4}!}
\label{eqn:newnewnewrep}
\end{equation}
analogous to \eqn{eqn:newnewrep}, where $c(q)$ cancels by momentum conservation
\eqn{eqn:mand}, and the integration measure is defined by
\begin{align}
\int_{123}^{4} &= \int_{0}^{1} \dd x_4 \int_{0}^{1} \dd x_3 \int_{0}^{x_3} \dd x_2 \int_{0}^{x_2} \dd x_1\, \delta(x_1)
\label{eqn:int_bdy_31} \\
&=\int_{0}^{1} \dd x_4 \, \delta(x_4)\, \bigg( \int_{0}^{1} \dd x_3 \int_{0}^{x_3} \dd x_2 \int_{0}^{x_2} \dd x_1
+ {\rm cyc}(x_1,x_2,x_3) \bigg) \; . \notag
\end{align}
One can check through the change of variables $x_i \rightarrow 1-x_i$ and
symmetry properties of the Green function that the measure $ \int_{132}^{4} $
with $x_2$ and $x_3$ interchanged yields the same result for the integral
\eqn{eqn:newnewnewrep}. Hence, one can equivalently employ the simpler measure
\eqn{eqn:int_bdy_np} with $x_i \in [0,1]$ tailored to the color structure $
{\rm Tr}(t^1 t^2) {\rm Tr}(t^3 t^4)$ and rewrite 
\begin{equation}
I_{123|4}(q) = \frac{1}{2} \int_{12}^{34} 
\prod_{1\leq i<j}^3 \sum_{n_{ij}=0}^{\infty}  \frac{ (s_{ij} \GLarg{1}{ 0 }{x_{ij}} )^{n_{ij}}  }{n_{ij}!}
\prod_{j=1}^3 \sum_{n_{j4}=0}^{\infty}   \frac{ \left(s_{j4} \GLarg{1}{ \tauh}{x_{j4} }  \right)^{n_{j4} } }{ n_{j4}!} \ .
\label{eqn:evennewnewnewrep}
\end{equation}
In the remainder of this appendix, we will follow the steps of
\subsecref{ssec:explicit} to expand the integral $ I_{123|4}(q)$ to the order
$\ap^3$, where the representation \eqn{eqn:evennewnewnewrep} is most convenient
for practical purposes. While the elliptic iterated integrals in this expansion lead to teMZVs for each
monomial in $s_{ij}$, the final results for the orders $\ap^{\leq 3}$ turn out
to comprise eMZVs only.

\subsection{Structure of the leading orders $\ap^{\leq 3}$}

We start by classifying the inequivalent integrals w.r.t.\ the cycle structure
of ${\rm Tr}(t^1 t^2 t^3 )\, {\rm Tr}( t^4)$ which occur at the orders
$\ap^{\leq 3}$ of \eqn{eqn:newnewnewrep}: With the shorthands
$P_{ij}=P(x_{ij})$ and 
\begin{equation}
\hat Q_{ij}=Q(x_{ij}) - c(q) = \GLarg{1}{ \tauh}{x_{ij} } \; ,
\label{eqn:hatQ}
\end{equation}
we have two inequivalent cases at the first order,
\begin{equation}
e_1^1= \int_{123}^{4}  P_{12} \co e_2^1= \int_{123}^{4}  \hat Q_{14} \, ,
 \label{31npl9}
\end{equation}
six cases at the second order,
\begin{align}
e_1^2 &= \frac{1}{2} \int_{123}^{4}  P_{12}^2 \co e^2_3 = \int_{123}^{4} P_{12} P_{13} \co e^2_5 = \int_{123}^{4} P_{12} \hat Q_{34} \label{31npl10} \\
e_2^2 &= \frac{1}{2} \int_{123}^{4}\hat Q_{14}^2 \co e^2_4 = \int_{123}^{4} \hat Q_{14} \hat Q_{24} \co e^2_6 = \int_{123}^{4} P_{12} \hat Q_{14} \ , \notag
\end{align}
and fourteen cases at the third order:
\begin{align}
 e_1^3 &= \frac{1}{6} \int_{123}^{4} P_{12}^3 \co \ \ \ \ \ \ \ \ \hspace{0.8mm}  e_6^3= \frac{1}{2} \int_{123}^{4} \hat Q_{14}^2 \hat Q_{24} \ , & e_{11}^3=   \int_{123}^{4} P_{12} P_{13} \hat Q_{14}  \ \phantom{.} \, \notag \\
  e_2^3 &= \frac{1}{6} \int_{123}^{4} \hat Q_{14}^3 \co \ \ \ \ \ \ \ \ \hspace{0.3mm} e_7^3= \frac{1}{2} \int_{123}^{4} P_{12}^2 \hat Q_{14} \ , & e_{12}^3=   \int_{123}^{4} P_{12} P_{13} \hat Q_{24}  \ \phantom{.} \, \notag \\
   e_3^3 &= \frac{1}{2} \int_{123}^{4} P_{12}^2 P_{13} \co \ \ \ \ \hspace{0.3mm} e_8^3= \frac{1}{2} \int_{123}^{4} P_{12}^2 \hat Q_{34} \ ,&e_{13}^3=   \int_{123}^{4} P_{12} \hat Q_{14} \hat Q_{24} \ \phantom{.} \label{31npl11} \\
    e_4^3 &=  \int_{123}^{4} P_{12} P_{13} P_{23} \co \ \ \ \! e_9^3=   \frac{1}{2} \int_{123}^{4} P_{12} \hat Q_{14}^2 \ , & e_{14}^3=   \int_{123}^{4} P_{12} \hat Q_{14} \hat Q_{34}  \ \phantom{.} \notag \\
    e_5^3 &=  \int_{123}^{4} \hat Q_{14} \hat Q_{24} \hat Q_{34} \co e_{10}^3=  \frac{1}{2} \int_{123}^{4} P_{12} \hat Q_{34}^2 \; .  \notag
\end{align}

\subsection{teMZVs at orders $\ap^{\leq 3}$}

As a next step, we evaluate the above $e^i_j$ in terms of teMZVs, using the
equivalence of the measures $\int_{123}^4$ and $\frac{1}{2}\int^{34}_{12}$
noted in \eqns{eqn:newnewnewrep}{eqn:evennewnewnewrep}.  By largely recycling
the results of \secref{ssec:explicit}, we obtain the following expressions at the first order,
\begin{align}
e_{1}^1	&= \frac{1}{2}d^1_1 =\frac{1}{2}  \omwb{1,0}{0,0}  
 \ , \ \ \ \ \ \ 
e_{2}^1 	=\frac{1}{2}d^1_2 \, \big|_{c(q)\rightarrow 0}=  \frac{1}{2}  \omwb{1,0}{\tauh,0}  \; ,
\label{31eqn:d12}
\end{align}
the following ones at the second order,
\begin{align}   
e^2_1 &= \frac{1}{2}  d^{2}_{1}  \, \big|_{c(q)\rightarrow 0} = \frac{1}{2} \omwb{1 , 1 , 0 }{0 , 0 , 0} 
\ , \ \ \ \ \ \ \ \ \ e^2_{3}= \frac{1}{2}  d^{2}_{4}  \, \big|_{c(q)\rightarrow 0} =   \frac{1}{2} \omwb{1 ,  0 }{0 ,  0}^2
\label{31eqn:d21} \\[8pt]
e^2_2 &= \frac{1}{2}  d^{2}_{2}  \, \big|_{c(q)\rightarrow 0} =  \frac{1}{2}   \omwb{1 , 1 , 0 }{ \tauh, \tauh, 0} 
\ , \ \ \ \ \ \ \ \  \! e^2_4= \frac{1}{2}  d^{2}_{5}  \, \big|_{c(q)\rightarrow 0} =   \frac{1}{2} \omwb{1 , 0 }{\tauh, 0 }^2
\label{31eqn:d22} \\[8pt]
%
%
e^2_5 &= \frac{1}{2}  d^{2}_{3}  \, \big|_{c(q)\rightarrow 0} =\frac{1}{2} \omwb{1 , 0 }{0, 0 }\omwb{1 , 0 }{\tauh, 0 } = e^2_6 \; ,
\label{31eqn:d25}  
\end{align}
and the following ones at the third order:
\begin{align}
e^{3}_{1} & = \frac{1}{2} d^3_1=  \frac{1}{2}  \omwb{1 , 1 , 1 , 0 }{0 , 0 , 0, 0} 
\label{31eqn:d31} \\[8pt]
e^{3}_{2} & = \frac{1}{2} d^3_2  \, \big|_{c(q)\rightarrow 0} =  \frac{1}{2}   \omwb{1 , 1 , 1 , 0}{ \tauh, \tauh, \tauh, 0} 
\label{31eqn:d32} \\[8pt]
e^{3}_{3} &=   \frac{1}{2} d^3_5   =  \frac{1}{2} \omwb{1 , 1 , 0 }{0 , 0 , 0} \omwb{1 ,  0 }{0 ,  0} 
\label{31eqn:d33} \\[8pt]
e^{3}_{4} &=\frac{1}{2} \omwb{1,0}{0,0}^3 - \zeta_2 \, 
\omwb{0 , 1 , 0 , 0 }{0 , 0 , 0 , 0}+ \frac{1}{6} \, 
\omwb{0 , 3 , 0 , 0 }{0 , 0 , 0 , 0}
\label{31eqn:d34} \\[8pt]
e^{3}_{5} &= \frac{1}{2} d^3_{12}  \, \big|_{c(q)\rightarrow 0}   =   \frac{1}{2}   \omwb{1  , 0}{ \tauh , 0}^3
\label{31eqn:d35} \\[8pt]
e^{3}_{6} &= \frac{1}{2} d^3_7  \, \big|_{c(q)\rightarrow 0}  =  \frac{1}{2} \omwb{1 , 1 , 0 }{\tauh, \tauh, 0} \omwb{1 ,  0 }{\tauh,  0} 
\label{31eqn:d37} \\[8pt]
e^3_{7}  &= \frac{1}{2} d^3_3  \, \big|_{c(q)\rightarrow 0}  =  \frac{1}{2}  \omwb{1 , 1 ,  0 }{0 , 0 , 0} \omwb{1  , 0}{ \tauh , 0} = e^3_{8}
\label{31eqn:d38} \\[8pt]
e^3_{9} &= \frac{1}{2} d^3_4  \, \big|_{c(q)\rightarrow 0}  = \frac{1}{2}  \omwb{1 , 1 ,  0 }{\tauh, \tauh, 0} \omwb{1  , 0}{ 0 , 0} = e^3_{10}
\label{31eqn:d39} \\[8pt]
e^3_{11} &= \frac{1}{2} d^3_8  \, \big|_{c(q)\rightarrow 0}  = \frac{1}{2} \omwb{1  , 0}{ 0 , 0}^2 \omwb{1  , 0}{ \tauh , 0} = e^3_{12}
\label{31eqn:d311} \\[8pt]
e^3_{13} &= \frac{1}{2} d^3_{11} \, \big|_{c(q)\rightarrow 0}  = \frac{1}{2} \omwb{1,0}{\tauh,0}^2 \omwb{1,0}{0,0} + \frac{1}{6} \, 
\omwb{0 , 3 , 0 , 0 }{0 , 0 , 0 , 0}
\label{31eqn:d313} \\[8pt]
e^3_{14} &= \frac{1}{2} d^3_9  \, \big|_{c(q)\rightarrow 0}  =  \frac{1}{2} \omwb{1,0}{\tauh,0}^2 \omwb{1,0}{0,0}
 \ .
\label{31eqn:d314}
\end{align}
Expressions for $d_{11}^3$ (cf.~\eqn{eqn:d311}) and $e^3_4$ are derived in  \appref{app:deleven}.

\subsection{Assembling the orders $\ap^{\leq 3}$}

Similar to \eqn{npl12}, momentum conservation\footnote{We have already
  exploited the equalities 
$e^2_5=e^2_6 \, , \ \
e^3_7=e^3_8 \, , \ \  
e^3_9=e^3_{10}$ and $e^3_{11}=e^3_{12}$ 
in simplifying $I_{123|4}(q)$ to the expression in \eqn{simp31int}.} leaves the
following contributions to the integral \eqn{eqn:newnewnewrep} at the orders
$\ap^{\leq 3}$,
\begin{align}
I_{123|4}(q) = \frac{1}{2}\ &+\ (s_{12}^2 + s_{12}s_{23} + s_{23}^2) (2 e^2_1 + 2 e^2_2 - e^2_3 - e^2_4 )
\label{simp31int} \\
 & + \  s_{12}s_{23}(s_{12}+s_{23})\,(-3 e^3_1  - 3 e^3_2 + 
 3 e^3_3 - e^3_4 - e^3_5 + 3 e^3_6  - 3 e^3_{13} + 3 e^3_{14}) + \ {\cal O}(\ap^4) \; ,
 \notag
\end{align}
in agreement with the $\ap^{\leq 1}$ results of \rcite{Hohenegger:2017kqy}.  The
combinations of $e^i_j$ can be expressed in terms of eMZVs
\begin{align}
I_{123|4}(q) \, \big|_{s_{12}^2 + s_{12}s_{23} + s_{23}^2} &=  \omwb{1,1 ,0}{0,0,0} - \frac{1}{2} \omwb{1,0}{0,0}^2 +  \omwb{1,1,0}{\tauh,\tauh ,0} - \frac{1}{2} \omwb{1,0}{\tauh,0}^2 \notag \\
&= \frac{ 7\zeta_2 }{12} + \omega(0,0,2) 
\label{eqn:stuffA}
\\
I_{123|4}(q) \, \big|_{s_{12}s_{23}(s_{12}+s_{23})} &=
\frac{3}{2} \omwb{1,0}{0,0}\omwb{1,1,0}{0,0,0}
-\frac{3}{2} \omwb{1,1,1,0}{0,0,0,0}
 -\frac{1}{2} \omwb{1,0}{0,0}^3 \notag \\
 &\ \ \ \ +\frac{3}{2} \omwb{1,0}{\tauh,0}\omwb{1,1,0}{\tauh,\tauh,0}
-\frac{3}{2} \omwb{1,1,1,0}{\tauh,\tauh,\tauh,0}
 -\frac{1}{2}  \omwb{1,0}{\tauh,0}^3 \notag \\
& \ \ \ \ + \zeta_2 \,\omega(0,1,0,0) -\frac{2}{3}\, \omega(0,3,0,0)  \notag \\
&=  2\, \zeta_2 \,\omega(0,1,0,0) - \frac{5}{6}\, \omega(0,3,0,0)
+\frac{\zeta_3}{4} \; ,
\label{eqn:stuffB}
\end{align}
where we have used the teMZV relations of \appref{app:temzvcomputation}.

\subsection{Summary of the orders $\ap^{\leq 3}$}

Once we insert the simplified expressions \eqns{eqn:stuffA}{eqn:stuffB} for the
combinations of $e^i_j$, the leading low-energy orders of the integral
\eqn{simp31int} boil down to the following eMZVs:
\begin{align}
I_{123|4}(q) & = \frac{1}{2} +(s_{12}^2 + s_{12}s_{23} + s_{23}^2)  \Big( \frac{ 7\zeta_2 }{12} + \omega(0,0,2)  \Big) \notag \\
&+
s_{12}s_{23}(s_{12}+s_{23}) \Big( 2\, \zeta_2 \,\omega(0,1,0,0) - \frac{5}{6}\, \omega(0,3,0,0)
+\frac{\zeta_3}{4}
\Big) + {\cal O}(\ap^4)  \; .
\label{31result}
\end{align}
From the constant terms of the eMZVs gathered in \eqn{exlength3}, our result
\eqn{31result} is consistent with the tree-level expression
\begin{align}
  I_{123|4}(q) &= - \frac{1}{\pi^2} \bigg[
\frac{ \Gamma(s_{12}) \Gamma(s_{23}) }{\Gamma(1+s_{12}+s_{23})} + {\rm cyc}(1,2,3)
 \bigg]
+ {\cal O}(q)\\
&= \frac{1}{2} + \frac{1}{4}\zeta_2 (s_{12}^2 + s_{12}s_{23} + s_{23}^2) + \frac{1}{2} \zeta_3 s_{12}s_{23}(s_{12}+s_{23})
+ {\cal O}(q,\ap^4) \; ,
 \notag 
\end{align}
which is known to arise at the $q^0$ order of the integral $I_{123|4}(q) $ \cite{Green:1981ya}.


\section{Sample integrals in non-planar string amplitudes}
\label{app:deleven}

In this appendix, we will derive the results \eqns{eqn:d311}{31eqn:d34} for
$d^3_{11}$ and $e^3_4$, respectively.  These integrals are the most difficult
cases at the orders $\ap^{\leq 3}$ because the Green functions form closed
subcycles such as $P_{ij}P_{ik}P_{jk}$ and $P_{ij}Q_{ik}Q_{jk}$. Identities
between elliptic iterated integrals will be seen to yield answers in terms of
teMZVs, and the extensions of these manipulations to all weights and lengths
\cite{Broedel:2014vla} guarantee that each term in the low-energy expansion of
the integrals \eqns{eqn:newnewrep}{eqn:newnewnewrep} can be expressed in terms
of teMZVs.

\subsection{The $d^3_{11}$ integral from $P_{ij}Q_{ik}Q_{jk}$}

The contributions to $d^3_{11}$ with at least one factor of $c(q)$ are
equivalent to those in $d^3_9$, so it is sufficient to study
\begin{align}
\int_{12}^{34}P_{34}Q_{13}Q_{14} \, \big|_{c(q)\rightarrow 0} &= 2 \int^1_0 \dd x_4 \int^{x_4}_0 \dd x_3 \, \GLarg{1}{ \tauh}{x_3} \GLarg{1}{ \tauh}{x_4}  \,\bigg( - \int^{x_4}_{x_3} \dd u\ f^{(1)}(u-{x_4}) \bigg) \notag \\
&= - 2 \int^1_0 \dd x_4  \GLarg{1}{ \tauh}{x_4}  \int^{x_4}_0 \dd u \  f^{(1)}(u-{x_4}) \int^u_0  \dd x_3\GLarg{1}{ \tauh}{x_3}
\label{appG1}
\\
&= - 2 \int^1_0 \dd x_4  \GLarg{1}{ \tauh}{x_4} \GLarg{1 &0 &1}{x_4 &0 &\tauh}{x_4}\; . \notag
\end{align}
In the first step, we have exploited that the integration regions with $0\leq
x_3 \leq x_4 \leq 1$ and $0\leq x_4 \leq x_3 \leq 1$ yield the same
contributions by the symmetry of the integrand $P_{34}Q_{13}Q_{14}$ under
exchange of $x_3$ and $x_4$. Moreover, we have used the integral representation
$P_{34}=- \int^{x_4}_{x_3} \dd u\ f^{(1)}(u-{x_4}) $ of the Green function,
reparametrized the integration domain with $0 \leq x_3 \leq u \leq x_4$ and
applied the definition \eqn{eqn:defGell} of elliptic iterated integrals.

The elliptic iterated integral $\GLarg{1 &0 &1}{x_4 &0 &\tau/2 }{x_4}$ in the
last line of \eqn{appG1} is not yet suitable for integration over $x_4$ in its
present form due to the twofold appearance of the integration variable. As
explained in \cite{Broedel:2014vla}, Fay relations among the weighting
functions allow to derive a differential equation in $x_4$ whose integration
yields the alternative representation
\begin{align}
\GLarg{1 &0 &1}{x_4 &0 &\tauh}{x_4}&=
\GLarg{0 &0 &2}{0 &0 &0 }{x_4}
+\GLarg{0 &0 &2}{0 &0 &\tauh}{x_4}
+\GLarg{0 &2 &0}{0 &\tauh&0}{x_4} \notag \\
& \ \ \ \ \ \ \ \
-\GLarg{0 &1 &1}{0 &\tauh&\tauh}{x_4}
-\GLarg{0 &1 &1}{0 &\tauh&0 }{x_4} \ .
\label{appG2}
\end{align}
After shuffle multiplication with $\GLarg{1}{ \tau/2 }{x_4}$, the integral over
$x_4$ in \eqn{appG1} can be readily performed, e.g.
\begin{align}
  \int^1_0 \dd x_4 & \GLarg{1}{ \tauh}{x_4} \GLarg{0 &1 &1}{0 &\tauh&0 }{x_4}  
  \nnl
  &= \int^1_0 \dd x_4  \, \bigg[ \GLarg{1 &0 &1 &1}{\tauh&0 &\tauh&0 }{x_4}  
+2\GLarg{0 &1 &1 &1}{0 &\tauh&\tauh&0 }{x_4}  
+\GLarg{0 &1 &1 &1}{0 &\tauh&0 &\tauh}{x_4}   \bigg] \notag \\
&= \GLarg{0 &1 &0 &1 &1}{0 &\tauh&0 &\tauh&0 }{1}  
+2\GLarg{0 &0 &1 &1 &1}{0 &0 &\tauh&\tauh&0 }{1}  
+\GLarg{0 &0 &1 &1 &1}{0 &0 &\tauh&0 &\tauh}{1}  
\notag \\
&=\omwb{1}{\tauh}
\omwb{1, 1, 0, 0}{0, \tauh, 0, 0 }
 - \omwb{1, 1, 0, 0, 1}{0, \tauh, 0, 0, \tauh}
 \label{appG3}
\end{align}
for the rightmost term in \eqn{appG2}. The shuffle operation in the last step
of \eqn{appG3} together with $\omwb{1}{\tau/2}=0$ reduces the number of terms
and leads to the following end result
\begin{align}
\int_{12}^{34}P_{34}Q_{13}Q_{14} \, \big|_{c(q)\rightarrow 0} &= 2 \omwb{2, 0, 0, 0, 1}{0, 0, 0, 0, \tauh} + 
 2 \omwb{2, 0, 0, 0, 1}{\tauh, 0, 0, 0, \tauh}+ 
 2 \omwb{0, 2, 0, 0, 1}{0, \tauh, 0, 0, \tauh} \notag \\
 & \ \ \ \ \    -  2 \omwb{1, 1, 0, 0, 1}{0, \tauh, 0, 0, \tauh} - 
 2 \omwb{1, 1, 0, 0, 1}{\tauh, \tauh, 0, 0, \tauh} \; .
\label{appG4}
\end{align}
Finally, the expression for $d^3_{11}$ in \eqn{eqn:d311} results from the eMZV
relation
\begin{align}
 & 2 \omwb{2, 0, 0, 0, 1}{0, 0, 0, 0, \tauh} + 
 2 \omwb{2, 0, 0, 0, 1}{\tauh, 0, 0, 0, \tauh}+ 
 2 \omwb{0, 2, 0, 0, 1}{0, \tauh, 0, 0, \tauh} \label{appG4emzv} \\
 & \ \ \ \ \ \ \ \ \ \   -  2 \omwb{1, 1, 0, 0, 1}{0, \tauh, 0, 0, \tauh} - 
 2 \omwb{1, 1, 0, 0, 1}{\tauh, \tauh, 0, 0, \tauh} 
 =
\omwb{1,0}{0,0}\omwb{1,0}{\tauh,0}^2 + \frac{1}{3} \, 
\omwb{0 , 3 , 0 , 0 }{0 , 0 , 0 , 0}  \; .
\notag
\end{align}


\subsection{The $e^3_{4}$ integral from $P_{ij}P_{ik}P_{jk}$}

A similar strategy applies to $e^3_4$ in \eqn{31npl11},
\begin{align}
\int_{123}^{4}P_{12}P_{13}P_{23}&=  \int^1_0 \dd x_3 \int^{x_3}_0 \dd x_2 \, \GLarg{1}{ 0 }{x_2} \GLarg{1}{ 0 }{x_3}  \,\bigg( - \int^{x_3}_{x_2} \dd u\ f^{(1)}(u-{x_3}) \bigg) \notag \\
&= -  \int^1_0 \dd x_3  \GLarg{1}{ 0 }{x_3}   \GLarg{1 &0 &1}{x_3 &0 &0 }{x_3}  \ ,
\label{appG6}
\end{align}
where the relevant identity among elliptic iterated integrals reads
\begin{align}
\GLarg{1 &0 &1}{x_3 &0 &0 }{x_3}=
2\GLarg{0 &0 &2}{0 &0 &0 }{x_3}
+\GLarg{0 &2 &0}{0 &0 &0}{x_3}
-2\GLarg{0 &1 &1}{0 &0 &0 }{x_3}
+\zeta_2 \GLarg{0}{0}{x_3}  \ .
\label{appG8}
\end{align}
Note the extra term $\zeta_2 \GLarg{0}{0}{x_3}$ in comparison to the analogous
identity \eqn{appG2}. Upon insertion into \eqn{appG6}, we obtain 
\begin{align}
  \int_{123}^{4}P_{12}P_{13}P_{23}&= 2 \omm(2, 0, 0, 0, 1) + 
  \omm(0, 2, 0, 0, 1)   -  2 \omm(1, 1, 0, 0, 1) - \frac{1}{2} \zeta_2 \omm(1 ,  0 )
 \; ,
\label{appG9}
\end{align}
which translates into the expression \eqn{31eqn:d34} for $e^3_4$ by the eMZV
relation
\begin{align}
  &2
  \omm(2, 0, 0, 0, 1)
  +\omm(0, 2, 0, 0, 1) 
  -  2 \omm(1, 1, 0, 0, 1)
  - \frac{1}{2} \zeta_2 \omm(1 ,  0)
   \notag \\
 & \ \ \ =\frac{1}{2} \omm(1,0)^3
  - \zeta_2  \omm(0,1,0,0) + \frac{1}{6}  \omm(0,3,0,0) \; .
\label{fromdatabase}
\end{align}


\section{Some all-order contributions}
\label{app:allorder}

For certain contributions to the integral \eqn{eqn:itintB} in the non-planar
open-string amplitude, closed-form expressions in terms of \temzv{}s can be
given to all orders in the $\alpha'$-expansion.  In the context and notation of
\subsecref{ssec:explicit}, we find 
\begin{align}
d_{n;1} &= \frac{1}{n!} \int_{12}^{34} P(x_{12})^n = \frac{ 1 }{ n! } \int_0^1 \dd x_2 \prod_{i=1}^n \int_0^{x_2} \dd y_i \; f^{(1)}(y_i) 
= \omwb{1, \dots , 1  ,0}{ 0,\dots , 0 ,0} \smash{\hspace{-1.65cm} \underbrace{\phantom{ \big ( abcd \, } }_{n \text{ times}}}
\label{eqn:dn1} \\[10pt]
d_{n;2} &= \frac{1}{n!} \int_{12}^{34} Q(x_{13})^n = 
\frac{ 1 }{ n! }  \int_{12}^{34} \left( \sum_{r=0}^n \binom{ n }{ r } c(q)^{n-r} \GLarg{1}{ \tauh}{x_3}^r \right) \notag \\
& = \sum_{r=0}^n \frac{1}{(n-r)!} c(q)^{n-r}
\omwb{1, \dots , 1  ,0}{ \tauh,\dots , \tauh,0} \smash{\hspace{-2.05cm} \underbrace{\phantom{ \big( \qquad \quad \, } }_{r \text{ times}}}
\label{eqn:dn2} \\[10pt]
d_{n,m;1} &= \frac{1}{n! m!} \int_{12}^{34} P(x_{12})^n P(x_{34})^m = d_{n;1} d_{m;1} \notag  \\
&= 
\omwb{1, \dots , 1  ,0}{ 0,\dots , 0 ,0} \smash{\hspace{-1.65cm} \underbrace{\phantom{ \big( abcd \, } }_{n \text{ times}}}  \qquad
\omwb{1, \dots , 1  ,0}{ 0,\dots , 0 ,0} \smash{\hspace{-1.65cm} \underbrace{\phantom{ \big( abcd \, } }_{m \text{ times}}}
\label{eqn:dnm1} \\[10pt]
d_{n,m;2} &= \frac{1}{n! m!} \int_{12}^{34} P(x_{12})^n  Q(x_{13})^m = d_{n;1} d_{m;2} \notag \\
&= \omwb{1, \dots , 1  ,0}{ 0,\dots , 0 ,0} \smash{\hspace{-1.65cm} \underbrace{\phantom{ \big( abcd \, } }_{n \text{ times}}}
\quad \; \; \sum_{r=0}^m \frac{1}{(m-r)!} c(q)^{m-r}
\omwb{1, \dots , 1  ,0}{ \tauh,\dots , \tauh,0} \smash{\hspace{-2.05cm} \underbrace{\phantom{ \big( \qquad \quad \,  } }_{r \text{ times}}}
\label{eqn:dnm2} \\[10pt]
d_{n,m;3} &= \frac{1}{n! m!} \int_{12}^{34} Q(x_{13})^n  Q(x_{14})^m = d_{n;2} d_{m;2}
\label{eqn:dnm3} \\[10pt]
d_{n,m,p;1} &= \frac{1}{n! m! p!} \int_{12}^{34} P(x_{12})^n P(x_{34})^m  Q(x_{13})^p = d_{n;1} d_{m,p;2} = d_{n;1} d_{m;1} d_{p;2} 
\label{eqn:dnmp1} \\[10pt]
d_{n,m,p;2} &= \frac{1}{n! m! p!} \int_{12}^{34} P(x_{12})^n Q(x_{13})^m  Q(x_{14})^p = d_{n;1} d_{m,p;3} = d_{n;1} d_{m;2} d_{p;2} \ .
\label{eqn:dnmp2}
\end{align}

\bibliographystyle{nplan}
\bibliography{nplan}

\end{document}